\documentclass{ieeeaccess}
\usepackage{cite}
\usepackage{amsmath,amssymb,amsfonts} 
\usepackage{graphicx}
\usepackage{textcomp}
\usepackage{caption}
\usepackage[ruled]{algorithm2e}
\usepackage{multirow}
\usepackage{algpseudocode}
\usepackage{setspace}
\usepackage{multicol, blindtext}
\usepackage{float}
\usepackage{breqn}
\usepackage{blindtext} 
\usepackage{placeins} 
\usepackage{subfigure}  
\usepackage{csquotes}     
\usepackage{setspace}
\usepackage{dblfloatfix}  
\usepackage{xurl} 
\def\BibTeX{{\rm B\kern-.05em{\sc i\kern-.025em b}\kern-.08em
    T\kern-.1667em\lower.7ex\hbox{E}\kern-.125emX}}

\def\BibTeX{{\rm B\kern-.05em{\sc i\kern-.025em b}\kern-.08em
    T\kern-.1667em\lower.7ex\hbox{E}\kern-.125emX}}

\begin{document}
\history{Date of publication xxxx 00, 0000, date of current version xxxx 00, 0000.}
\doi{10.1109/ACCESS.2017.DOI}

\title{From Data to Decisions: The Power of Machine Learning in Business Recommendations}

\author{\uppercase{Kapilya Gangadharan}\authorrefmark{1}, ~\uppercase{Anoop Purandaran}\authorrefmark{2}, ~\uppercase{K. Malathi}\authorrefmark{3}, ~\uppercase{Barathi Subramanian}\authorrefmark{4}, ~\uppercase{Rathinaraja Jeyaraj}\authorrefmark{5}, and ~\uppercase{Soon Ki Jung}\authorrefmark{6}}

\address[1,3]{Saveetha School of Engineering, Saveetha Institute of Medical and Technical Sciences, Chennai, India. (E-mail: kapilya@gmail.com, malathi.infotech@gmail.com)} 
\address[2]{Lowes Companies Inc, Charlotte, NC, USA. (Email: anooppurandaran@gmail.com)} 
\address[4,6]{Kyungpook National University, Daegu, South Korea. (Email: barathi.sn93@gmail.com, skjung@knu.ac.kr)} 
\address[5]{University of Huston-Victoria, Texas, USA; (Email: jrathinaraja@gmail.com)} 
%\tfootnote{This paragraph of the first footnote will contain support 
%information, including sponsor and financial support acknowledgment. For 
%example, ``This work was supported in part by the U.S. Department of 
%Commerce under Grant BS123456.''}
 
\corresp{Corresponding author: Soon Ki Jung (E-mail: skjung@knu.ac.kr).} 

\begin{abstract}
This research aims to explore the impact of machine learning (ML) on the evolution and efficacy of recommendation systems (RS), particularly in the context of their growing significance in commercial business environments. Methodologically, the study delves into the role of ML in crafting and refining these systems, focusing on aspects such as data sourcing, feature engineering, and the importance of evaluation metrics, thereby highlighting the iterative nature of enhancing recommendation algorithms. The deployment of recommendation engines (RE), driven by advanced algorithms and data analytics, is explored across various domains, showcasing their significant impact on user experience and decision-making processes. These REs not only streamline information discovery and enhance collaboration, but also accelerate knowledge acquisition, which is vital in navigating the digital landscape for businesses. They contribute significantly to sales, revenue, and the competitive edge of enterprises by offering improved recommendations that align with the individual needs of the customer. The research identifies the growing expectations of users for a seamless and intuitive online experience, where content is personalized and dynamically adapted to changing preferences. Future research includes exploring advances in deep learning models, ethical considerations in the deployment of RS, and addressing scalability challenges. This study emphasizes the indispensability of comprehending and using ML in RS for researchers and practitioners to tap into the full potential of personalized recommendation in commercial business prospects.
\end{abstract}


\titlepgskip=-30pt

\maketitle

\begin{figure*}[b!] 
	\centering 
	\includegraphics[width=0.65\textwidth]{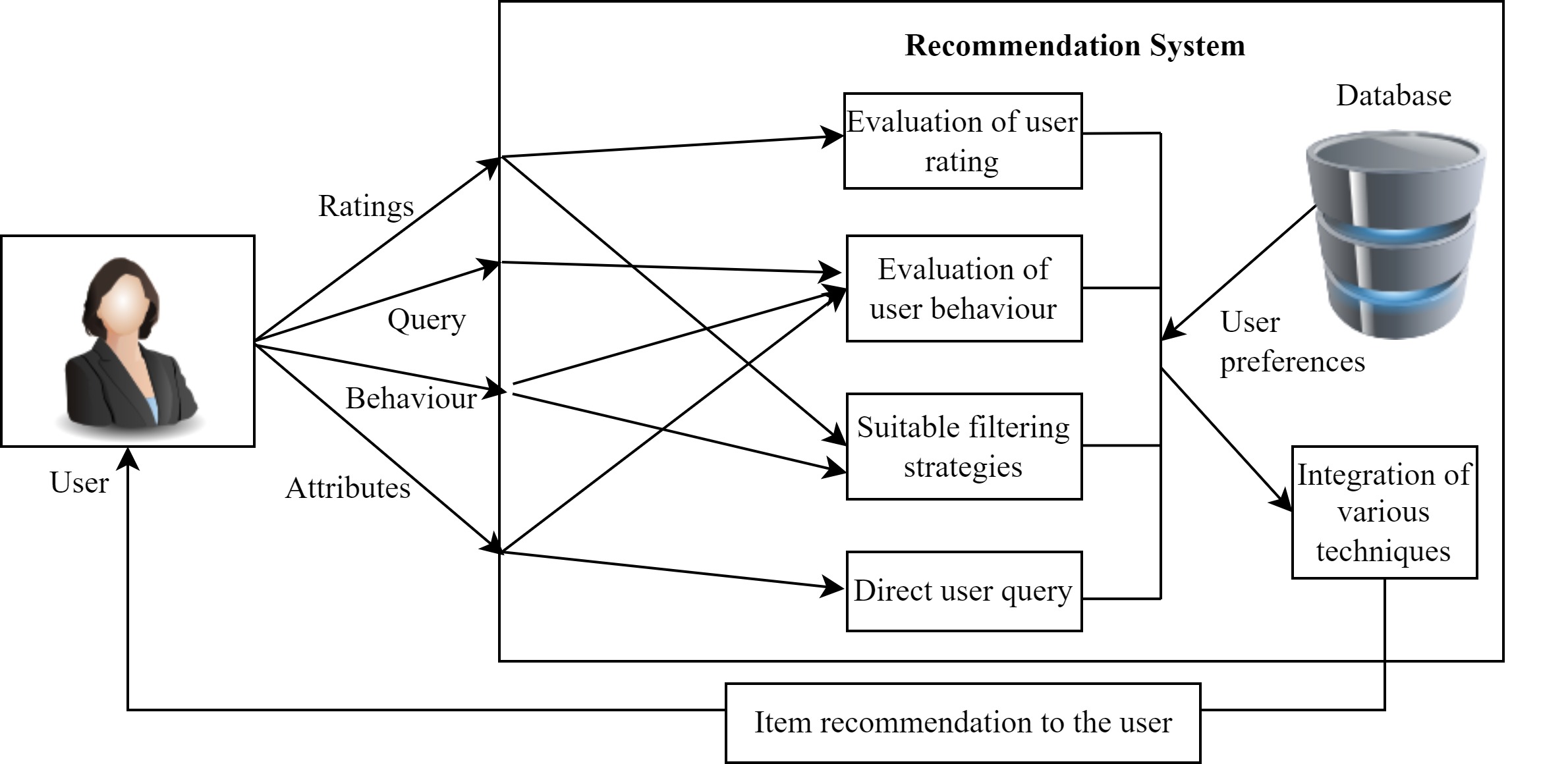}
	\\
	\caption{\textrm{Structure of a typical RS}}
	\label{fig:1}       
\end{figure*}

\section{Introduction}
\label{intro} 
Due to the growing popularity of the Internet, changing personalization trends, and evolving user behaviors, recommendation systems (RS) have become highly effective tools for filtering and recommending online content. These systems are a subclass of information filtering systems that aim to predict user preferences or recommend items [1] (such as products, services or content) based on past behavior and patterns. It plays a crucial role in helping users discover and engage with the most relevant and appealing content. Recommendation engines (RE) are designed in such a way that simplifies the process of finding relevant information on the Internet and reduces the user's effort and time required for searching. The field of pervasive computing has seen significant advancements in recent years, leading to an online data overload problem. Pervasive computing, which involves embedding computational intelligence into everyday objects and environments, [2] has generated vast amounts of data from various sources, including sensors, devices, and applications. This data overload has made it challenging for users to find relevant and useful content on the Internet. To address this issue, several procedures and technologies with lower computational requirements have been developed. Such systems aid users in managing the overwhelming amount of online data by providing tailored content recommendations aligned with individual interests and preferences. By doing so, they enhance the user experience, [3] increase user engagement, and help users discover the most relevant content. Among the solutions to tackle data overload and help users discover content, RS has gained significant attention. The RS have become increasingly popular across a range of sectors, including e-commerce, content streaming, social media, movie [4] and retail [5]. Top examples of how RE are used to improve user experience and propel corporate success are Netflix [6] [7], Amazon [8], YouTube [9], etc., key details regarding their application in these industries are as follows:   \\[0.1cm]
\textit{Personalized recommendations} - It is a core component of many online platforms and services. They involve analyzing user data, including past purchase history, search queries, browsing behavior, demographic information, and preferences to provide personalized product recommendations. This enhances the user experience by helping customers discover relevant products and make informed purchasing decisions based on online historical interactions. They have proven useful not only in commercial markets like e-commerce and retail but also in a wide range of other industries, including psychology, mathematics, Academic Research [10], E-Learning [11], Job Matching [12], tourism [13] and many more.  \\[0.1cm]
\textit{Increased user engagement} - By offering personalized recommendations, companies can keep users engaged on their platforms for longer periods. Users are more likely to find products of interest [14], leading to potentially more sales. It can lead to higher user satisfaction, longer user retention, and ultimately, more successful businesses. \\[0.1cm]
\textit{Enhanced sales} -  RE can [15] significantly impact a company's bottom line by driving sales. When customers are presented with items, they are more likely to purchase, conversion rates increase, and the average order value may rise as well. Sales often require a multifaceted approach that combines various strategies and continuous optimization to meet the ever-changing needs and preferences of the target audience. It's important to  [16] monitor the results of the efforts and adapt the sales strategies accordingly to achieve sustained growth. \\[0.1cm]
\textit{Inventory management} -  It is a critical aspect of running a successful business. Properly managing the inventory can have a significant impact on profitability, customer satisfaction, and overall business efficiency. RS can also help the [17] business manage its inventory more efficiently. By promoting specific products based on demand and user preferences, companies can optimize their stock levels and reduce overstock or understock situations. \\[0.1cm]
\textit{User retention} - It is an essential component of building a successful and sustainable business, especially in the digital era. Retaining existing customers is often more cost-effective than acquiring new ones, and loyal customers can become brand advocates. The RS offers an enhanced [18] and an engaging shopping experience. Satisfied customers are more likely to return and make repeat purchases, contributing to long-term customer loyalty. \\[0.1cm]
\textit{Continuous improvement} - It involves a combination of data management, algorithm development, user feedback analysis, and technical optimization. Regular evaluation and adaptation are essential to ensure the system remains relevant and valuable [19] to users while meeting business objectives. For RS to function, users must engage with items, providing them with ratings or feedback. These ratings may be gathered using explicit or implicit techniques, or occasionally both. Explicit ratings are obtained directly from users through their explicit actions, such as selecting a star rating (e.g., 1 to 5 stars). Implicit ratings are collected indirectly from user interactions with items, often without the user explicitly assigning a rating. Many modern RSs use hybrid approaches that combine both explicit and implicit data. By blending explicit and implicit feedback [20], these systems aim to provide more accurate and personalized recommendations. Most companies are constantly refining their RS based on user feedback. They use machine learning (ML) techniques and data analysis to update and improve their recommendation algorithms, ensuring the recommendations remain accurate and relevant. \par 

The working principle of RS, as depicted in Figure 1, involves a sequence of steps beginning with user interactions within the system. These interactions are processed by various system components, such as user behavior analysis, collaborative filtering (CF), and the analysis of ratings and direct queries. Notably, some user inputs are processed by multiple components simultaneously. These components collectively contribute to building a user model [21], which Is key in predicting the user's preferences across a range of items. Some models, like CF or content-based filtering, additionally utilize information about other users and additional data to enhance recommendation accuracy. The system integrates the models  [22] to formulate precise user recommendations. In certain scenarios, the system may prioritize certain models over others based on their effectiveness in certain contexts. For example, in situations with a limited number of users, CF might be less preferred, demonstrating the system's adaptive capability to various user and context-specific scenarios. \par 

ML-powered RS has completely changed how consumers find goods, services, and content in the digital era. These algorithms, which provide users with individual recommendations for products on e-commerce websites, movies [23] on streaming services, or relationships on social networks, have become essential to many online platforms. Efficiency is a crucial consideration, especially for real-time or large-scale RS. Depending on the algorithm and data source, there may be trade-offs between computational resources, recommendation quality, and scalability. Therefore, it's essential to choose an algorithm and data source that aligns with the goals and constraints of the specific RS. Additionally, ongoing monitoring, evaluation, and optimization [24]  are important to ensure that the system continues to perform efficiently and effectively. The way that businesses interact with their clients has been drastically enhanced by these platforms. By providing individualized content or products, they not only improve the user experience but also boost customer satisfaction, conversion rates, and user engagement. These systems continue to evolve, incorporating more advanced algorithms and techniques [25] to provide even more accurate and effective recommendations. Considering this wide range of research importance, this article covers the commercial business opportunities for implementing ML-based RS and gives an overview of the research and insights into these systems. Moreover, the assessment of recommendation algorithms has gained significance in recent times, particularly with the accessibility of extensive datasets such as MovieLens. Therefore, this study offers a thorough comparison of several recommendation algorithms by evaluating how efficiently they perform on key metrics including F1-score, accuracy, precision, and recall. These criteria were selected because they offer an accurate evaluation of the algorithms' stability and effectiveness in various conditions. This study attempts to provide useful perspectives regarding the viability of various models to perform specific recommendation tasks by focusing on these criteria.   \par 

The article is organized as follows, as shown in Figure 2: Section 2 briefly describes the most important methods and related work followed in previously published research papers on RS.  Section 3 describes the various strategies for constructing RS. Section 4 briefs how data is prepared for ML-powered RE. Section 5 gives tips on how RE has become an indispensable tool for business. Section 6 explains the Impact of Data on how RE Propelled Business Expansion. Section 7 explains the future hold of the RS and Section 8 Concludes the research insights on RS. Appendix A contains a list of abbreviations utilized throughout this article.  

\begin{figure}[t!] 
	\centering 
	\includegraphics[width=0.48\textwidth]{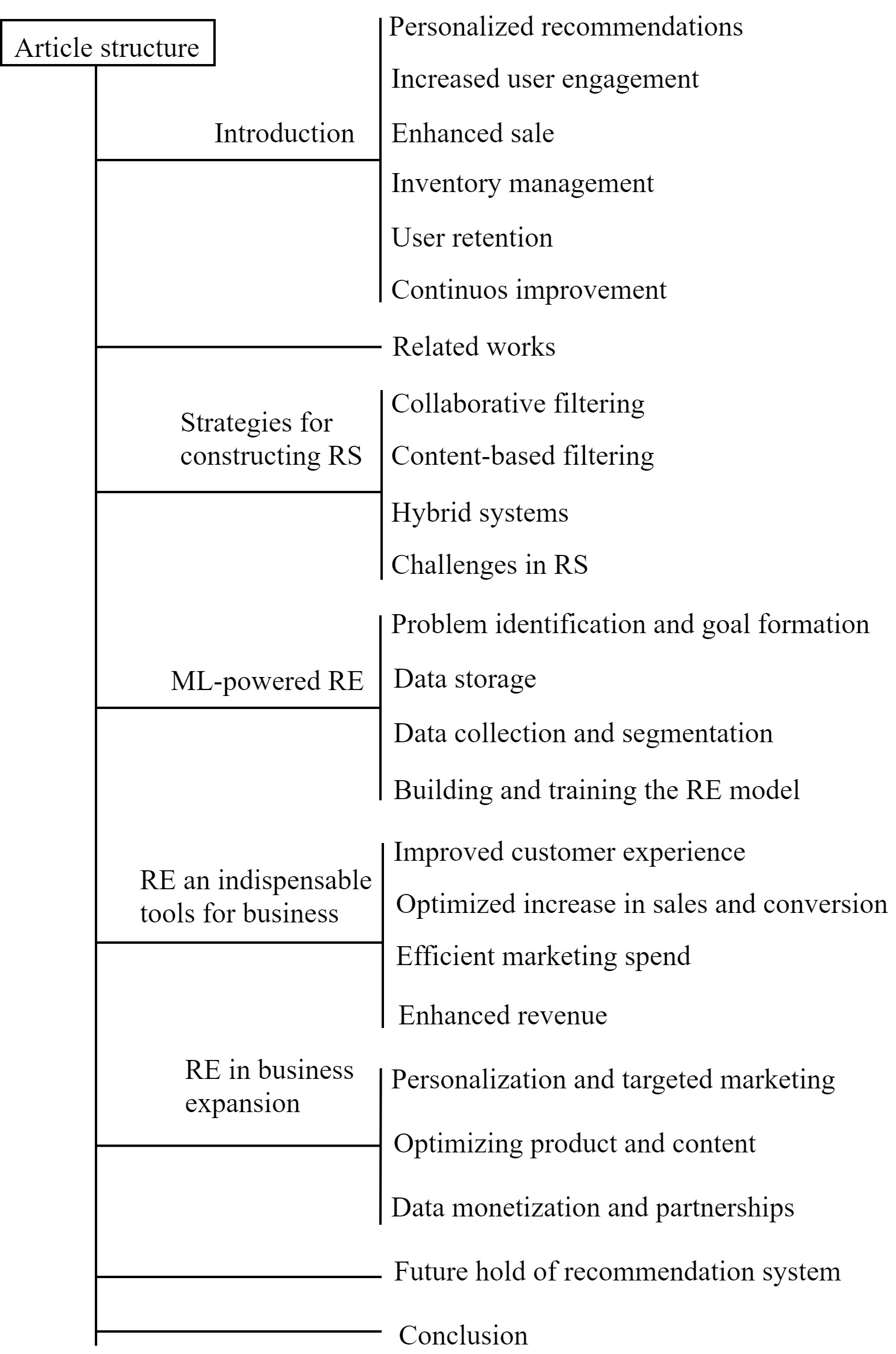}
	\\
	\caption{\textrm{Structure of the article}}
	\label{fig:2}       
\end{figure}

\section{Related works}
In this section, we discuss some of the significant key research findings related to ML-based RS from a business perspective. Pal and Counts et al. (2011) [26] performed a study employing features such as the count of original tweets, conversational tweets, and re-tweets by a tweeter to assess their authority across various topics. A Gaussian Mixture Model was utilized to calculate the authority score for each tweeter. The authority score, as defined in the paper, facilitated the establishment of a total ordering under certain conditions for all users. To validate their findings, the researchers surveyed to obtain user ratings on the authority of tweeters, utilizing this information as the ground truth for authority ranking. This study highlights the progressive improvement of business growth by illustrating the use of social media analytics to discern influential voices within various domains. Lorenzi et al. (2011) [27] introduced an \enquote{assumption-based multiagent} for the tourism industry represents a sophisticated ML approach to personalizing travel package recommendations by considering user preferences within the tourism industry. This system effectively executed various tasks, including the discovery, filtering, and integration of specific information to create better travel packages that aligned with user requirements. Hu et al. (2012) [28] designed a system to suggest songs that align with the user's preferences, introducing novel tracks to the user, and matching their listening patterns. This is achieved by assessing the freshness of a song through the application of the \enquote{Forgetting Curve} and evaluating user preferences based on their activity log. The user's listening pattern is carefully analyzed to gauge their level of interest in the upcoming song recommendation. This approach ensures that the recommendations not only cater to the user's established preferences but also introduce a refreshing element to enhance their overall music-listening experience, thereby driving user retention and business revenue.   \par 

Bobadilla et al. 2013, [29] characterized RS as the ability to enhance user experience contribute to overall enterprise revenue and decision-making and hold a pivotal role in the realm of technology leading to higher sales and competitive advantage. Aher and Lobo et al. 2013 [30] developed an e-learning system that recommends courses, leveraging data mining techniques including simple K-means clustering and the association rule mining (ARM) algorithm. The effectiveness of the proposed system was successfully demonstrated within the context of massive open online courses (MOOCs) exemplifying the application of ML in personalized educational content. This personalization leads to better user engagement and satisfaction, which, in turn, enhances the reputation and adoption of e-learning platforms, contributing to the educational sector's business growth. Yang et al. 2014 [31], Lu et al. (2015) [32], [33] these systems have garnered significant attention from researchers, leading to numerous reviews. They delve into diverse aspects of RS, ranging from their features to the algorithms employed, reflecting the comprehensive exploration and analysis conducted in this dynamic field facilitating advancements in personalized user experiences and, consequently, business models based on user-centric services. Liu et al. (2014) [34] implemented a novel route recommendation model to provide personalized and real-time route suggestions for self-driven tourists, aiming to minimize queuing time and navigate around traffic congestions at popular tourist destinations. The recommendations are enhanced based on individual user preferences it not only improve customer satisfaction but also encourage repeat usage, benefiting businesses in the tourism and travel industry by attracting more customers seeking personalized travel experience. Diao et al. (2014) [35] proposed a joint probabilistic model, named JMARS, combining CF and topic modeling to capture user interest distribution and movie content distribution for improved movie recommendations showcasing the power of ML algorithms. By capturing user interest and movie content distributions more accurately, the system enhances user engagement with the platform, leading to increased viewership and potentially higher revenue for movie recommendation platforms. Li et al. (2015) [36] designed a RS based on fuzzy linguistic modeling to aid users in locating experts within the field of knowledge management systems. This system was successfully implemented in the aircraft industry, highlighting the impact of ML algorithms on professional networking and expertise location. This system's success in real-world scenarios illustrates how ML-based recommendations can streamline operations, enhance decision-making processes, and improve productivity, leading to business efficiencies and growth.

A recommender method was proposed by Lin and Jingtao (2015) [37] for online shopping platforms. This innovative approach utilizes various client contextual information, including access, click, read, and purchase data, to ascertain the preference degree for each item. Items with higher preference degrees are then recommended to the customer. Notably, the focus of this paper was primarily on purchase and click actions, with other actions such as browsing, rating, and searching deemed less significant for consideration. It effectively tackles the sparsity problem in RS, improving personalized marketing efforts and potentially increasing sales by recommending items with higher preference degrees. However, the method's limitation in addressing other RS challenges suggests room for further enhancement in personalization and user engagement strategies. Wen et al. (2015) [38] identified that Individuals with similar interests in social networks are thought to be more likely to buy things connected to those interests. Existing solutions, on the other hand, often presume a static and fixed structure for the social network and users' buying preferences. This contrasts with the dynamic character of genuine social networks, in which major changes occur over time and members' buying interests alter in response to changing preferences. This observation underlines the importance of adapting RS to the fluid nature of user interests and social connections, offering businesses insights into leveraging social network dynamics for targeted marketing and sales strategies, thereby enhancing customer relationship management, and driving sales through social commerce. Li et al. (2016) [39] developed a movie-based RS leveraging user feedback gathered from microblogs and social networks. The system utilizes the sentiment-aware association rule mining algorithm to generate recommendations, drawing insights from frequent program patterns, program metadata similarity, and program view logs. This approach not only enhances the accuracy of recommendations by incorporating user sentiments but also provides entertainment platforms with a sophisticated tool to increase viewer engagement and satisfaction, directly contributing to customer retention and revenue generation through personalized content offerings. Kang and Lee et al. (2017) [40] proposed a model that has the idea of a user modeling framework that transforms a user's social media output into relevant categories within news media platforms. The framework utilizes Wikipedia as a knowledge source to build a comprehensive user profile that captures the individual's different interests and preferences. This model uses ML to understand and categorize user interests across various content platforms, enabling news and media businesses to offer highly personalized content. This strategy can significantly improve user engagement and loyalty, leading to increased viewership and potential advertising revenue. \par 

\begin{table*}[t!]\centering
\caption{\textrm{Related works on RS}}
\small
\label{tab:1}       % Give a unique label 
% \resizebox{\linewidth}{!}{%
\begin{tabular}{|c|l|l|l|l|l|}
\hline
\textbf{Article} & \multicolumn{1}{c|}{\textbf{Focus}}                                     & \multicolumn{1}{c|}{\textbf{Contribution}}                                         & \multicolumn{1}{c|}{\textbf{Method}}                                     & \multicolumn{1}{c|}{\textbf{Domain}} & \multicolumn{1}{c|}{\textbf{Outcome}}                                        \\ \hline
{[}26{]}         & \begin{tabular}[c]{@{}l@{}}Social media \\ authority\end{tabular}       & \begin{tabular}[c]{@{}l@{}}Gaussian mixture \\ model\end{tabular}                  & User survey                                                              & Social                               & \begin{tabular}[c]{@{}l@{}}User order \\ establishment\end{tabular}          \\ \hline
{[}36{]}         & \begin{tabular}[c]{@{}l@{}}Expert location \\ in KM\end{tabular}        & \begin{tabular}[c]{@{}l@{}}Fuzzy linguistic \\ modeling\end{tabular}               & \begin{tabular}[c]{@{}l@{}}Aircraft industry \\ use\end{tabular}         & Expert Rec.                          & \begin{tabular}[c]{@{}l@{}}Real-world \\ efficiency\end{tabular}             \\ \hline
{[}27{]}         & Travel packages                                                         & Multiagent system                                                                  & \begin{tabular}[c]{@{}l@{}}User preference \\ consideration\end{tabular} & Tourism                              & \begin{tabular}[c]{@{}l@{}}Integrated travel \\ information\end{tabular}     \\ \hline
{[}28{]}         & \begin{tabular}[c]{@{}l@{}}Music record \\ system\end{tabular}          & \begin{tabular}[c]{@{}l@{}}Forgetting curve, \\ activity log\end{tabular}          & \begin{tabular}[c]{@{}l@{}}User pattern \\ analysis\end{tabular}         & Music                                & \begin{tabular}[c]{@{}l@{}}Fresh track \\ records\end{tabular}               \\ \hline
{[}29{]}         & \begin{tabular}[c]{@{}l@{}}Record systems' \\ role\end{tabular}         & -                                                                                  & -                                                                        & General                              & \begin{tabular}[c]{@{}l@{}}User experience \\ and revenue boost\end{tabular} \\ \hline
{[}30{]}         & \begin{tabular}[c]{@{}l@{}}E-learning course \\ records\end{tabular}    & \begin{tabular}[c]{@{}l@{}}K-means, associat-\\ ion rule mining\end{tabular}       & \begin{tabular}[c]{@{}l@{}}MOOCs \\ application\end{tabular}             & E-learning                           & Course records                                                               \\ \hline
{[}31{]}         & Record system                                                           & \begin{tabular}[c]{@{}l@{}}Comprehensive \\ exploration\end{tabular}               & Literature review                                                        & General                              & \begin{tabular}[c]{@{}l@{}}Features and \\ algorithms insight\end{tabular}   \\ \hline
{[}34{]}         & \begin{tabular}[c]{@{}l@{}}Tourist route \\ records\end{tabular}        & \begin{tabular}[c]{@{}l@{}}Personalized \\ route model\end{tabular}                & \begin{tabular}[c]{@{}l@{}}User preference \\ analysis\end{tabular}      & Tourism                              & \begin{tabular}[c]{@{}l@{}}Queue and traffic \\ minimization\end{tabular}    \\ \hline
{[}35{]}         & \begin{tabular}[c]{@{}l@{}}Movie records \\ system\end{tabular}         & \begin{tabular}[c]{@{}l@{}}Collaborative \\ filtering and topic\end{tabular}       & \begin{tabular}[c]{@{}l@{}}Interest and \\ content analysis\end{tabular} & Movies                               & Improved records                                                             \\ \hline
{[}37{]}         & \begin{tabular}[c]{@{}l@{}}Online shopping \\ records\end{tabular}      & \begin{tabular}[c]{@{}l@{}}Contextual infor-\\ mation usage\end{tabular}           & \begin{tabular}[c]{@{}l@{}}Purchase and \\ click focus\end{tabular}      & E-commerce                           & \begin{tabular}[c]{@{}l@{}}Sparsity problem \\ address\end{tabular}          \\ \hline
{[}38{]}         & \begin{tabular}[c]{@{}l@{}}Social network \\ product rec.\end{tabular}  & \begin{tabular}[c]{@{}l@{}}Dynamic network \\ and preference\end{tabular}          & -                                                                        & Social                               & \begin{tabular}[c]{@{}l@{}}Social dynamic \\ adaptation\end{tabular}         \\ \hline
{[}39{]}         & \begin{tabular}[c]{@{}l@{}}Movie records \\ from feedback\end{tabular}  & \begin{tabular}[c]{@{}l@{}}Sentiment-aware \\ association rule mining\end{tabular} & \begin{tabular}[c]{@{}l@{}}Program patt-\\ ern analysis\end{tabular}     & Movies                               & \begin{tabular}[c]{@{}l@{}}Personalized \\ records\end{tabular}              \\ \hline
{[}40{]}         & \begin{tabular}[c]{@{}l@{}}News user \\ modeling\end{tabular}           & \begin{tabular}[c]{@{}l@{}}Social media to \\ news categories\end{tabular}         & \begin{tabular}[c]{@{}l@{}}Wikipedia for \\ profiles\end{tabular}        & News media                           & \begin{tabular}[c]{@{}l@{}}User interest \\ capture\end{tabular}             \\ \hline
{[}41{]}         & \begin{tabular}[c]{@{}l@{}}Music streaming \\ records\end{tabular}      & \begin{tabular}[c]{@{}l@{}}Activity-based \\ system\end{tabular}                   & ML features                                                              & Music                                & \begin{tabular}[c]{@{}l@{}}Personalized \\ music records\end{tabular}        \\ \hline
{[}42{]}         & \begin{tabular}[c]{@{}l@{}}Graph-based \\ link prediction\end{tabular}  & \begin{tabular}[c]{@{}l@{}}User-topic relatio-\\ nship model\end{tabular}          & \begin{tabular}[c]{@{}l@{}}Graph info \\ dimensions\end{tabular}         & Social                               & \begin{tabular}[c]{@{}l@{}}Link prediction \\ accuracy\end{tabular}          \\ \hline
{[}43{]}         & \begin{tabular}[c]{@{}l@{}}E-commerce \\ product records\end{tabular}   & \begin{tabular}[c]{@{}l@{}}Clustering repres-\\ entation\end{tabular}              & \begin{tabular}[c]{@{}l@{}}RNN and att-\\ ention\end{tabular}            & E-commerce                           & \begin{tabular}[c]{@{}l@{}}Scalability and \\ diversity\end{tabular}         \\ \hline
{[}44{]}         & \begin{tabular}[c]{@{}l@{}}Personality-\\ based product\end{tabular}    & \begin{tabular}[c]{@{}l@{}}Meta route and \\ interest mining\end{tabular}          & \begin{tabular}[c]{@{}l@{}}Deep model \\ comparison\end{tabular}         & E-commerce                           & \begin{tabular}[c]{@{}l@{}}Accurate \\ product records\end{tabular}          \\ \hline
{[}45{]}         & \begin{tabular}[c]{@{}l@{}}Dynamic goods \\ records system\end{tabular} & \begin{tabular}[c]{@{}l@{}}Reinforcement \\ learning\end{tabular}                  & \begin{tabular}[c]{@{}l@{}}Real-time user\\  preference\end{tabular}     & E-commerce                           & \begin{tabular}[c]{@{}l@{}}Responsive \\ user records\end{tabular}           \\ \hline
\end{tabular}
% }
\end{table*}

A.L Murciego et al. (2017) [41] introduced smartphone activity utilization, a music streaming RS to facilitate activity detection from mobile signals, the proposed system benefited from feature selection methodologies combined with ML techniques such as support vector machine (SVM), and random forest (RF), instance-based k-nearest neighbor (IBK), multilayer perception (MLP), and naive bayes (NB). This approach enabled the music streaming services to offer dynamic and context-aware recommendations, enhancing user experience and potentially increasing platform engagement and subscription rates by aligning music recommendations with the user's current activities. Zarrinkalam et al. (2019) [42] developed a graph-based link prediction scheme a representation model derived from three categories of information: user explicit and implicit contributions to topics, relationships between users, and the similarity among topics. This approach aimed to enhance link prediction accuracy by considering multiple dimensions of information in a graph-based framework. This model benefits businesses by enabling more accurate predictions of user connections and interests, facilitating targeted marketing strategies, and enhancing network analysis for social media platforms, ultimately driving user engagement and advertising revenue. Wang et al. (2020) [43] developed a personalized e-commerce product RS based on learning clustering representation. In contrast to the limitations of the traditional kNN approach in selecting nearby object sets, this system introduced a neighbor factor, a time function, and a dynamic selection model to enhance the choice of nearby object sets. The integration of recurrent neural networks (RNN) and attention mechanisms was employed to build the e-commerce product RS. This system addresses scalability and diversity issues in product recommendations, offering e-commerce platforms a powerful tool to improve personalization, increase conversion rates, and enhance customer satisfaction through personalized product suggestions. Dhelimet al. 2021 [44] introduced a product recommendation model based on personality including meta route discovery and user interest mining approaches. This model's findings outperformed session-based and deep-learning models, demonstrating its effectiveness in giving personalized and accurate product recommendations. This system significantly benefits online retailers by improving recommendation accuracy, reducing decision fatigue among consumers, and driving sales through highly personalized shopping experiences. Ke et al. (2021) [45] introduced a dynamic goods RS based on reinforcement learning. This system could learn from reduced entropy loss error in real-time applications, making it adaptable and responsive to evolving user preferences and environmental changes. This approach allows businesses to dynamically adjust recommendations, ensuring high relevance and user engagement. It is particularly beneficial for businesses in rapidly changing markets, as it supports the continuous learning and adaptation of recommendation strategies to maintain user interest and satisfaction.  These studies collectively highlighted the transformative impact of ML on RS across various industries, demonstrating how advanced ML techniques can be utilized to enhance business outcomes through improved personalization, user engagement, and operational efficiency. These methods specifically offer businesses valuable insights and tools to optimize marketing strategies, improve product offerings, and ultimately drive business growth.

Table 1 provides a concise summary of research studies on RS, presenting an overview of the publication year, central themes, principal contributions, methodologies employed, fields of application, and the results achieved, serving as a useful reference for a literature review. Based on the literature review discussed above, it examines several techniques for RS. Researchers used Gaussian Mixture Models and fuzzy language modeling to assess user authority and recommend experts. Other research focuses on individualized suggestions in a variety of fields, including trip packages, music, and e-commerce products. Some systems use social network data to improve suggestions, while others rely on sentiment-aware association rule mining. Recent improvements include dynamic RS that uses reinforcement learning as well as models based on user personality to provide more accurate and personalized product recommendations.

\section{Methodology}
This study adopts a comparative framework to evaluate the performance of ML-powered RS, focusing on CF, content-based filtering (CBF), and hybrid models. Each approach was assessed for its ability to address challenges like data sparsity, cold start problems, and scalability.

\subsection{Data collection}
The primary dataset used is the MovieLens dataset, which includes user-item interaction data such as ratings, timestamps, and metadata like genres. Key components include:
\begin{itemize}
    \item Users: Unique identifiers for individuals.
    \item Items: Metadata about movies (e.g., genres, release years).
    \item Ratings: User-provided scores (1–5).
\end{itemize} 
\subsection{Data Pre-processing}
The raw data underwent preprocessing to enhance its utility for modeling:
\begin{itemize}
    \item Segmentation: Users were grouped using K-means clustering based on behavioral patterns.
    \item Feature Extraction: Features like average user ratings, temporal trends, and item popularity were generated.
    \item Sentiment Analysis: Metadata text (e.g., reviews) was analyzed using NLP techniques.
\end{itemize}  

\subsection{Models and algorithms}
The study implemented three categories of recommendation models to generate personalized suggestions. CF leverages user-item interactions to generate personalized recommendations by analyzing patterns in user behavior. It uses two main approaches: user-based filtering and item-based filtering. CBF takes a different approach, focusing on the attributes of items to recommend those that are like what a user has previously liked. Hybrid Models combine the strengths of CF and CBF, using techniques like weighted voting to enhance recommendation accuracy and diversity. Advanced methods, such as NCF, are also integrated into hybrid models to address challenges like data sparsity and to improve scalability, creating a robust RS. To further refine the recommendation models, user feedback on recommended items was iteratively collected and incorporated into the training process. This mechanism enabled the models to adapt to evolving user preferences, enhancing accuracy and personalization.

\subsection{Evaluation metrics and experimental setup}
Model performance was evaluated using several metrics to comprehensively assess recommendation effectiveness. Precision, Recall, and the F1 Score were employed to measure the accuracy of recommendations, while root mean square error (RMSE) quantified prediction accuracy by comparing predicted and actual user ratings. Coverage was used to assess the diversity of recommended items. Additionally, mean average precision (MAP) was included to evaluate the ranking quality of recommendations, focusing on the precision of top-ranked items. Finally, the area under the receiver operating characteristic curve (AUROC) was used to measure the model’s ability to differentiate between relevant and irrelevant recommendations, providing an overall measure of classification performance. Together, these metrics offered a robust framework for evaluating accuracy, diversity, and ranking effectiveness. For training, in general, the dataset was split into 80\% for training and 20\% for testing and a five-fold cross-validation approach was employed to ensure robust evaluation. 

The methodology described above provides the foundation for evaluating the performance of RS. The next section introduces the main strategies for constructing RS, which form the core approaches are further examined in subsequent sections, where their implementation, performance, and implications are analyzed. This progression includes discussions on data preparation, experimental results, insights derived from the findings, and future directions for enhancing RS.

\section{Strategies for constructing RS}
The taxonomy of RS strategies is outlined in Figure 3, dividing them into three main categories: CF, Content-based filtering, and Hybrid systems, each with its unique approach to generating suggestions.  Alongside these, it maps out the challenges that RS encounters, including issues like data sparsity, the cold start problem, scalability, and the gray sheep problem. Other hurdles include shilling attacks, limited serendipity, the complexity of capturing user preferences and context dynamically, the risk of overspecialization, integration complexity, and the critical process [46] of algorithmic selection. These challenges represent the key obstacles that developers and researchers must address to advance the effectiveness and efficiency of RS.

\begin{figure}[t!] 
	\centering 
	\includegraphics[width=0.4\textwidth]{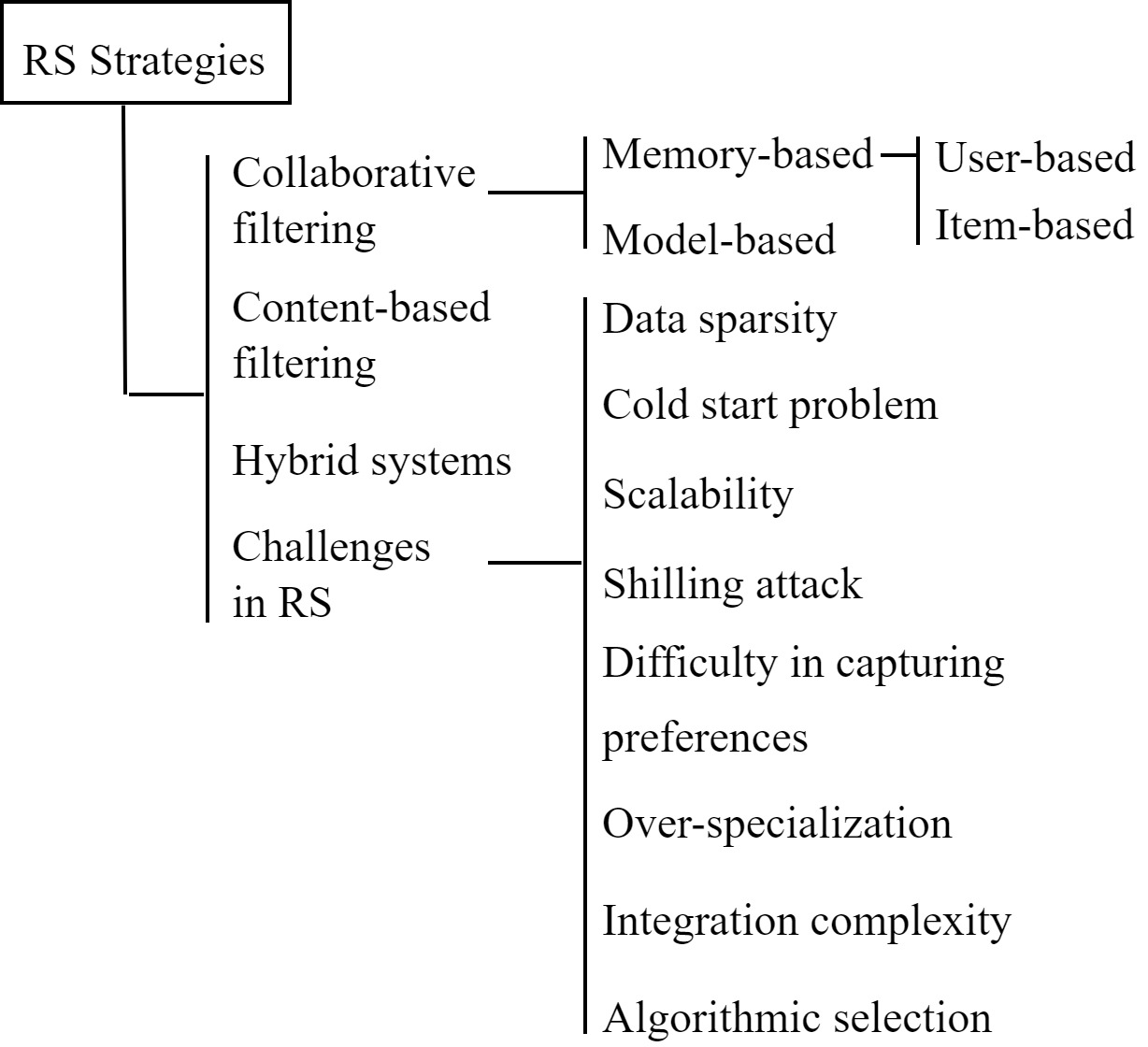}
	\\
	\caption{\textrm{Taxonomy of RS strategies}}
	\label{fig:3}       
\end{figure}

\subsection{Collaborative filtering (CF)}
CF is a recommendation strategy that employs user behavior, online activities, and preferences to anticipate and recommend things or information based on similarities with other users. This method  [47] assumes that users who have shown similar preferences in the past will continue to have similar preferences in the future. The process usually begins with the creation of a user-item matrix, in which rows represent users, columns represent objects (products, movies, series, etc.), and matrix entries indicate the users' historical interactions or preferences for the items. The similarity between users is then calculated based on these interactions, and predictions or recommendations are made for a user by considering the preferences of users with similar tastes. The RE indeed relies on various ML algorithms [48] to analyze customer preferences and make personalized product recommendations. In the process of CF within a movie RS, user ratings for movies are initially collected; for example, User A gives high ratings to Movies 1 and 2, while User B also rates Movies 1 and 2 highly, in addition to Movie 3. The system then identifies the similarity between Users A and B based on their shared preferences for Movies 1 and 2. As a result of their similar tastes, User A is regarded as a neighbor to User B. Influencing this relationship, the system predicts that User A will likely enjoy Movie 3, a title highly rated by User B but not yet seen by User A. Consequently, the system recommends Movie 3 to User A, applying the established patterns of similarity and user preferences to personalize content recommendations. Memory-based methods and model-based CF are the two main approaches within CF for building RS. The goal is to provide personalized recommendations [49] to users, leveraging the identified preferences and collaborative patterns. The system continually refines its understanding of user preferences based on ongoing interactions and feedback. 

\subsubsection{Memory-based methods}
It uses the historical interactions and preferences of users to make recommendations. And it relies directly on similarity measures [50] between users or items. There are two main types of memory-based filtering: user-based and item-based.

\begin{figure}[b!] 
	\centering 
	\includegraphics[width=0.35\textwidth]{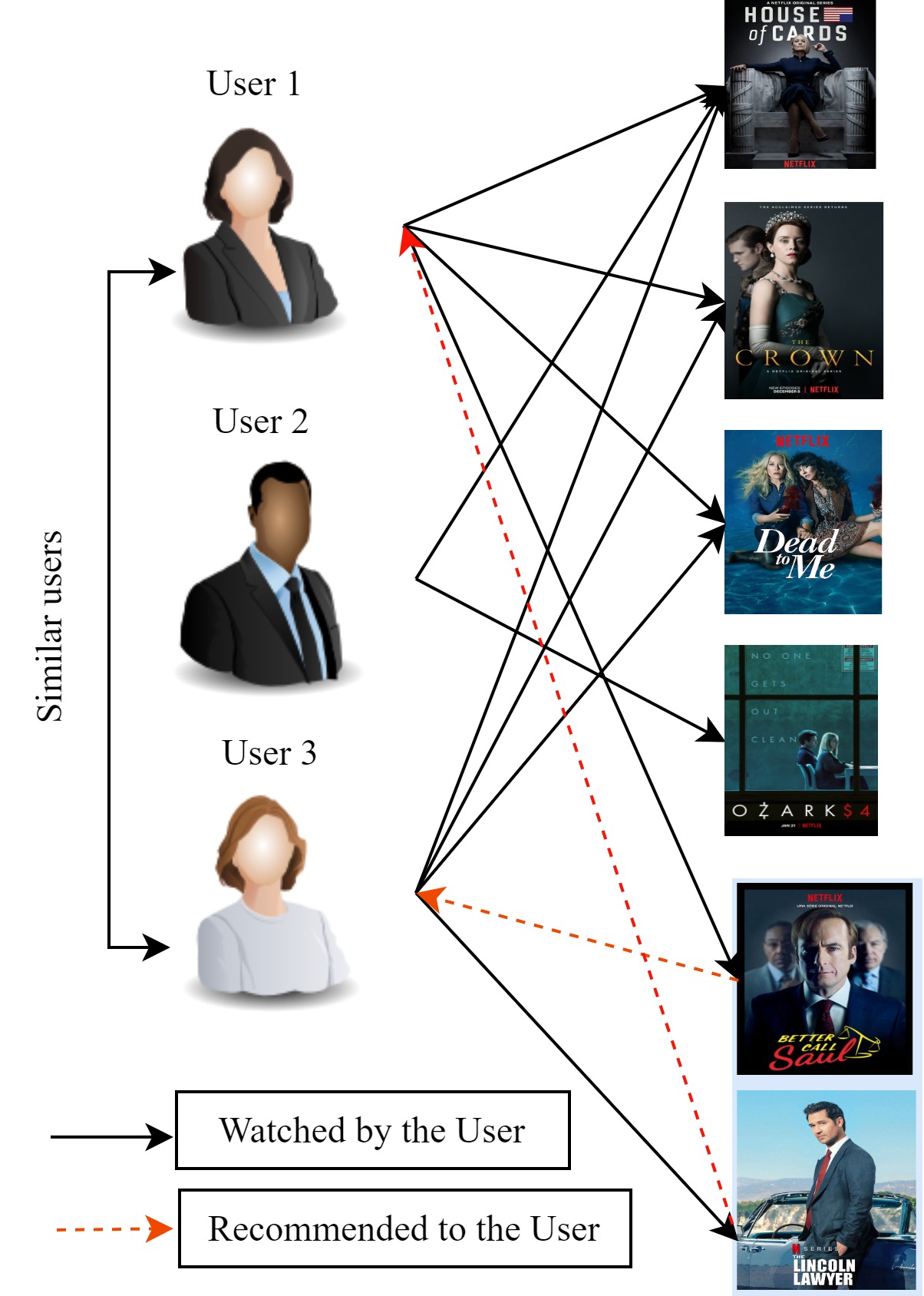}
	\\
	\caption{\textrm{User-based CF}}
	\label{fig:4}       
\end{figure}

\paragraph{User-based CF}
Recommends items to a user based on the preferences of users with similar tastes and performs explicit data collection. It is Simple and transparent. Explicit data collection involves directly asking users to provide information about their preferences [51] by compiling lists or rating products. Calculate the similarity between users based on their historical preferences. For a target user, identify users with similar tastes and recommend items that these similar users have liked or interacted with. Figure 4 visually demonstrates the mechanism of user-based CF, highlighting how a user's preferences are matched with those of similar users to generate recommendations. Where User 1 and User 3 have similar preferences, and both liked or watched a particular set of series under different genres, User 1 is more likely to receive recommendations based on User 3’s watch or like history and vice versa from the figure, it is evident that User 2 doesn’t receive any recommendation based on User 1 and 3 because of dissimilarity in the preference.

\paragraph{Item-based CF}
Recommends items to a user based on the likes or interactions with the other users and performs implicit data collection methods. Implicit data collection involves observing user interactions on social media platforms, often powered by AI, or monitoring user activities such as past purchases, likes, and views [52] of items by the users on e-commerce websites. The filtering calculates the similarity between items based on user interactions. For a target user, recommend items that are like those the user has shown interest in. Figure 5 shows an example of Item based filtering. If a user liked or watched a certain genre of series, item-based CF will recommend other series in the same genre. It addresses sparsity issues better than user-based CF. Effective for recommending niche items.

\begin{figure}[b!] 
	\centering 
	\includegraphics[width=0.33\textwidth]{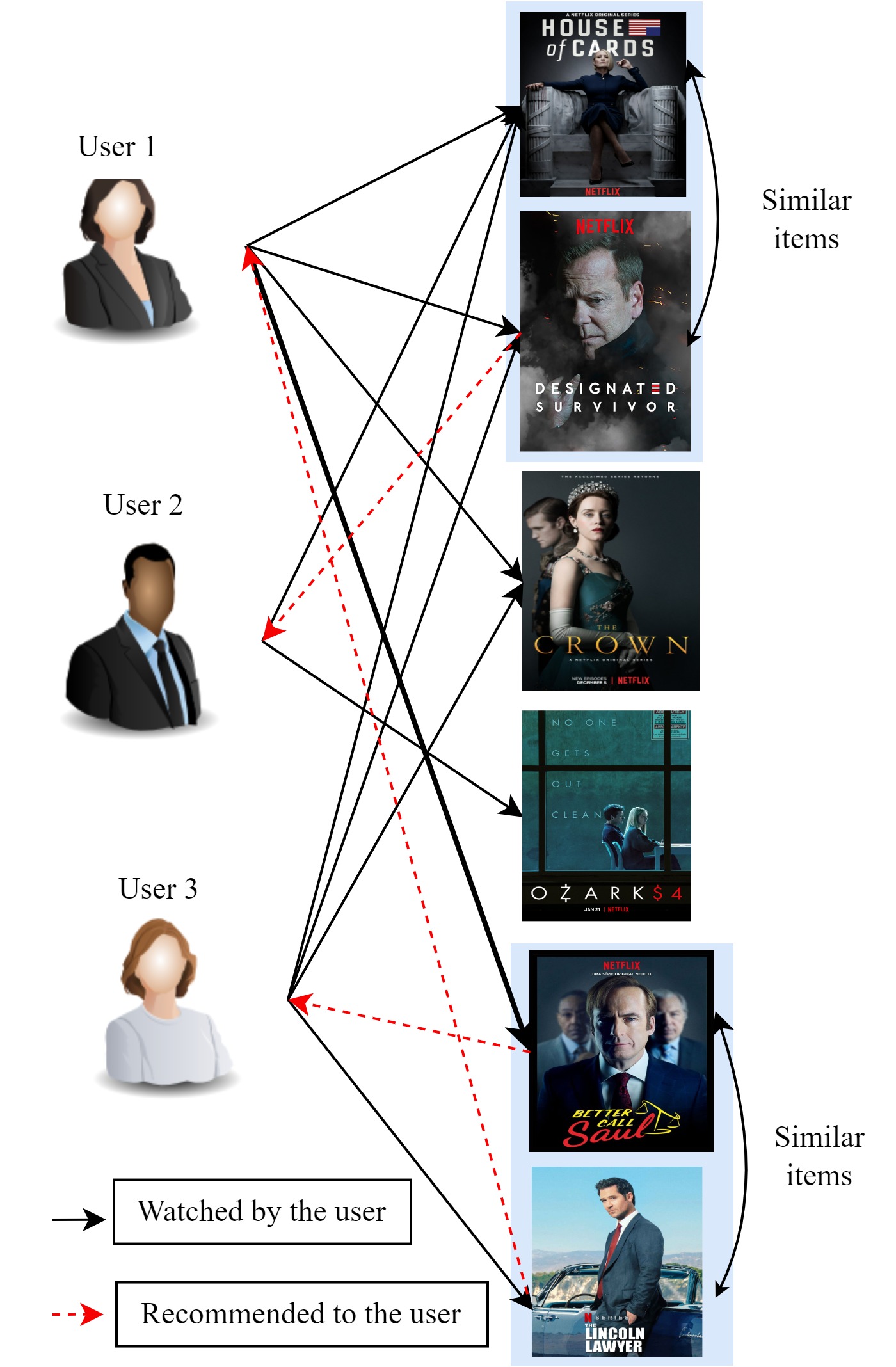}
	\\
	\caption{\textrm{Item-based CF}}
	\label{fig:5}       
\end{figure}
\begin{figure}[t!] 
	\centering 
	\includegraphics[width=0.44\textwidth]{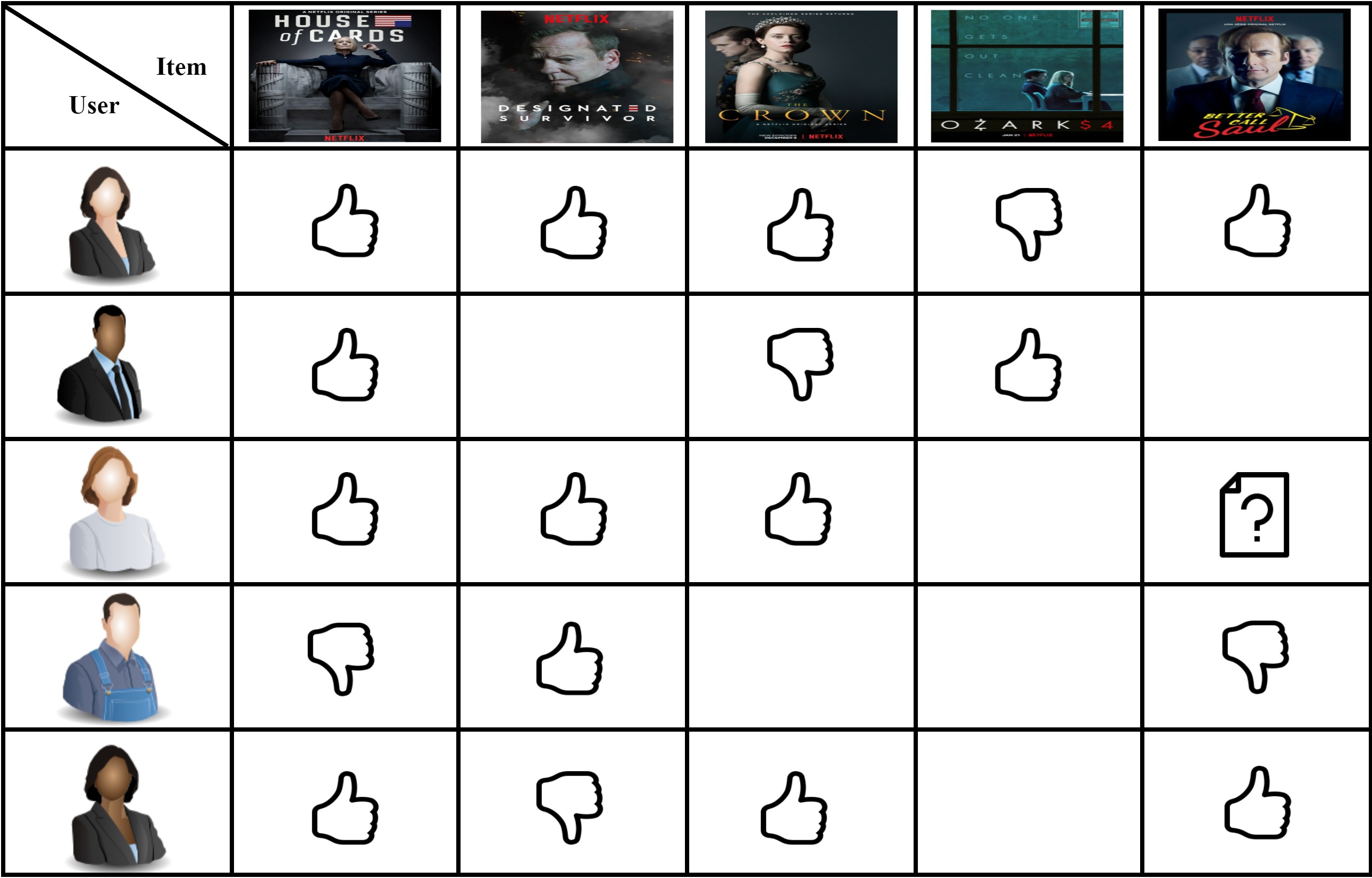}
	\\
	\caption{\textrm{User-item interaction}}
	\label{fig:6}       
\end{figure}
\subsubsection{Model-based method}
Model-based CF is an approach where a predictive model is created to make recommendations based on user-item interactions. This method involves the use of ML algorithms and techniques to learn patterns and relationships from the available data. Matrix factorization [53], singular value decomposition (SVD), clustering models, user-based k-nearest neighbors, Bayesian networks, and deep learning models are common techniques employed in model-based CF. This method is often preferred in scenarios where the amount of data is large and sparse, making it challenging for memory-based methods to handle effectively. Model learning and dimensionality reduction strategies address some of the more conventional issues with RS, like [54] scalability and sparsity. As depicted in Figure 6, these models are trained on historical user-item interactions, equipping them to generalize and make accurate predictions, even with sparse data. However, the trade-off is increased computational complexity and potential challenges in model interpretability.  \\[0.1cm]
\textit{\textbf{Research question (RQ) 1:}} How can CF techniques, encompassing memory-based and model-based methods, be optimized to enhance recommendation accuracy and efficiency in the face of challenges such as data sparsity, scalability, and maintaining the interpretability of ML models in user-item interaction scenarios? \\[0.1cm]
Optimizing CF techniques to enhance recommendation accuracy and efficiency, especially in sparse data environments, involves leveraging both memory-based and model-based methods. A deep CF algorithm integrating convolutional neural networks with CF is established in the research [55] showcasing significant accuracy improvements. Performance comparison of prediction methods is performed in [56] to address sparsity, with Boosted Double Means Centering emerging as a promising solution. Discussed about MKR-Bine model in [57]  that employs a neural network graph to enhance higher-order interactions, enhancing performance on sparse datasets. Furthermore, the authors in [58] improved memory-based CF by discovering potential similarity relationships, and  [59] presented a hybrid approach combining SVD with ontology-based methods to improve accuracy and efficiency. These advancements suggest that a blend of innovative ML frameworks and algorithms can effectively tackle the challenges of CF in RS.

\subsection{Content-based filtering}
A CBF system is a type of RS that suggests items to users based on the characteristics or features of the items themselves and the preferences expressed by the user. Unlike CF, which relies on user-item interactions, it focuses on the attributes of items and the user's preferences for those attributes. When a user interacts with the system by providing a positive rating for a particular item as represented in  [60], it indicates a preference for that item. Interaction includes both explicit feedback, such as ratings, and implicit feedback, including clicks, views, or purchases, all of which inform the system of user preferences. The steps involve initially, constructing detailed user profiles to understand individual preferences, then developing a content similarity index to assess the similarities between various items and employing a neighbor finder algorithm to identify users with similar interests. Further, [61] generates a list of potential items for recommendation and finally, determining the weight and similarity of each item to ensure a diverse and relevant set of recommendations.  \par

To explain in detail, all items present in the item profile of the positively rated item are aggregated to build a user profile. The user profile is represented as a vector in the same feature space as the items, where each feature corresponds to a specific attribute. The system considers various attributes of items, such as price, category, and features defined by keywords and tags. In the context of an e-commerce platform, a product's features could include its brand, specifications, and other relevant attributes. For a movie, features could include the director, actors, genre, and plot keywords. User preferences are inferred from their historical purchases and related feedback (reviews, ratings, etc.) as shown in Figure 7. The system calculates the similarity between the user profile vector and the vectors representing other items. Items with the highest similarity scores are recommended to the user. As mentioned in [62] the user profile is updated dynamically as the user interacts with the system over time, reflecting evolving preferences.    \par 

\begin{figure}[t!] 
	\centering 
	\includegraphics[width=0.33\textwidth]{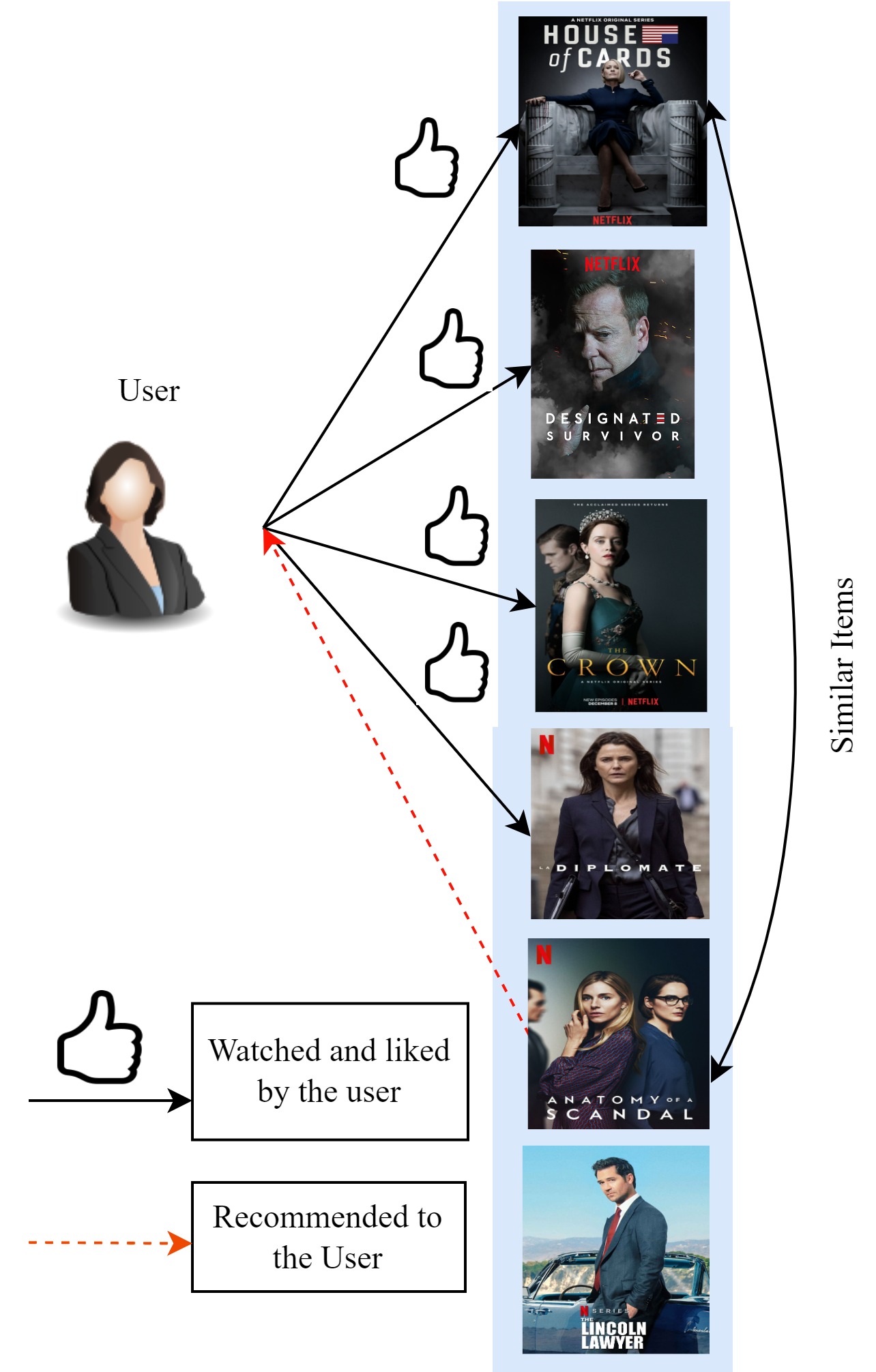}
	\\
	\caption{\textrm{Content-based filtering}}
	\label{fig:7}       
\end{figure}

The system utilizes an ML algorithm, such as Bayesian classifiers, decision trees, clustering, or other techniques, to analyze customers' purchase patterns. The algorithm learns to identify patterns and relationships between item characteristics and user preferences. The algorithm investigates historical purchase patterns to identify correlations between certain item features and positive user feedback. It learns which features contribute to a positive user experience and satisfaction. Based on the learned patterns as demonstrated in [63], the system recommends other products that share similar features with those previously bought and positively reviewed. Recommendations are improved to the individual user's preferences, emphasizing items with characteristics that align with their historical choices. RS has the benefit of being customized to each user's tastes. For first-time users, the system can offer precise recommendations. It provides transparency because of the explicit item features. \\[0.1cm]
\textit{\textbf{RQ2:}} How effective are CBF systems in providing accurate and personalized recommendations by analyzing item characteristics and user preferences in various domains such as e-commerce and multimedia content platforms? \\[0.1cm]
CBF systems, which recommend items based on their features and user preferences, have shown varying degrees of effectiveness across different domains like e-commerce and multimedia content platforms. Their effectiveness hinges on the accurate construction of user profiles, often derived from interactions such as ratings, views, or purchases. These profiles are crucial for comparing items using a content similarity index, where the system assesses similarities based on item attributes like brand or genre. The incorporation of ML algorithms by the authors [64], [65] like Bayesian classifiers or decision trees, further refines these systems by identifying patterns correlating item characteristics with user preferences. This process allows for dynamic updating of user profiles, ensuring that recommendations remain relevant over time. However, the effectiveness of these systems can vary, especially in their ability to personalize recommendations for new users with limited interaction history as mentioned in [66] or domains with highly specific item characteristics from [67]. Additionally, the transparency of these systems in the recommendation process and their impact on user satisfaction are critical factors that influence their overall effectiveness. While CBF systems offer a high degree of customization and can adapt to user preferences, the extent of their effectiveness can depend significantly on the domain and the specific implementation of the filtering algorithms.

\begin{table*}[t!]\centering
\caption{\textrm{Strategies for constructing RS}}
\footnotesize
\label{tab:2}       % Give a unique label 
\resizebox{\linewidth}{!}{%
\begin{tabular}{|c|l|l|l|l|l|l|}
\hline
\textbf{Article}                                                        & \multicolumn{1}{c|}{\textbf{Method}}                             & \multicolumn{1}{c|}{\textbf{Characteristics}}                & \multicolumn{1}{c|}{\textbf{Strength}} & \multicolumn{1}{c|}{\textbf{Challenges}}                                                 & \multicolumn{1}{c|}{\textbf{Limitations}}                      & \multicolumn{1}{c|}{\textbf{Usability}} \\ \hline
\begin{tabular}[c]{@{}c@{}}{[}47{]}, {[}48{]}, \\ {[}49{]}\end{tabular} & Collaborative                                                    & Behavioral                                                   & Personalized                           & \begin{tabular}[c]{@{}l@{}}Cold start, sparsity, \\ scalability\end{tabular}             & Bias                                                           & Community                               \\ \hline
{[}50{]}, {[}51{]}                                                      & Memory-based                                                     & Historical                                                   & Direct                                 & \begin{tabular}[c]{@{}l@{}}Scalability, new user \\ adaptation\end{tabular}              & Volume                                                         & Simple                                  \\ \hline
\begin{tabular}[c]{@{}c@{}}{[}53{]}, {[}54{]}, \\ {[}55{]}\end{tabular} & Model-based                                                      & Predictive                                                   & Scalable                               & \begin{tabular}[c]{@{}l@{}}Computational comp-\\ lexity, interpretability\end{tabular}   & \begin{tabular}[c]{@{}l@{}}Data intens-\\ iveness\end{tabular} & Large-scale                             \\ \hline
\begin{tabular}[c]{@{}c@{}}{[}60{]}, {[}61{]}, \\ {[}62{]}\end{tabular} & Content-based                                                    & \begin{tabular}[c]{@{}l@{}}Attribute-\\ centric\end{tabular} & Customized                             & \begin{tabular}[c]{@{}l@{}}Limited content anal-\\ ysis, overspecialization\end{tabular} & Narrow                                                         & Niche                                   \\ \hline
\begin{tabular}[c]{@{}c@{}}{[}68{]}, {[}69{]}, \\ {[}70{]}\end{tabular} & Hybrid                                                           & Integrated                                                   & Comprehensive                          & \begin{tabular}[c]{@{}l@{}}Integration complexity, \\ algorithm selection\end{tabular}   & Dependent                                                      & Adaptive                                \\ \hline
{[}68{]}, {[}69{]}                                                      & Meta-level                                                       & Input-Output                                                 & Accurate                               & Complex integration                                                                      & Complex                                                        & Multi-model                             \\ \hline
{[}70{]}                                                                & \begin{tabular}[c]{@{}l@{}}Feature-augme-\\ ntation\end{tabular} & Enriching                                                    & Enhanced                               & Scalability                                                                              & Expansive                                                      & Complementary                           \\ \hline
{[}71{]}                                                                & \begin{tabular}[c]{@{}l@{}}Feature comb-\\ ination\end{tabular}  & Merging                                                      & Informed                               & Feature tuning                                                                           & Selective                                                      & Multi-feature                           \\ \hline
{[}72{]}                                                                & Mixed                                                            & Diverse                                                      & Varied                                 & Complexity in balancing                                                                  & Multifaceted                                                   & Versatile                               \\ \hline
{[}73{]}                                                                & Cascade                                                          & Sequential                                                   & Refined                                & Latency                                                                                  & Delayed                                                        & Stepwise                                \\ \hline
{[}74{]}                                                                & Switching                                                        & Adaptive                                                     & Responsive                             & Real-time processing                                                                     & Dynamic                                                        & Flexible                                \\ \hline
{[}75{]}                                                                & Weighted                                                         & Weighted                                                     & Customized                             & Weighting different signals                                                              & Calibration                                                    & Enhanced                                \\ \hline
\end{tabular}
}
\end{table*}

\subsection{Hybrid systems}
Hybrid RS intensify the strengths of different recommendation techniques like CF and CBF as shown in Figure 8, addressing the limitations of individual approaches [68]. This integration aims to enhance overall performance and accuracy in recommender applications. These systems strive to provide users with more personalized, accurate, and diverse recommendations. The choice of a specific hybridization approach [61] depends on the nature of the application, available data, and the desired balance between different recommendation strategies. Experimentation and fine-tuning are typically necessary to optimize hybrid RS for specific contexts. The common hybridization methods are: \\[0.1cm]
\begin{figure}[t!] 
	\centering 
	\includegraphics[width=0.5\textwidth]{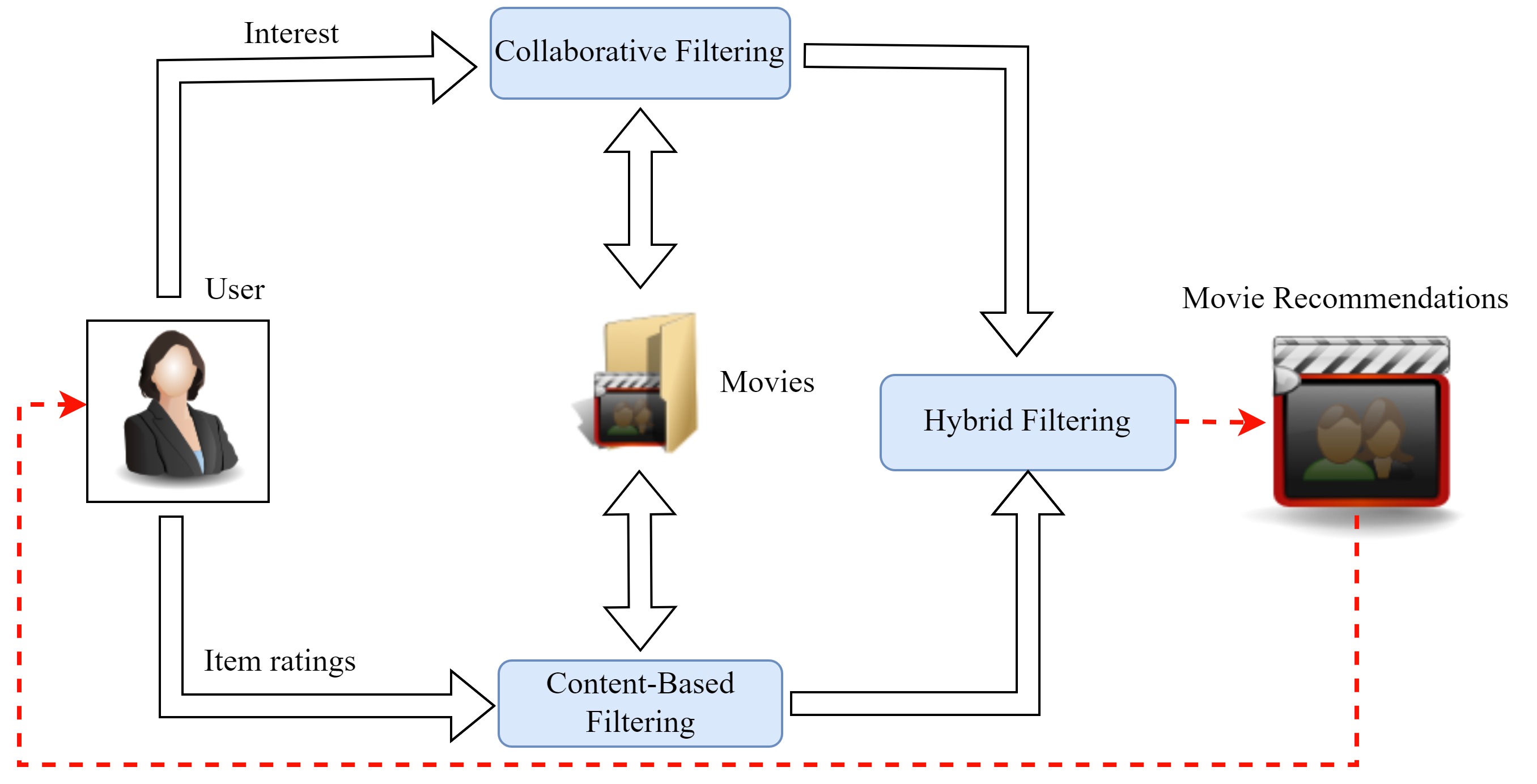}
	\\
	\caption{\textrm{Hybrid filtering}}
	\label{fig:8}       
\end{figure}
\textit{Meta-Level Hybridization} [69] - Uses the output of one technique as input for another, enhancing the recommendation process. As it incorporates insights from multiple models, improving recommendation accuracy. This approach enriches the recommendation experience by supplementing it with a series of images that narratively depict the respective plot, thereby enhancing user engagement and providing a richer contextual understanding of the recommendations. Improves accuracy by leveraging insights from multiple models. It requires complex data processing and integration.  \\[0.1cm]
\textit{Feature-Augmentation Hybridization} [70] - Augments features from one technique with features from another to create a more comprehensive set of recommendation attributes. It combines complementary information and enhances recommendation quality by combining complementary information. However, the primary challenge lies in managing the expanded feature space, which may lead to scalability issues as the complexity of the data increases.\\[0.1cm]
\textit{Feature-Combination Hybridization} [71] - It combines features from different techniques during the recommendation process and creates a comprehensive set of features for more informed recommendations. Creates a rich feature set for recommendations, informing better decision-making.  \\[0.1cm]
\textit{Mixed Hybridization} [72] - Simultaneously provides recommendations from both collaborative and content-based approaches. It offers diverse suggestions by considering multiple perspectives. It integrates user-specified group activities with conventional CF methods, utilizing a mixed hybridization approach. The complexity of balancing inputs from various recommendation approaches presents a significant challenge, requiring sophisticated algorithms to harmonize the recommendations effectively.\\[0.1cm]
\textit{Cascade Hybridization} [73] - Uses the output of one technique to influence or filter the recommendations of another technique sequentially. It provides a structured refinement process, allowing for fine-tuned suggestions. Emphasizing their utility in merging two distinct elements, each with their unique strengths, enhances the system's overall performance. Introducing sequential steps can lead to increased latency in delivering recommendations, impacting user experience.\\[0.1cm]
\textit{Switching Hybridization} [74] - Dynamically switches between different recommendation techniques based on user behavior or system conditions. That automatically adapts to changing user preferences and system states. This flexibility requires real-time analysis and processing capabilities, posing challenges in continuously monitoring user interactions and system performance to switch methods effectively.\\[0.1cm]
\textit{Weighted Hybridization} [75] - Combines recommendations from different techniques using weighted averages or other mathematical operations. It allows for flexible blending of recommendation signals, emphasizing one approach over the other based on predefined factors. In weighted hybridization systems, weights are typically assigned to different components based on their importance or reliability in predicting user preferences. These weights can be static, determined by domain experts or through empirical analysis, or dynamic, adjusted continuously based on the system's performance and user feedback. Determining the appropriate weights for different recommendation signals is a challenging task.\\[0.1cm]
Table 2 presents a concise view of the strategies for constructing RS, capturing the essence of each with a single descriptor. The overview of the strategies for constructing RS, their unique characteristics, main strengths, primary challenges they face, limitations, the contexts in which they are most usable, and the associated references for each method. \\[0.1cm]
\textit{\textbf{RQ 3:}} What optimization strategies enhance the performance of hybrid RS, and how do different hybridization methods, such as meta-level, feature-augmentation, feature-combination, mixed, cascade, switching, and weight, contribute to the accuracy and diversity of recommendations in user-item interaction platforms? \\[0.1cm]
Recent research [76] has identified several optimization strategies that enhance the performance of hybrid RS, each contributing uniquely to the accuracy and diversity of recommendations. Weighted hybridization exhibits different model outputs, like SVD and kNN, through weighted sums to boost the recommendation quality. Mixed Hybridization uses a blend of strategies as stated in [77], potentially with semantic ontologies, to surpass single-method limitations and heighten the personalization. Evolutionary Optimization [78] employs algorithms following the Strength of Pareto Optimization to finely balance accuracy, diversity, and novelty. Multi-objective optimization, like bacterial foraging optimization [79], targets several goals simultaneously, thus enhancing RE efficiency. Particle swarm optimization (PSO) [80] focuses on diversifying recommendations without compromising accuracy. Pareto-efficient Strategies propose accurate, diverse, and novel items through either efficient ranking [81] or weighted algorithm combinations. Algorithmic Blending [82] introduces novel methods like modified denoising for specific domains, such as e-learning, to improve system performance. Lastly, multi-objective evolutionary algorithms (MOEAs) such as MOEA-PRCP for crowd-funding platforms  [83], aim to optimize both utility-accuracy and topic-diversity for profitable and varied recommendations. These methodologies are pivotal for overcoming challenges like cold-start and data sparsity, user context changes, and evolving preferences, thereby ensuring an optimized user experience through more precise and varied suggestions. Nonetheless, empirical validation of these strategies requires an in-depth analysis of full-text research studies.

\begin{table*}[t!]\centering
\caption{\textrm{Summary of challenges in RS}}
\footnotesize
\label{tab:3}       % Give a unique label 
\resizebox{\linewidth}{!}{%
\begin{tabular}{|c|l|l|l|l|}
\hline
\textbf{Article}                                              & \multicolumn{1}{c|}{\textbf{Challenge}}                                  & \multicolumn{1}{c|}{\textbf{Nature of challenge}}                                        & \multicolumn{1}{c|}{\textbf{Impact on systems}}                                                 & \multicolumn{1}{c|}{\textbf{Potential solutions}}                                          \\ \hline
{[}84{]}                                                      & Data sparsity                                                            & \begin{tabular}[c]{@{}l@{}}Incomplete user-item \\ interactions\end{tabular}             & \begin{tabular}[c]{@{}l@{}}Reduced accuracy in \\ predictions\end{tabular}                      & \begin{tabular}[c]{@{}l@{}}Advanced algorithms to \\ handle sparse data\end{tabular}       \\ \hline
{[}85{]}                                                      & Cold start problem                                                       & \begin{tabular}[c]{@{}l@{}}Lack of historical data \\ for new users/items\end{tabular}   & \begin{tabular}[c]{@{}l@{}}Difficulty in providing \\ personalized recommendations\end{tabular} & \begin{tabular}[c]{@{}l@{}}Utilizing demographic \\ data or content features\end{tabular}  \\ \hline
{[}86{]}                                                      & Scalability                                                              & \begin{tabular}[c]{@{}l@{}}Handling increased\\ computational demands\end{tabular}       & \begin{tabular}[c]{@{}l@{}}Affects real-time \\ recommendation efficiency\end{tabular}          & \begin{tabular}[c]{@{}l@{}}Scalable algorithms and \\ distributed   computing\end{tabular} \\ \hline
{[}87{]}                                                      & Gray sheep users                                                         & \begin{tabular}[c]{@{}l@{}}Users with moderately \\ divergent preferences\end{tabular}   & \begin{tabular}[c]{@{}l@{}}Challenges in accurate \\ preference capture\end{tabular}            & \begin{tabular}[c]{@{}l@{}}Diverse recommendation \\ techniques\end{tabular}               \\ \hline
{[}88{]}                                                      & Shilling attack                                                          & \begin{tabular}[c]{@{}l@{}}Malicious manipulation \\ of system\end{tabular}              & \begin{tabular}[c]{@{}l@{}}Compromises recommendation \\ integrity\end{tabular}                 & \begin{tabular}[c]{@{}l@{}}Robust anomaly detection \\ methods\end{tabular}                \\ \hline
{[}89{]}                                                      & Limited serendipity                                                      & \begin{tabular}[c]{@{}l@{}}Over-reliance on past \\ preferences\end{tabular}             & \begin{tabular}[c]{@{}l@{}}Restricts exposure to diverse \\ content\end{tabular}                & \begin{tabular}[c]{@{}l@{}}Incorporating exploratory \\ factors in algorithms\end{tabular} \\ \hline
\begin{tabular}[c]{@{}c@{}}{[}90{]}, \\ {[}91{]}\end{tabular} & Over-specialization                                                      & \begin{tabular}[c]{@{}l@{}}Excessive focus on past \\ preferences\end{tabular}           & Leads to a filter bubble effect                                                                 & \begin{tabular}[c]{@{}l@{}}Balancing personalization \\ with novelty\end{tabular}          \\ \hline
{[}92{]}                                                      & \begin{tabular}[c]{@{}l@{}}Capturing dynamic \\ preferences\end{tabular} & \begin{tabular}[c]{@{}l@{}}Adapting to changing \\ user needs\end{tabular}               & \begin{tabular}[c]{@{}l@{}}Less contextually relevant \\ recommendations\end{tabular}           & \begin{tabular}[c]{@{}l@{}}Real-time monitoring and \\ adaptive models\end{tabular}        \\ \hline
{[}93{]}                                                      & Integration complexity                                                   & \begin{tabular}[c]{@{}l@{}}Combining different \\ recommendation techniques\end{tabular} & \begin{tabular}[c]{@{}l@{}}Harmonization and compatibility \\ challenges\end{tabular}           & \begin{tabular}[c]{@{}l@{}}Unified models with \\ seamless integration\end{tabular}        \\ \hline
{[}94{]}                                                      & Algorithmic selection                                                    & Choosing suitable algorithms                                                             & Suboptimal recommendations                                                                      & \begin{tabular}[c]{@{}l@{}}Strategic algorithm selection \\ and adaptation\end{tabular}    \\ \hline
\end{tabular}

}
\end{table*}

\subsection{Challenges in RS}
RS faces several challenges, ranging from data-related issues to algorithmic complexities and user experience considerations. CF RS, which relies on user-item interactions and similarities between users or items, face specific challenges that impact their effectiveness and user satisfaction. Content-based RS encounter several challenges that impact their performance. One significant hurdle lies in the limited serendipity of recommendations, as these systems heavily rely on item features and past user preferences, potentially restricting users to a narrow range of content aligned with their historical choices. Combining collaborative and content-based methods necessitates addressing the cold start problem for both new users and items, as well as devising effective strategies to handle data sparsity. The complexity of hybrid models raises scalability concerns, especially as the user and item base grows. Maintaining diversity in recommendations while avoiding over-specialization remains a challenge, as hybrid systems must navigate the fine line between personalization and introducing users to novel content. Few major challenges faced by the existing RS are listed below and the summary is presented in Table 3.

\subsubsection{Data sparsity}
Data Sparsity refers to the situation where the user-item interaction matrix exhibits a substantial number of missing values. This indicates that users have not engaged with most available items [84], leading to incomplete historical interaction data. The challenge arises when attempting to make accurate predictions or generate meaningful recommendations based on such limited or sparse data. CF methods, which rely heavily on historical user-item interactions, are particularly impacted by data sparsity.

\subsubsection{Cold start problem }
The cold start problem in RS [85] refers to the challenge encountered when dealing with new users or items that have limited or no historical interaction data. This lack of past behavior makes it difficult for the system to provide accurate and personalized recommendations. The cold start problem can manifest in two forms: user cold start, where there is insufficient data about a new user's preferences, and item cold start, where a new item lacks historical usage information. 

\subsubsection{Scalability}
Scalability in terms of collaborative and content-based systems [86] refers to the system's ability to handle increased computational demands effectively as the volume of users, items, or data grows. As user bases and item catalogs expand, the computational complexity of recommendation algorithms becomes a critical consideration. Achieving scalability is essential for ensuring that the system can deliver timely and efficient recommendations, especially in real-time scenarios where users expect quick responses. It can affect the efficiency of real-time recommendation generation, especially for large-scale systems.

\subsubsection{Gray sheep users }
Gray sheep users, in the context of collaborative systems [87], refer to a subset of users who exhibit preferences that are neither strongly aligned with most users nor distinctly unique to the individual. Unlike \enquote{white sheep} users who align closely with the preferences of the majority and \enquote{black sheep} users who have highly unique tastes, gray sheep users fall in between, making their preferences more challenging to discern accurately. These users pose a unique challenge for recommendation algorithms as traditional CF methods may struggle to effectively capture their divergent preferences. 

\subsubsection{Shilling attack }
Shilling attack [88] refers to a malicious attempt to manipulate or influence the recommendations provided by the system. The attacker aims to promote certain items, products, or content by artificially inflating their perceived popularity or desirability. Shilling attacks can be detrimental to the integrity and effectiveness of RS, as they compromise the trustworthiness of the recommendations and may lead to biased or inaccurate suggestions.

\subsubsection{Limited serendipity}
Content-based systems [89] may struggle to recommend items that are outside a user's known preferences or characteristics. Limited serendipity refers to a constraint within RS where the algorithm tends to provide suggestions that closely align with a user's past preferences or behaviors, potentially restricting exposure to diverse or unexpected content. In such cases, the system may prioritize items that share similarities with the user's historical interactions, leading to a lack of surprise or discovery of novel items outside the user's interests.

\subsubsection{Over-specialization}
Content-based systems [90] may lead to over-specialization where the system becomes excessively focused on users' past preferences, potentially limiting the diversity of recommendations. When a recommendation algorithm is overly enriched to a user's historical behavior, it may result in a narrow set of suggestions that closely align with what the user has already experienced. This phenomenon can lead to a \enquote{filter bubble} effect  [91], where users are predominantly exposed to content like their established preferences, hindering the system's ability to introduce them to new or varied items. 

\subsubsection{Difficulty in capturing user preferences  }
Adapting to changes in user preferences [92] over time can be challenging for content-based systems. If user preferences evolve, the system may take time to adjust, leading to less accurate recommendations in the short term. Content-based systems may find it challenging to adapt recommendations to dynamic contexts, such as changing user needs or environmental factors. Recommendations may not be as contextually relevant in situations where user preferences vary based on time, location, or other dynamic factors.

\subsubsection{ Integration complexity}
Integration complexity in the context of RS refers to the challenges and difficulties associated with combining or integrating different recommendation techniques or models within a unified system. As RS evolves [93], there is a growing interest in creating hybrid models that advance the strengths of multiple approaches, such as CF, content-based filtering, and knowledge-based methods. However, integrating these diverse techniques poses several complexities. Issues related to integration complexity may include harmonizing disparate data sources, reconciling different feature representations, and aligning the output of various recommendation algorithms. Additionally, ensuring seamless coordination between different models, especially in real-time or dynamic environments, requires careful consideration of algorithmic compatibility, computational efficiency, and the overall system architecture.

\begin{figure*}[t!] 
	\centering 
	\includegraphics[width=0.82\textwidth]{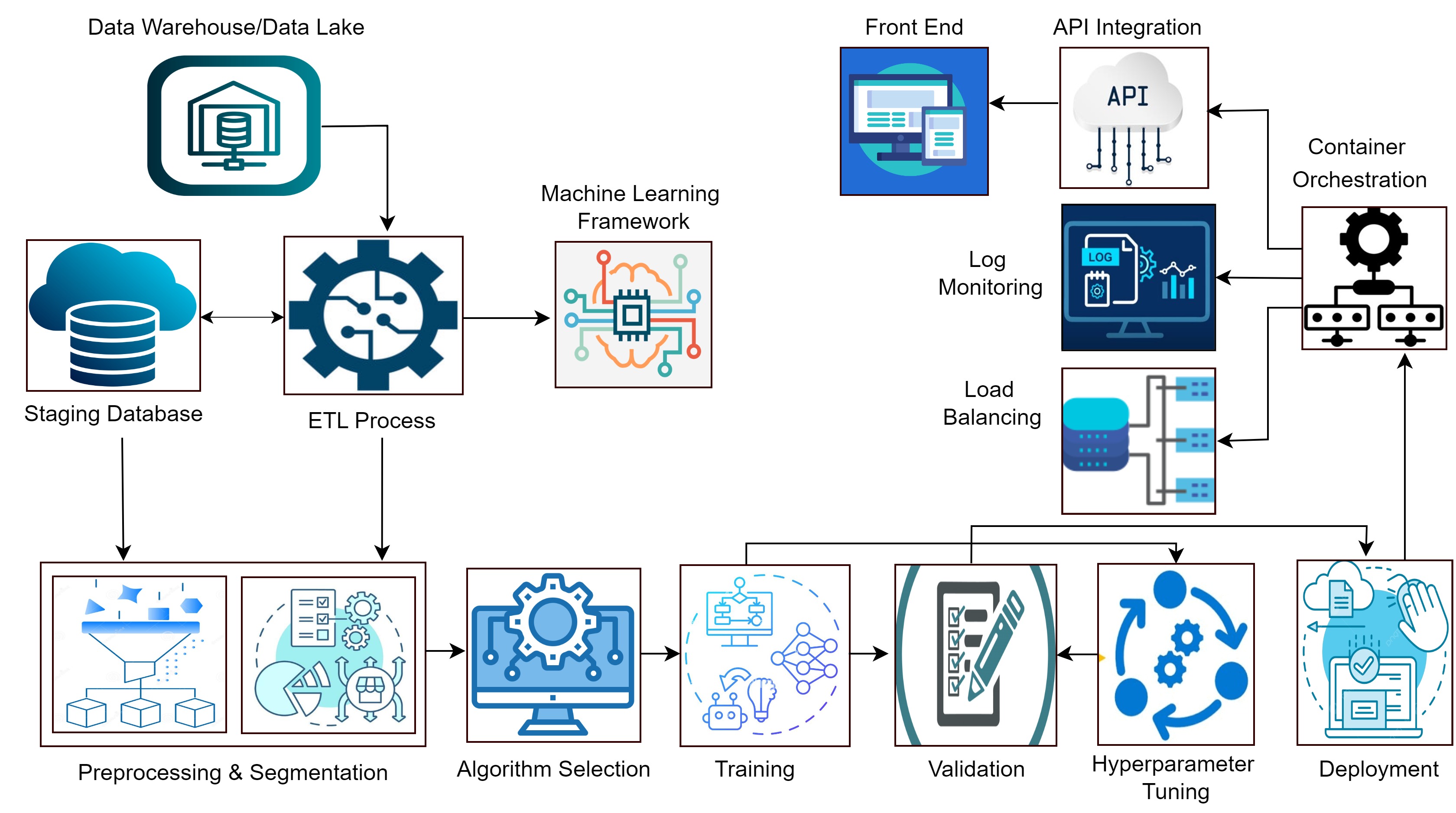}
	\\
	\caption{\textrm{Pipeline for user-item interaction-based RS}}
	\label{fig:9}       
\end{figure*}

\begin{table*}[b!]\centering
\caption{\textrm{Overview of database technologies}}
\small
\label{tab:4}       % Give a unique label 

\begin{tabular}{|l|l|l|l|l|}
\hline
\multicolumn{1}{|c|}{\textbf{Database type}} & \multicolumn{1}{c|}{\textbf{Traditional SQL}}                                                    & \multicolumn{1}{c|}{\textbf{NoSQL}}                                                        & \multicolumn{1}{c|}{\textbf{Data warehouses}}                                                       & \multicolumn{1}{c|}{\textbf{Data lakes}}                                                       \\ \hline
Data type                                    & Structured                                                                                       & \begin{tabular}[c]{@{}l@{}}Semi-structured/ \\ Unstructured\end{tabular}                   & Structured/semi-structured                                                                          & Raw/Unstructured                                                                               \\ \hline
Technologies                                 & \begin{tabular}[c]{@{}l@{}}MySQL, Postgre\\ SQL, Oracle\end{tabular}                             & \begin{tabular}[c]{@{}l@{}}MongoDB, Redis, \\ Cassandra\end{tabular}                       & \begin{tabular}[c]{@{}l@{}}Amazon Redshift, Google \\ Big Query, Snowflake\end{tabular}             & \begin{tabular}[c]{@{}l@{}}Apache Hadoop, \\ Amazon S3, Azure \\ Data Lake\end{tabular}        \\ \hline
Advantages                                   & \begin{tabular}[c]{@{}l@{}}Efficient for struc-\\ tured data and ACID \\ compliance\end{tabular} & \begin{tabular}[c]{@{}l@{}}Schema flexibility \\ and handles large \\ volumes\end{tabular} & \begin{tabular}[c]{@{}l@{}}Optimized for analytics and \\ centralized data integration\end{tabular} & \begin{tabular}[c]{@{}l@{}}Scalable and supports \\ batch/real-time \\ processing\end{tabular} \\ \hline
Scalability                                  & Moderate                                                                                         & High                                                                                       & High                                                                                                & Very High                                                                                      \\ \hline
Performance                                  & \begin{tabular}[c]{@{}l@{}}High for structured \\ queries\end{tabular}                           & \begin{tabular}[c]{@{}l@{}}High for unstruct-\\ ured data\end{tabular}                     & High for analytical queries                                                                         & \begin{tabular}[c]{@{}l@{}}High for large data \\ sets\end{tabular}                            \\ \hline
Complexity                                   & Moderate                                                                                         & Low-moderate                                                                               & High                                                                                                & High                                                                                           \\ \hline
Use cases                                    & \begin{tabular}[c]{@{}l@{}}User profiles, \\ product catalogs\end{tabular}                       & \begin{tabular}[c]{@{}l@{}}User interactions, \\ behavior tracking\end{tabular}            & \begin{tabular}[c]{@{}l@{}}Advanced user analysis, \\ complex querying\end{tabular}                 & \begin{tabular}[c]{@{}l@{}}Big data analytics, \\ diverse data types\end{tabular}              \\ \hline
\end{tabular}

\end{table*}

\subsubsection{Algorithmic selection}
Algorithmic selection is a critical aspect of RS development, involving the strategic choice of algorithms and determining the appropriate times to switch between them. This process is far from trivial, as poor algorithmic selection can lead to suboptimal recommendations [94], ultimately diminishing user satisfaction. The challenge lies in identifying the most suitable algorithms for a given recommendation task, considering factors such as the nature of the data, user behavior, and the system's objectives. Different algorithms excel under varying circumstances, and their performance can be influenced by factors like data sparsity, user diversity, and the presence of new items or users (cold start scenarios). Moreover, the dynamics of user preferences and the evolving nature of content require a dynamic approach to algorithmic selection. RS often needs to adapt and switch between algorithms to stay effective over time. The challenge is not only in selecting algorithms that perform well individually but also in orchestrating them to achieve a synergistic effect.\\[0.1cm]
\textit{\textbf{RQ 4:}} How can RS effectively address the cold start problem for both new users and items, ensuring accurate and personalized recommendations despite the absence of historical interaction data?\\[0.1cm]
To effectively handle the cold start problem in RS, numerous novel ways have been proposed by the researchers. In [95] used pre-trained language models to perform sentiment analysis, translating user profiles and item features into correct recommendations in the absence of historical data. The PUPP framework developed in [96] utilizes soft clustering and active learning to accurately recommend items for new users and products. The study in [97] applies word2vec and genre2Vec algorithms, alongside blockchain technology, to predict ratings for new items effectively, addressing the item-side cold-start issue. To address the item-side cold start issue, [98] leverages online social network reviews to crowdsource recommendations for new businesses. Moreover, [98] incorporates Local Collective Embeddings in matrix factorization for reliable new item recommendations. These approaches mark a trend toward more flexible, context-aware systems that may generate reliable suggestions with less starting data. \\[0.1cm]
\textit{\textbf{RQ 5:}} What approaches can hybrid RS adopt to maintain an optimal balance between personalization and the introduction of novel content, thereby avoiding over-specialization and enhancing user discovery?\\[0.1cm]
Hybrid RS are experimenting with innovative techniques to establish a balance between personalization and the introduction of new content, avoiding over-specialization. [99] introduced objective criteria such as content difference and genre accuracy and used algorithms like JohnsonMax for addressing data sparsity and building weak ties for serendipitous recommendations. The author in [100] proposed a two-stage hybrid deep learning-based CF system that focuses on recording communication between items and users for increased customization and discussed a hybrid Bayesian stacked auto-denoising encoder (HBSADE) model that recognizes latent user interests and evaluates contextual reviews to provide suggestions without over-specialization. These strategies work together to help hybrid systems achieve their goal of providing a richer and varied user experience. 

\section{Preparing the data for ML powered RE}
ML powered RE work by leveraging algorithms to analyze patterns and preferences in user behavior, to provide recommendations. Regardless of whether they are based on collaborative or content-based filtering, ML-powered RE typically adheres to a multi-stage pipeline. This pipeline serves as a systematic framework [2] designed to seamlessly process and transform both product and customer data into personalized suggestions. The distinct stages within this pipeline contribute to the system's ability to comprehend user preferences, providing recommendations that cater to individualized needs and preferences. These systems are widely used in various industries such as e-commerce, streaming services, social media, and more. The overall architecture of the pipeline is shown in Figure 9. 

\subsection{Problem identification and formulation}
The key aspects of the entire system depend on identifying the core problem RE aims to solve. In this case, it's enhancing user experience and driving sales by providing personalized product recommendations. The elements to be considered while building the system are to understand user needs and pain points [101], define the scope and scale of the RE and assess the impact on user engagement and business metrics. Then formulate clear and specific goals aligned with the identified problem. For example: 
\begin{itemize}
    \item Increase user engagement by $X\%$ and improve conversion rates on product recommendations and enhance customer satisfaction through personalized suggestions. 
    \item Define measurable metrics and key performance indicators (KPIs) to evaluate the success of the RS which includes click-through rates (CTRs) on recommendations [102], conversion rate on recommended products and revenue generated from recommendations. 
    \item Specify the types of data needed to achieve the goals and consider how recommendations will be presented to users (e.g., on a website, mobile app, or through email). 
    \item Decide on the frequency and timing of recommendations. 
\end{itemize}

\subsection{Data storage}
Depending on the kind of data that a RS needs to assess [103], data sets should be combined into an appropriate repository. NoSQL databases and data lakes offer horizontal scalability, making them well-suited for handling large volumes of complex data and growing user bases. These include user interactions, reviews, and diverse content, in addition to conventional SQL databases, which are designed to efficiently store structured data. Data lakes provide flexibility in handling various data formats, making them suitable for scenarios where data structures might evolve and serve as adaptable repositories to take in any type of data. Data governance and organization are crucial to prevent turning into \enquote{data swamps} [104] which requires careful metadata management. Data warehouses are optimized for fast query performance, making them suitable for recommendation scenarios where quick access to relevant data is crucial and excel at integrating [105] data from multiple sources, providing a unified view for analysis. Table 4 shows the overview of different database technologies. 

\begin{figure*}[t!] 
	\centering 
	\includegraphics[width=0.82\textwidth]{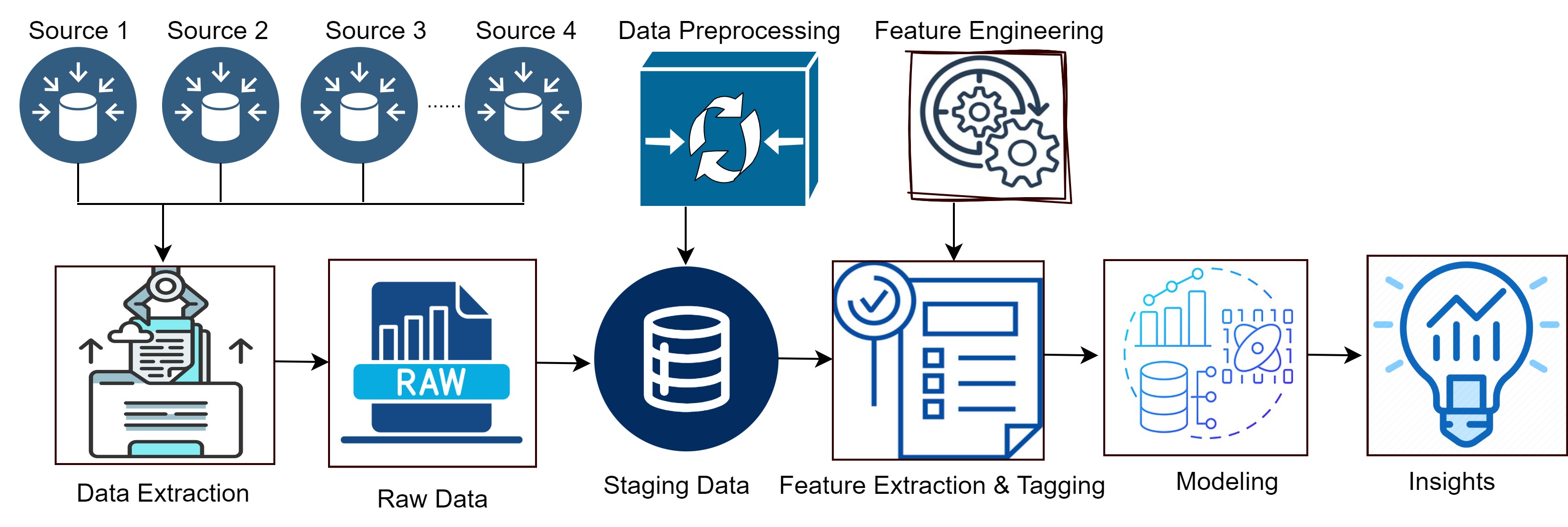}
	\\
	\caption{\textrm{Optimizing customer segmentation }}
	\label{fig:10}       
\end{figure*}

\subsection{Data collection and segmentation}
Data collection and segmentation are fundamental processes in the operation of RS, and they play pivotal roles in enhancing the accuracy and personalization of the recommendations provided to users. The first step involves collecting data on user behavior. This data can include user preferences, browsing history, purchase history, ratings, and more  [106]. The more data available, the better the system can understand user preferences. Gathering diverse and extensive datasets is crucial for training a robust ML model. As shown in Figure 9, after collecting data, the next step in the data processing pipeline is often the extract, transform, load (ETL) operation. ETL is a process used to gather data from different sources, transform it into a desired format, and then load it into a destination (database) for further analysis and insights. To segment the customers, i.e., group them into a specific buyer type or personality based on their traits, and then target them with relevant ideas, an ML system requires huge data sets [107]. Purchasing history, content usage, browsing behavior, product reviews, access devices, and many other factors can all be considered relevant metrics.  \par 

\begin{figure}[b!] 
	\centering 
	\includegraphics[width=0.5\textwidth]{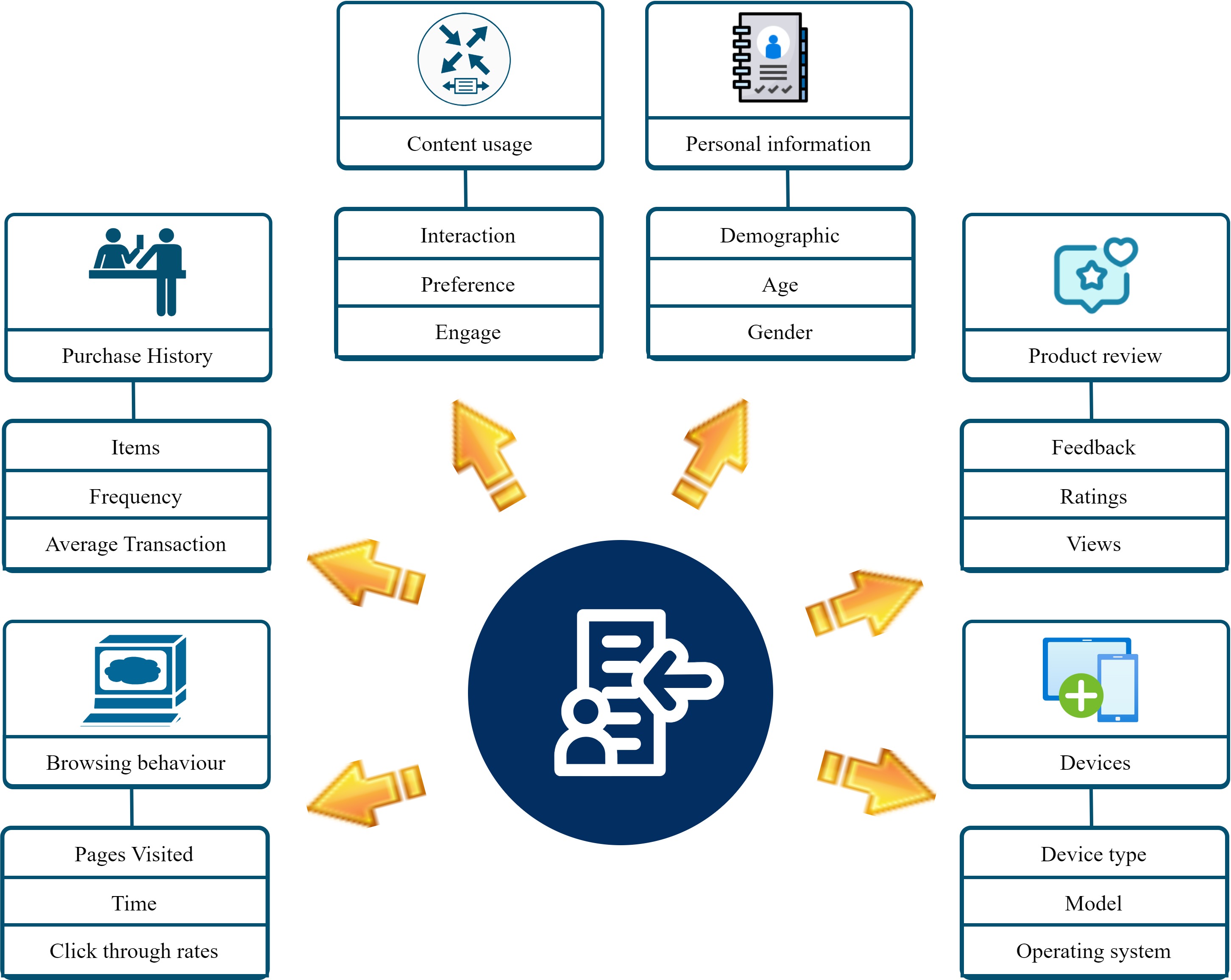}
	\\
	\caption{\textrm{Customer segmentation metrics. }}
	\label{fig:11}       
\end{figure}

The first step involves collecting data on user behavior. This data can include user preferences, browsing history, purchase history, ratings, and more  [106]. The more data available, the better the system can understand user preferences. Gathering diverse and extensive datasets is crucial for training a robust ML model. As shown in Figure 10, after collecting data, the next step in the data processing pipeline is often the extract, transform, load (ETL) operation. ETL is a process used to gather data from different sources, transform it into a desired format, and then load it into a destination (database) for further analysis and insights. To segment the customers, i.e., group them into a specific buyer type or personality based on their traits, and then target them with relevant ideas, an ML system requires huge data sets [107]. Purchasing history, content usage, browsing behavior, product reviews, access devices, and many other factors can all be considered relevant metrics. The detailed customer segmentation metrics are shown in Figure 11.

ML techniques [106] have become increasingly popular in various industries, particularly in marketing and customer relationship management, for automating and optimizing customer segmentation. It is the process of dividing a company's customer base into distinct groups with similar characteristics or behaviors. It enables the business to adjust its marketing tactics and product offerings to various client categories, resulting in more effective and personalized marketing campaigns. By considering criteria such as demographics, behavior, geography, and psychographics as shown in Figure 12, companies can improve their marketing strategies  [108] to specific customer segments, ultimately leading to improved customer satisfaction and increased profitability. 

Effective data collection and segmentation [109][110] lay the foundation for personalized customer experiences, targeted marketing, and improved customer satisfaction. These practices, when implemented thoughtfully, contribute to a more customer-centric approach to business strategies. Customer segments may evolve due to changes in customer behavior, market trends, or external factors. However, it is imperative to acknowledge that customer segments are not static entities; they can evolve due to shifts in consumer behavior, market dynamics, or external influences. As such, the practice of continuous monitoring and the regular updating of segmentation models is vital to ensure that businesses remain adaptive and responsive to these changes, thereby sustaining their customer-centric focus and staying competitive in today's dynamic marketplace.

\subsubsection{Data pre-processing } 
Raw data is often messy and needs preprocessing. This step involves cleaning the data, handling missing values, and transforming it into a format suitable for the ML algorithms. Feature engineering involves  [111] selecting, combining, or creating new features that can enhance the model's predictive capabilities. For instance, creating features like \enquote{average time spent on site} or \enquote{frequency of purchases in the last month}, \enquote{items purchased}, \enquote{time of purchase}, \enquote{ratings given}, \enquote{genres liked}, etc. 

\begin{figure}[t!] 
	\centering 
	\includegraphics[width=0.5\textwidth]{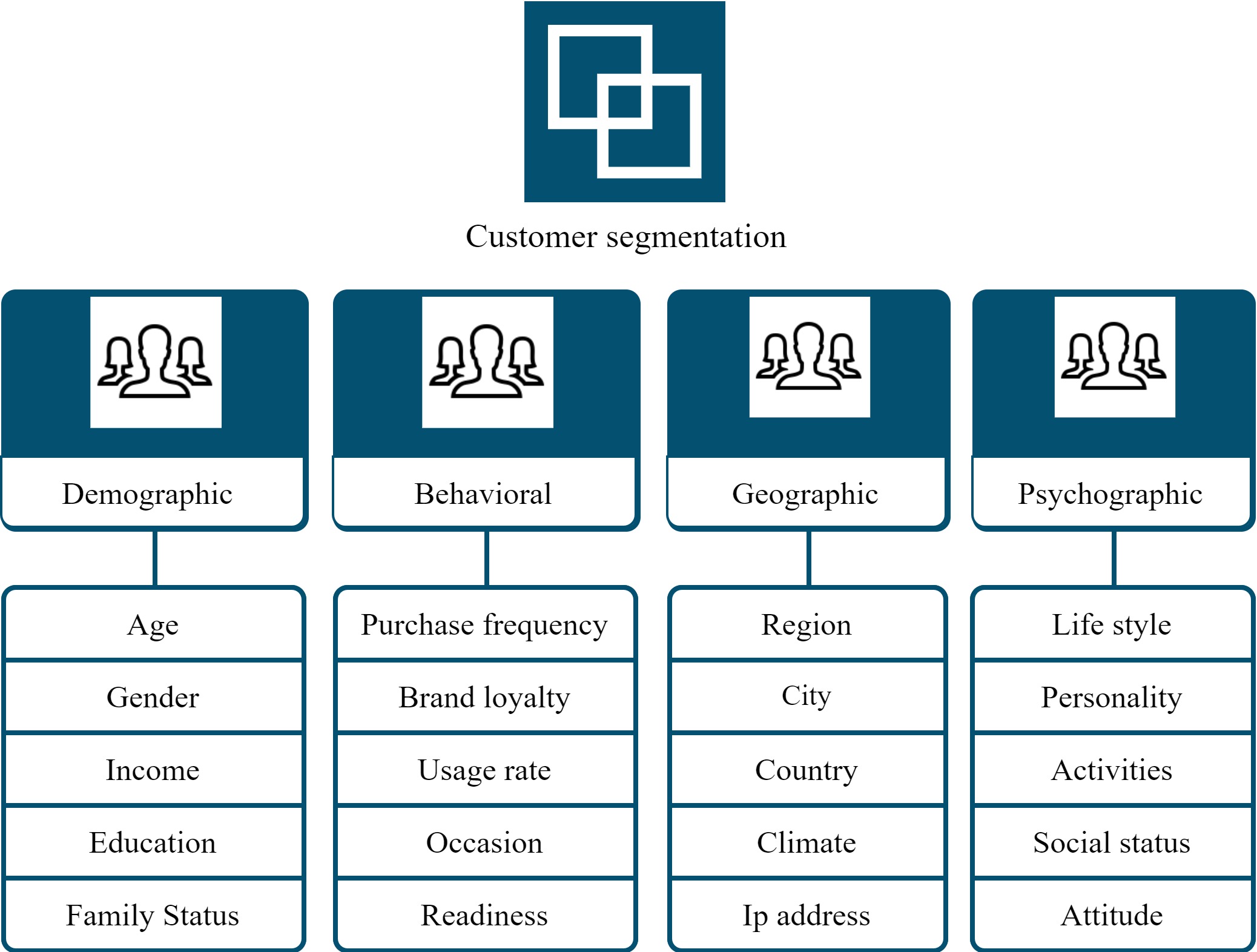}
	\\
	\caption{\textrm{Optimizing customer segmentation. }}
	\label{fig:12}       
\end{figure}

\subsubsection{Tagging and product features}
Tagging and product features [112] are crucial components in the context of building RS and customer segmentation. It involves assigning descriptive labels or metadata to items in a dataset. Tagging products with relevant attributes allows the system to understand the characteristics of each item.  Tags and features may evolve over time [113],  and it's essential to dynamically update them as new items are added or as user preferences change. In some systems, users may contribute to tagging by providing their own labels or feedback. This user-generated data can enrich the dataset and improve the system's understanding of user preferences. To effectively tag products, we can improvise techniques such as natural language processing (NLP) and ML. NLP helps us extract meaningful information  [114] from product descriptions, reviews, and other textual data. By understanding the context and sentiment behind the text, we can assign appropriate tags to each product.

One popular method  [115] is using pre-trained NLP models, such as word embeddings or transformers. These models have been trained on vast amounts of text data and can effectively capture the semantics and context of words. By applying these models to product descriptions or reviews [116], we can extract relevant tags and features. Another technique is using rule-based approaches. These approaches involve defining a set of rules or patterns that match specific tags or features. For example, we can define rules to identify product colors based on keywords or patterns in the text. While rule-based approaches may not be as sophisticated as ML models, they can still provide valuable insights and complement other techniques. Furthermore, CF [117] can be used to extract product features. CF analyzes the behavior and preferences of similar users to make recommendations. By identifying patterns and similarities between users, we can infer their preferences for certain product features. For example, if users with similar tastes often purchase products with a specific color, we infer that color is a relevant feature.

\subsubsection{Data analysis and decision making}
Data analysis and decision-making are fundamental components of effective RS. These systems process massive volumes of data to understand user preferences, behaviors, and interactions, thereby enabling personalized content recommendations.  The effectiveness of a RS depends on its ability to analyze data accurately and make informed decisions that are aligned with the preferences of the users. It requires not only sophisticated algorithms but also an in-depth understanding of user behavior and preferences. It is important to deal with sparse data, ensure privacy, maintain scalability, and adapt to changing user interests. As RS are complex, decision-making balances algorithmic precision with user-centric considerations, aimed at improving user experience, engagement, and satisfaction.

\paragraph{Exploratory data analysis}
Exploratory data analysis (EDA) is a dynamic process  [118] that involves employing a variety of algorithms and visualization techniques to uncover meaningful insights in the data. By gaining a deep understanding of user-item interactions, preferences, and patterns, data scientists can make informed decisions during subsequent stages of building and optimizing RE. It plays a crucial role in uncovering user preferences, item popularity, and interactions between users and items. Key Goals of EDA for RE are Analyze user profiles and item attributes to understand the diversity and distribution of preferences. Explore patterns in user-item interactions  [119], such as popular items, user engagement, and temporal trends. Identify outliers or anomalies in the data that may impact the performance of recommendation algorithms. Assess the sparsity of the user-item interaction matrix [120], a common challenge in RS and explore associations and co-occurrences between items to identify potential cross-selling opportunities. It is not a one-time activity but an iterative process that evolves as the data and system requirements change.  \\[0.1cm]
\textit{Descriptive statistics} [121] - It is used for understanding the characteristics of data in RE. They provide a summary of key features, central tendencies, and variations within the dataset. In the context of RE, descriptive statistics are employed to gain insights into user behavior, item popularity, and the distribution of ratings. Calculate basic statistics (mean, median, standard deviation) for numerical features, such as user ratings and item popularity. Summarize categorical variables, such as item categories or user demographics.\\[0.1cm]
\textit{Data visualization} [122] - In EDA, it helps making data-driven decisions, communicating findings, and uncovering actionable insights. The choice of visualization techniques depends on the nature of the data and the specific questions to be answered about user behavior, item characteristics, and overall system performance. It allows us to gain insights into patterns, trends, and relationships within the data using various visualization techniques, including:
\begin{itemize}
    \item histograms that display the distribution of numerical features.
    \item box-plots that identify outliers and distribution characteristics.
    \item heat-maps that visualize correlations between variables.
    \item scatter plots that explore relationships between numerical features. 
\end{itemize}
\textit{User-item interaction matrix} [123] - It is a fundamental component in the EDA process for RE. This matrix represents the interactions between users and items, where rows correspond to users, columns correspond to items, and the entries represent some form of interaction (e.g., ratings, clicks, purchases). Analyzing this matrix provides insights into user preferences, item popularity, and data sparsity. \\[0.1cm]
\textit{Temporal analysis} [124] - Analyze how user interactions with items evolve. Use time series plots to identify temporal trends, seasonality, or changes in user behavior. It is not only about understanding past behavior but also about informing predictions and strategies for the future. By examining how user-item interactions change over time, RE can adapt to evolving user preferences and market dynamics. It's an iterative process that involves continuous monitoring and adjustment based on temporal insights.\\[0.1cm]
\textit{Association rule mining} [125] - Use association rule mining methods (Apriori) are used to find associations and frequently occurring items. Use association rules visualization to find products that users commonly co-purchase. It helps identify interesting relationships, patterns, and associations between different items in a dataset. In the context of RE [126], association rule mining can reveal co-occurrences and dependencies between items, leading to insights that can be used for personalized recommendations. By revealing patterns of co-occurrence, RE can enhance their ability to suggest relevant items to users based on their historical interactions.

\paragraph{Feature engineering}
Choosing the ideal characteristics to train the ML model [127] is known as feature engineering. It involves transforming raw data from multiple sources into meaningful features that capture the essence of the products and user preferences. By engineering features, we can uncover hidden patterns and relationships [128] that can improve the accuracy of our recommendations. One common technique in feature engineering is one-hot encoding. This technique converts categorical features, such as product categories or user demographics, into binary vectors. Each category is represented as a separate binary variable [129], indicating whether the product or user belongs to that category. One-hot encoding allows the RE to easily compare and analyze different categories, enabling more accurate recommendations. Another technique is numerical feature scaling. In some cases, numerical features, such as product prices or user ratings, may have different scales or ranges. Scaling these features ensures that they are on a similar scale, preventing any bias in the recommendation process. Common scaling techniques include standardization and normalization [130], which transform the data to have zero mean and unit variance or to a specific range, respectively. Additionally, feature engineering can involve creating new features based on existing ones. For example, we can calculate the average rating of a product based on user reviews or the popularity of a product based on the number of purchases. These engineered features provide additional insights and can improve the performance of the RS.

\paragraph{Data normalization}
Data normalization refers to the process of transforming and scaling the input data to a standard format or range. The goal [131] is to ensure that different features or variables used by the RE are on a similar scale, preventing certain features from dominating others due to differences in their magnitudes. Normalization is particularly important when using algorithms that are sensitive to the scale of input features, such as many ML models. Normalize the data to ensure that it is on a consistent scale. This helps prevent any bias or distortion during model training. \\[0.1cm]
\textit{Min-max scaling} [132] - This method scales the data to a specific range (commonly [0, 1]). Each data point is subtracted by the minimum value in the feature and divided by the range (difference between the maximum and minimum values), as given in Eq. (1). 
\begin{equation}
\label{eq:1} 
X_N=\frac{X-X_{Min}}{X_{Max}-X_{Xmin}},
\end{equation} 
\begin{itemize}
    \item $X_N$ represents the normalized value of an observation,
    \item $X$ is the original value of an observation,
    \item $X_{Min}$ is the minimum value of the feature across all observations, and
    \item $X_{Max}$ is the maximum value of the feature across all observations.
\end{itemize}
\textit{Z-Score} [133] - Standardization transforms the data to have a mean ($\mu$) of 0 and a standard deviation ($\sigma$) of 1. It involves subtracting $\mu$ and dividing by $\sigma$.
\begin{equation}
\label{eq:2} 
X_S=\frac{X-\mu}{\sigma},
\end{equation} 
\begin{itemize}
    \item $X_S$ is the Z-score,
    \item $X$ is the original value of an observation,
    \item $\mu$ is the mean of the population or sample, and 
    \item $\sigma$ is the standard deviation of the population or sample.
\end{itemize} 
\textit{Robust scaling} [134] - Robust scaling is like standardization but uses the median and interquartile range (IQR) instead of the mean and standard deviation. It is less sensitive to outliers.
\begin{equation}
\label{eq:3} 
X_R=\frac{X-Median}{IQR},
\end{equation} 
\begin{itemize}
    \item $X_R$ is robustly scaled value,
    \item $X$ is the original value of an observation,
    \item Median is the median of the feature across all observations, and
    \item IQR is the interquartile range of the feature, calculated as the difference between the $75^{th}$ percentile (Q3) and the $25^{th}$ percentile (Q1) of the data.
\end{itemize}  
\textit{Log transformation} [135] - Log transformation is useful when the data has a skewed distribution. It can help in making the data more symmetrical and reducing the impact of extreme values.
\begin{equation}
\label{eq:4} 
X_{Log}=Log(X),
\end{equation} 
\begin{itemize}
    \item $X_{Log}$ is the transformed value,
    \item $X$ is the original value of an observation, and
    \item $Log$ is the logarithm function with base $b$ (common choices for $b$ are 10, the natural logarithm $e$ or 2).
\end{itemize}  	 
\textit{Normalization for sparse data} [136] - In RE with sparse data (user-item interactions), it is common to normalize data within each user or item independently. This ensures that the recommendations are based on relative preferences within each user or item context	

\subsubsection{Customer segmentation based on RFM model}
Customer segmentation using the RFM model is a popular marketing and customer relationship management strategy. RFM [137]  stands for Recency, Frequency, and Monetary, and it is a data-driven strategy to consumer segmentation based on previous behavior and interactions with the business. In RFM analysis [109], the quantitative values of R, F, and M qualities are usually determined by rating or scoring clients based on their purchasing behavior.
\begin{itemize}
    \item Recency (R): It indicates how recently a consumer made a purchase. A lower suggests a recent purchase, implying greater involvement and interest.
    \item Frequency (F): How frequently a customer purchases. A higher score indicates more frequent purchases and therefore more loyalty.
    \item Monetary (M): How much money does a customer spend on purchases? A higher score indicates a larger monetary worth and therefore a higher spender.
\end{itemize} 
\begin{table}[t!]\centering
\caption{\textrm{RFM segmentation table with quantitative values representing recency, frequency, and monetary contributions}}
\footnotesize
\label{tab:5}       % Give a unique label 
\resizebox{\linewidth}{!}{%
\begin{tabular}{|l|c|c|c|l|}
\hline
\multicolumn{1}{|c|}{\textbf{Segment}}                        & \textbf{Recency} & \textbf{Frequency} & \textbf{Monetary} & \multicolumn{1}{c|}{\textbf{Description}}                                                                            \\ \hline
\begin{tabular}[c]{@{}l@{}}Best \\ customers\end{tabular}     & 1                & 1                  & 1                 & \begin{tabular}[c]{@{}l@{}}Bought most recently, most \\ often, and spend the most.\end{tabular}                     \\ \hline
\begin{tabular}[c]{@{}l@{}}Loyal \\ customers\end{tabular}    & 2                & 2                  & 3                 & \begin{tabular}[c]{@{}l@{}}Buy on a regular basis. \\ Responsive to promotions.\end{tabular}                         \\ \hline
\begin{tabular}[c]{@{}l@{}}Big \\ spenders\end{tabular}       & 1                & 4                  & 1                 & \begin{tabular}[c]{@{}l@{}}Spend big money but do so \\ infrequently.\end{tabular}                                   \\ \hline
Almost lost                                                   & 3                & 2                  & 4                 & \begin{tabular}[c]{@{}l@{}}Haven’t purchased for some \\ time but spent a lot when did.\end{tabular}                 \\ \hline
\begin{tabular}[c]{@{}l@{}}Lost \\ customers\end{tabular}     & 4                & 1                  & 5                 & \begin{tabular}[c]{@{}l@{}}Haven’t purchased for the \\ longest time, but spent a lot \\ when they did.\end{tabular} \\ \hline
\begin{tabular}[c]{@{}l@{}}Inactive \\ customers\end{tabular} & 5                & 5                  & 5                 & \begin{tabular}[c]{@{}l@{}}Last purchased a long time \\ ago and spent little.\end{tabular}                          \\ \hline 
\end{tabular}

}
\end{table}
In Table 5, the segments are ordered from the most valuable (best customers) to least (lost cheap customers), with \enquote{1} indicating the highest/best score and \enquote{5} the lowest/worst for the respective RFM measures. However, RFM alone may not be sufficient for deeper consumer segmentation because it does not consider other potentially relevant characteristics such as customer preferences, product categories, seasonality, and customer lifetime value. As a result, more variables may be incorporated to fine-tune the segmentation and make it more predictive and effective. For example, adding demographic data, channel preferences, or even psychographic information can result in more refined segmentation. As more variables are added to the study, the complexity grows, necessitating automated analysis. Automated analysis may use ML models capable of handling multidimensional information, revealing patterns that traditional RFM analysis does not. In the research work [138] author developed a clustering technique that incorporates the RFM model and formal idea analysis to overcome standard clustering restrictions. Their approach produces both explicit and implicit knowledge of actual marketing tactics. In [139] author employed the K-means clustering algorithm with RFM attributes to categorize consumers into groups and rank them according to Customer Lifetime Value, demonstrating how RFM-oriented features may be used for effective segmentation. The author presented a multi-behavior RFM model in [140] that employs a self-organizing map algorithm to assess the weight correlations between user behaviors and optimize techniques. \par
\noindent
\textit{\textbf{RQ 6:}} How does the choice of database technology (SQL vs. NoSQL vs. Data Warehouses vs. Data Lakes) impact on the efficiency and scalability of RE? \\[0.1cm]
RE benefits from various technologies, including SQL for data integrity and transactional consistency, NoSQL for scalability and flexibility, Data Warehouses for powerful analytics, and Data Lakes for a scalable repository for diverse datasets, all of which contribute to the overall performance and scalability of RE. \\[0.1cm]
\textit{\textbf{RQ 7:}} What are the best practices in data preprocessing, tagging, and feature engineering for ensuring the effectiveness of ML-powered RE? \\[0.1cm]
To assure consistency, data preprocessing includes cleaning the data, dealing with missing values, and scaling or normalizing characteristics. To improve the RE comprehension of item attributes, precise item labeling with metadata is necessary for effective tagging, especially in systems such as content-based filters. To convert unstructured data into meaningful features that represent user behavior and preferences, feature engineering is essential. This process involves methods such as numerical feature scaling, one-hot encoding for categorical data, and building composite features that represent user interactions and preferences. To create a model that can anticipate user preferences with accuracy and improve customization, several procedures are necessary. \\[0.1cm]
\textit{\textbf{RQ 8:}} How does ML improve the process of customer segmentation and recommendation in comparison to traditional methods? \\[0.1cm]
ML improves client segmentation and recommendation procedures dramatically over traditional approaches by using advanced clustering models and ensemble-based systems [141] to provide individualized product recommendations. It uses extensive data analysis, including feedback, to produce accurate personal credit and size estimates. In a variety of industries [142], including healthcare and energy management, ML algorithms assess client behavior and environmental elements more effectively. Bayesian statistics, CF, and association rule mining improve recommendation accuracy and customization. ML techniques [143] [144] also incorporate feature extraction from customer assessments, resulting in better suggestions, exceeding traditional methods in terms of precision, adaptability, and user engagement.\\[0.1cm]
\textit{\textbf{RQ 9:}} What role do advanced ML techniques play in automating and optimizing the customer segmentation process, especially in the context of the RFM model? \\[0.1cm]
Advanced ML approaches automate and optimize customer segmentation, notably in the RFM model, by allowing for customer group clustering, complicated data analysis, and actionable insights for focused marketing and improved customer relationships. A hybrid approach that combines RFM and ML analyzes client segments more effectively. [145] this strategy helps firms understand consumer characteristics better, target the correct customers, keep them, and boost revenue. Furthermore, ML approaches extract features [146] [147] from existing properties and recommend marketing tactics for customer retention.  \\[0.1cm]
\textit{\textbf{RQ 10:}} How can additional variables beyond RFM attributes be incorporated for more refined customer segmentation? \\[0.1cm]
More precise client segmentation is possible when recommendation algorithms incorporate characteristics beyond RFM qualities. These extra variables may consist of user interaction data (clicks, views, engagement time), psychographic information (lifestyle, preferences), demographic data (age, gender, and location), and product category preferences. A multi-dimensional analysis is made possible by integrating various data, which results in a more thorough client profile. Then by analyzing these various data sets, ML techniques like clustering algorithms can find subtle trends and client base segments. By considering the unique requirements and interests of various consumer groups, this method improves the accuracy of personalized advice and marketing strategies.

\begin{figure*}[t!] 
	\centering 
	\includegraphics[width=0.8\textwidth]{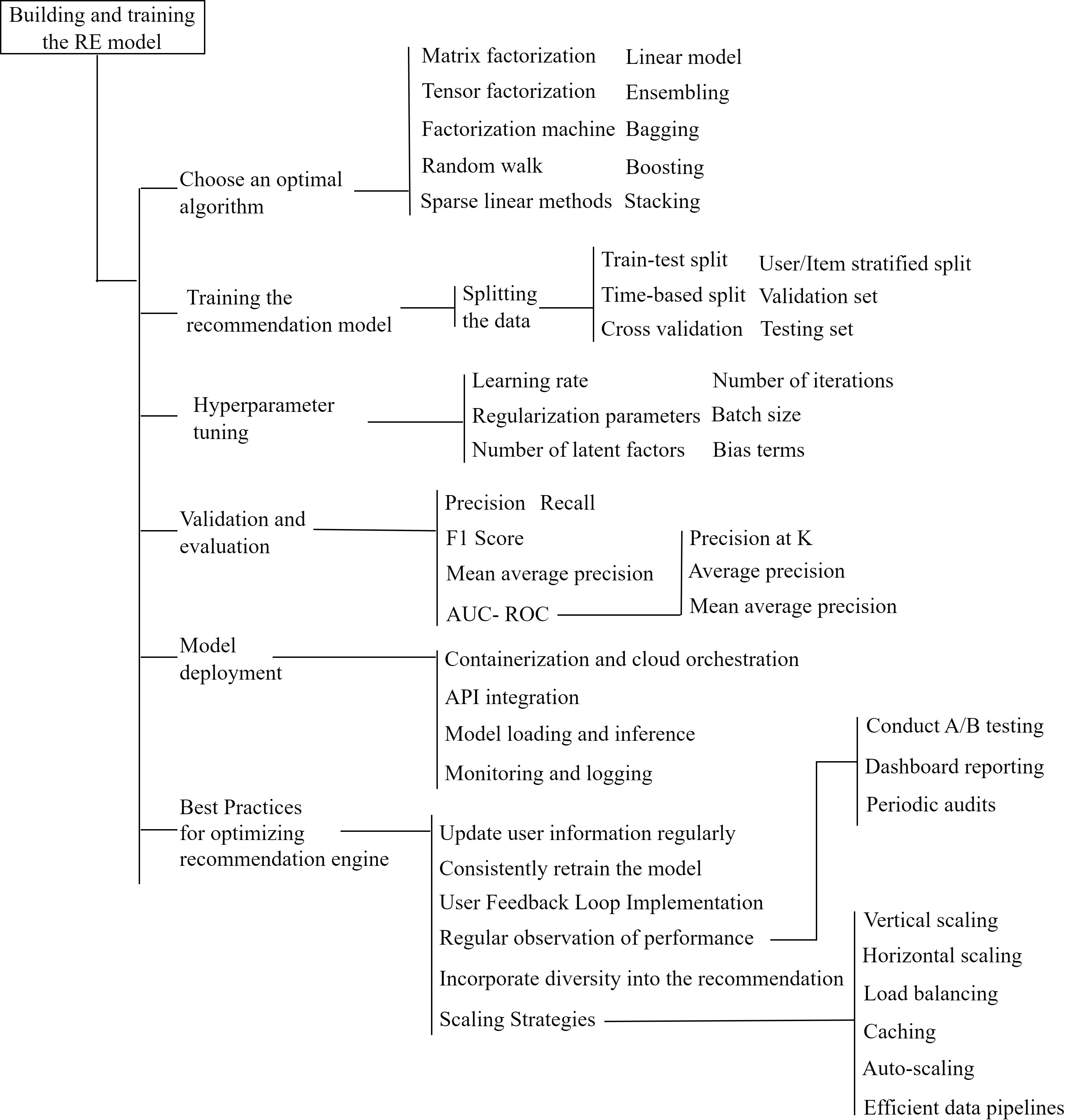}
	\\
	\caption{\textrm{Step-by-step-structured process of building and training RE model}}
	\label{fig:13}       
\end{figure*}

\subsection{Building and training the RE Model}
Building and training a RE model is a structured process as shown in Figure 13 that begins with the selection of a suitable algorithm—options include matrix factorization, random walks, or ensemble methods like bagging and boosting. After choosing the optimal algorithm, the model is trained with data that has been split into training, validation, and testing sets. Hyperparameter tuning is conducted to refine the model further, focusing on parameters such as learning rate, regularization factors, and the number of iterations. The performance of the model is evaluated using metrics such as precision, recall, F1 Score, and AUC-ROC. Once validated, the model is deployed with considerations for containerization, cloud deployment, API integration, and monitoring. To optimize the RE, best practices such as regular updates of user information, feedback loop implementation, performance metric reviews, diversity incorporation, and appropriate scaling strategies are employed. Additionally, techniques like A/B testing and dashboard reporting are used to ensure the model's effectiveness in practical applications. 

\subsubsection{Choosing an optimal algorithm}
The optimal algorithm for building an RE depends on various factors, including the nature of the data, the size of the dataset, the availability of features, and the specific use case. Based on the requirements and data characteristics, choose the most suitable recommendation algorithm. Consider factors like data sparsity, scalability, and personalization level. 

\paragraph{Matrix factorization}
Matrix Factorization has proven to be effective in RE, especially in situations where the user-item interaction matrix is sparse  [148], and it helps in capturing latent patterns that might not be apparent in the original data. This technique has been used in CF methods to provide personalized recommendations based on user preferences and item characteristics.

\begin{equation}
\label{eq:5}, 
R\approx UV^T.
\end{equation} 
\begin{itemize}
    \item $R$ is the user-item interaction matrix. Rows represent users, columns represent items, and entries contain the user-item interactions (e.g., ratings, purchases, views),
    \item $U$ is the user matrix. Each row represents a user, and the entries are the user's latent factors, and 
    \item $V$ is the item matrix. Each row represents an item, and the entries are the item's latent factors.
\end{itemize} 
		
The objective of training is to determine the optimal values for $U$ and $V$ such that the product $UV^T$ approximates $R$ as closely as possible. This is achieved by minimizing a loss function, often based on the difference between predicted and actual user-item interactions. The optimization problem is typically formulated as follows:

\begin{equation}
\label{eq:6} 
Min_{U,V}\sum_{(i,j)\epsilon R}\left ( R_{ij}-\left ( UV^T \right ) \right )^{2}+\lambda (\left \| U \right \|^{2}+\left \| V \right \|^{2}).
\end{equation} 
\begin{itemize}
    \item $(i,j)$ represents the indices of observed user-item interactions in $R$,
    \item $R_{ij}$ is the actual user-item interaction value in $R$,
    \item $\left (UV^T  \right )_{ij}$ the predicted interaction value obtained by multiplying the corresponding user and item vectors,
    \item $ \lambda $ is a regularization parameter to control the model's complexity and prevent overfitting, and
    \item $\left \| U^2 \right \|$ and $\left \| V^2 \right \|$ are regularizing terms to penalize large values in U and V.
\end{itemize} 
	  
The training process involves adjusting the values in $U$ and $V$ iteratively to minimize the loss function. Various optimization techniques, such as stochastic gradient descent (SGD) or alternating least squares (ALS), are commonly used for this purpose [149]. Different variants of matrix factorization may have additional terms or modifications in the objective function, depending on the specific algorithm used. 

\paragraph{Tensor factorization}
Tensor factorization [150] is an extension of matrix factorization designed to handle higher-dimensional data structures known as tensors. In RE, tensors can represent interactions between users, items, and additional dimensions such as time, location, or context. The basic idea of Tensor Factorization is to decompose a higher-order tensor into a combination of lower-dimensional latent factors. The equation for Tensor Factorization can be expressed as:

\begin{equation}
\label{eq:7} 
T \approx \sum_{k-1}^{K}U_k  \otimes V_k  \otimes W_k.
\end{equation} 
\begin{itemize}
    \item $T$ is the observed tensor, representing user-item interactions in a multi-dimensional space,
    \item $U_k$, $V_k$ and $W_k$ are factor matrices for each mode (dimension) of the tensor, 
    \item $\otimes$ denotes the tensor product, and
    \item $K$ represents the number of latent factors. 
\end{itemize} 
	
During training, the objective is to identify the optimal values for the factor matrices ($U_k$, $V_k$, $W_k$) that minimize the difference between the predicted tensor $U_k \otimes V_k \otimes W_k$ and the observed tensor $T$. The optimization process involves adjusting the factor matrices iteratively. Tensor Factorization is particularly useful when dealing with multi-dimensional data, providing a more detailed understanding of user-item interactions by capturing complex patterns in various contexts or conditions. For example, in RE, additional dimensions could include time (capturing how preferences change over time) or context (capturing preferences in different situations). The optimization techniques used for Tensor Factorization can include variations of gradient descent, alternating least squares, or other tensor factorization algorithms depending on the characteristics of the data and the goals of the RE. 

\paragraph{Factorization machine}
Factorization machines are flexible models and can be applied to various recommendation scenarios  [151], providing accurate predictions, and capturing intricate patterns in user-item interactions. They have been used alongside traditional CF and content-based methods to enhance RE performance.

The factorization machines are very effective in dealing with high-dimensional data, making them suitable for scenarios where the feature space is large, such as in RS with numerous users and items. It can handle sparse data efficiently, making it well-suited for situations where the interaction matrix between users and items is sparse. This is common in RE where users interact with only a small subset of items. They are designed to model second-order interactions between features. In the context of RE, this means capturing how the preferences for certain items may depend on combinations of user and item features. The process is robust in the presence of incomplete data. In RS, where not all users interact with all items, factorization machines can provide accurate predictions even when data is missing and computationally efficient which makes them suitable for large-scale recommendation tasks. The basic equation for a Factorization Machine of order 2 (FM2) is as follows:

\begin{equation}
\label{eq:8} 
\hat{y}(x)=w_0+\sum_{i+1}^{n}w_ix_i+\sum_{i=1}^{n}\sum_{j=i+1}^{n}<v_i,v_j>x_ix_j.
\end{equation} 
\begin{itemize}
    \item $\hat{y}(x)$ is the predicted output,
    \item $w_0$ is the global bias term,
    \item $w_i$ are the weights for individual features,
    \item $v_i$ are the latent vectors for feature interactions, and
    \item In RS, $x_i$ and $x_j$ can represent user and item features, $v_i$ and $v_j$ capture the interactions between the features.
\end{itemize} 

\paragraph{Random walk }
Random Walk is indeed a graph-based algorithm [152][153] that explores connections between items or users in a network. It involves navigating the network by moving from one node to another based on random decisions, capturing similarity between items or users in the process. This captured similarity can then be used to make recommendations. The RE data is represented as a graph, where nodes represent users or items, and edges represent connections or interactions between them. For example, users who have similar preferences or items that are frequently co-purchased could be connected. The algorithm starts with a seed node (e.g., a user or an item) and performs random walks by traversing edges in the graph. At each step, the algorithm randomly chooses a neighboring node to move to. As the random walk progresses, it captures similar information between nodes based on the paths taken. Nodes that are frequently visited together during random walks are considered more similar. The captured similarity information is used to make recommendations. For example, items that are like the items frequently visited by a user in the random walk can be recommended to that user. In a simplified form, the similarity between two nodes (items or users) $i$ and $j$ based on random walks could be calculated as the frequency with which they co-occur during the walks:

\begin{equation}
\label{eq:9} 
Sim(i,j)=\frac{No. \ of \ RW \ where \ i \ and \ j \ cooccur}{Total \ number \ of \ RW}.
\end{equation}

\paragraph{Sparse linear methods}
The sparse linear methods (SLIM) algorithm [154][155] involves factorizing the item-item similarity matrix to capture underlying patterns in item relationships. The objective function of SLIM includes a sparsity-inducing regularization term to handle the inherent sparsity in recommendation matrices. Below is a simplified version of the SLIM objective function.
\begin{equation}
\label{eq:10} 
Min_{W,H}\sum_{i=1}^{N}\sum_{j=1}^{N}(S_{ij}-(W^TH)_{ij})^{2}+\lambda(\left \| W \right \|_1+\left \| H \right \|)_1.
\end{equation} 
\begin{itemize}
    \item $N$ is the number of items,
    \item $S$ is the item-item similarity matrix,
    \item $W$ is a matrix representing the weights for each item's contribution to others.
    \item $H$ is a matrix representing the linear combination of items,
    \item $(W^TH)_{ij}$ is the dot product of the $i$-th row of $W$ and the $j$-th column of $H$, representing the predicted similarity between items $i$ and $j$,
    \item $\lambda$ is a regularization parameter to control the sparsity of the solution, and
    \item $\left \| W \right \|_1$ and $\left \| L \right \|_1$ are L1 norms of the matrices and respectively, introducing sparsity to the solution.
\end{itemize} 
The objective function is minimized using optimization techniques, such as stochastic gradient descent or alternating least squares. The regularization term encourages the model to find a sparse solution, which is advantageous in the context of RS where the user-item interaction matrix is inherently sparse. This simplified version of the SLIM objective function illustrates the key components of the algorithm. In practice, various enhancements and optimizations may be introduced, and the specific details of the regularization term or optimization method may vary depending on the implementation or variant of SLIM being used.

\paragraph{Linear model}
A linear model, in the context of RE  [156], is a type of model that predicts user preferences for items based on linear relationships between features. These models assume that the relationship between user and item features ($F$) is linear, meaning the predicted preference is a weighted sum of the features, possibly with an additional bias term. Linear model for predicting user-item preferences is represented as:
\begin{equation}
\label{eq:11} 
\hat{r}_{ui}=w_0+w_1\cdot F_1+w_2\cdot F_2+ \cdots + w_n\cdot F_n.
\end{equation} 
\begin{itemize}
    \item $\hat{r}_{ui}$ is the predicted preference for user $u$ and $i$,
    \item $w_0$ is the bias term, representing the baseline preference,
    \item $w_1, w_2, \cdots, w_n$ are the weights assigned to the features, and
    \item $F_1, F_2, \cdots, F_n$ are the features associated with the user-item pair.
\end{itemize}  
Linear models are computationally efficient, easy to interpret, and quick to train. They serve as a fundamental building block in ML and are often used as baseline models. However, their simplicity comes with limitations, particularly in capturing complex and nonlinear relationships in data. In RE where user preferences can be influenced by intricate patterns, more advanced models, such as CF, matrix factorization, or deep learning, are often explored to capture nuanced user-item interactions.
	
\paragraph{Ensembling}
Ensembling refers to the technique of combining multiple models [157] to improve overall predictive performance or robustness. The idea is that by aggregating the predictions of multiple models, the ensemble can achieve better results than any individual model. Ensembling is a common practice in ML [132], and it can be applied to various types of models, including both linear and nonlinear models. In RE, ensembling can be applied to combine predictions from multiple recommendation models to form a final recommendation. 
\begin{equation}
\label{eq:12} 
\hat{r}^{ensemble}_{ui}= \sum_{k=1}^{n}w_k \cdot \hat{r}^{(k)}_{ui}.
\end{equation} 
\begin{itemize}
    \item $\hat{r}^{ensemble}_{ui}$ is the final ensemble prediction for user $u$ and item $i$,
    \item $w_k$ represents the weight assigned to the prediction from model $k$. The weights are typically determined during the training or validation phase and can be static or dynamic. The choice of weighting strategy depends on the characteristics of the models and the data, and
    \item $\hat{r}^{(k)}_{ui}$ is the predicted preference score from model $k$ for user $u$ and item $i$.
\end{itemize}   
The specific weighting strategy and the number of models involved in the ensemble would depend on the characteristics of the recommendation task, the diversity of the models, and the available data. The weights can be tuned to optimize overall recommendation performance.

\begin{table*}[t!]\centering
\caption{\textrm{Comparison of algorithm analysis based on performance}}
\footnotesize
\label{tab:6}       % Give a unique label 
% \resizebox{\linewidth}{!}{%
\begin{tabular}{|c|l|l|l|l|l|l|}
\hline
\textbf{Article}                                                     & \multicolumn{1}{c|}{\textbf{Algorithm}}                          & \multicolumn{1}{c|}{\textbf{Application}}                                    & \multicolumn{1}{c|}{\textbf{Dataset}}                                               & \multicolumn{1}{c|}{\textbf{Method}}                                           & \multicolumn{1}{c|}{\textbf{Similarity Measure}}                               & \multicolumn{1}{c|}{\textbf{Solutions Addressed}} \\ \hline
{[}148{]}                                                            & \begin{tabular}[c]{@{}l@{}}Matrix \\ Factorization\end{tabular}  & \begin{tabular}[c]{@{}l@{}}Personalized \\ recommendations\end{tabular}      & User-item ratings                                                                   & \begin{tabular}[c]{@{}l@{}}Decomposes user-item \\ matrix into latent factors\end{tabular}      & \begin{tabular}[c]{@{}l@{}}Cosine similarity, \\ Euclidean distance\end{tabular} & \begin{tabular}[c]{@{}l@{}}Efficient for sparse \\ datasets\end{tabular}                         \\ \hline
{[}150{]}                                                            & \begin{tabular}[c]{@{}l@{}}Tensor \\ Factorization\end{tabular}  & \begin{tabular}[c]{@{}l@{}}Context-aware \\ recommendations\end{tabular}     & \begin{tabular}[c]{@{}l@{}}Multi-dimensional \\ user-item context data\end{tabular} & \begin{tabular}[c]{@{}l@{}}Extends matrix factoriza-\\ tion to higher dimensions\end{tabular}   & \begin{tabular}[c]{@{}l@{}}Tensor decomposition \\ techniques\end{tabular}       & \begin{tabular}[c]{@{}l@{}}Suitable for context-\\ aware recommendations\end{tabular}            \\ \hline
{[}151{]}                                                            & \begin{tabular}[c]{@{}l@{}}Factorization \\ Machine\end{tabular} & \begin{tabular}[c]{@{}l@{}}Diverse recomm-\\ endation scenarios\end{tabular} & \begin{tabular}[c]{@{}l@{}}High-dimensional \\ sparse data\end{tabular}             & \begin{tabular}[c]{@{}l@{}}Interacts with variables \\ using factorized parameters\end{tabular} & Factorized similarity                                                            & \begin{tabular}[c]{@{}l@{}}Efficient in high-dime-\\ nsional sparse data\end{tabular}            \\ \hline
\begin{tabular}[c]{@{}c@{}}{[}152{]}\\    \\  {[}153{]}\end{tabular} & \begin{tabular}[c]{@{}l@{}}Random \\ Walk\end{tabular}           & \begin{tabular}[c]{@{}l@{}}Similar item \\ recommendations\end{tabular}      & \begin{tabular}[c]{@{}l@{}}Graph-based user-\\ item networks\end{tabular}           & \begin{tabular}[c]{@{}l@{}}Navigates graph via \\ random steps\end{tabular}                     & Transition probabilities                                                         & \begin{tabular}[c]{@{}l@{}}Good for capturing net-\\ work-based similarities\end{tabular}        \\ \hline
\begin{tabular}[c]{@{}c@{}}{[}154{]}\\    \\  {[}155{]}\end{tabular} & \begin{tabular}[c]{@{}l@{}}Sparse Linear \\ Methods\end{tabular} & Large-scale   RS                                                             & \begin{tabular}[c]{@{}l@{}}User-item interaction \\ data\end{tabular}               & \begin{tabular}[c]{@{}l@{}}Learns sparse item-item \\ similarity matrix\end{tabular}            & \begin{tabular}[c]{@{}l@{}}Learned item \\ similarities\end{tabular}             & \begin{tabular}[c]{@{}l@{}}Effective in handling \\ sparse data\end{tabular}                     \\ \hline
{[}156{]}                                                            & Linear Model                                                     & \begin{tabular}[c]{@{}l@{}}Basic user-item \\ matching\end{tabular}          & \begin{tabular}[c]{@{}l@{}}Basic user-item\\ features\end{tabular}                  & Simple linear relationships                                                                     & Linear coefficients                                                              & Quick, easy to interpret                                                                         \\ \hline
{[}132{]}                                                            & Ensembling                                                       & Robust RS                                                                    & Diverse                                                                             & Combines multiple models                                                                        & \begin{tabular}[c]{@{}l@{}}Aggregated model \\ outputs\end{tabular}              & \begin{tabular}[c]{@{}l@{}}Robust, improved \\ accuracy\end{tabular}                             \\ \hline
{[}158{]}                                                            & Bagging                                                          & \begin{tabular}[c]{@{}l@{}}Enhancing model \\ performance\end{tabular}       & Any                                                                                 & \begin{tabular}[c]{@{}l@{}}Aggregate predictions from \\ bootstrap samples\end{tabular}         & Aggregated predictions                                                           & \begin{tabular}[c]{@{}l@{}}Reduces model \\ variance\end{tabular}                                \\ \hline
{[}159{]}                                                            & Boosting                                                         & \begin{tabular}[c]{@{}l@{}}Performance \\ optimization\end{tabular}          & Any                                                                                 & \begin{tabular}[c]{@{}l@{}}Sequential model building \\ on previous errors\end{tabular}         & Weighted aggregation                                                             & \begin{tabular}[c]{@{}l@{}}Improves weak \\ learners\end{tabular}                                \\ \hline
{[}160{]}                                                            & Stacking                                                         & Advanced RS                                                                  & Varied                                                                              & \begin{tabular}[c]{@{}l@{}}Meta-modeling over base \\ models\end{tabular}                       & \begin{tabular}[c]{@{}l@{}}Meta-learner \\ integration\end{tabular}              & \begin{tabular}[c]{@{}l@{}}Combines diverse \\ model predictions\end{tabular}                    \\ \hline
\end{tabular}
\end{table*}

\paragraph{Bagging (bootstrap aggregating)}
Bagging is a popular ensemble method that can be applied to RE. The idea behind bagging [158] is to train multiple instances of the same recommendation model on different subsets of the training data and combine their predictions to improve overall performance. Each model in the ensemble is trained independently, and the final prediction is often obtained by averaging (for regression tasks) or majority voting (for classification tasks).
\begin{equation}
\label{eq:13} 
\hat{r}^{ensemble}_{ui}= \frac{1}{n}\sum_{k=1}^{n}\hat{r}^{(k)}_{ui}.
\end{equation}
Here, $n$ is the number of models in the ensemble. Bagging can be applied to various recommendation models, including collaborative and content-based filtering. The choice of the base and the number of models in the ensemble depends on the characteristics of the data and the recommendation task. Bagging is a versatile technique that can be adapted to different recommendation scenarios, and it is often used in conjunction with other ensemble techniques such as boosting and stacking for further performance improvement.

\paragraph{Boosting}
It is an ensemble learning technique that can be applied to RE [159] to improve the predictive performance of models. Unlike bagging, boosting builds a sequence of models sequentially, with each model focusing on correcting the errors of its predecessor. It makes predictions for user-item pairs using the ensemble of models. The final prediction is often a weighted sum of the individual model predictions. Each subsequent model gives more attention to instances that were misclassified by the previous models, leading to an ensemble with improved overall accuracy. Popular boosting algorithms include AdaBoost, Gradient Boosting, and XGBoost. 
\begin{equation}
\label{eq:14} 
\hat{r}^{ensemble}_{ui}= \sum_{k=1}^{n}\alpha_k \cdot \hat{r}^{(k)}_{ui},
\end{equation}
$\alpha_k$ represents the weight assigned to the prediction from model $k$. The choice of model, learning rate, and number of boosting rounds are the key hyper-parameters to consider.

\paragraph{Stacking}
Stacking, also known as stacked generalization, is an ensemble learning technique that can be applied to RE. The key idea behind stacking is to train multiple base recommendation models and then use a meta-model to combine their predictions. The meta-model [160] is trained on the predictions of the base models, aiming to improve the overall predictive performance. Stacking allows the ensemble to benefit from the diverse strengths of different base models.
\begin{equation}
\label{eq:15} 
\hat{r}^{ensemble}_{ui}= g(\hat{r}^{(1)}_{ui}, \hat{r}^{(2)}_{ui}, \hat{r}^{(3)}_{ui}, \cdots, \hat{r}^{(n)}_{ui}).
\end{equation}
\begin{itemize}
    \item $g$ represents the meta-model.
    \item $\hat{r}^{(1)}_{ui}, \hat{r}^{(2)}_{ui}, \hat{r}^{(3)}_{ui}, \cdots, \hat{r}^{(n)}_{ui}$ are the predictions from the base models.
\end{itemize}   
Stacking is a flexible ensemble technique that can be adapted to different recommendation scenarios. The success of stacking depends on the diversity of base models and characteristics of the data. It is often used in conjunction with other ensemble methods to further improve overall performance. Table 6 represents an overview of various algorithm analyses concerning the effective performance of the system. The actual results and performance of these methods can vary based on specific implementation and data characteristics.

\begin{table*}[t!]\centering
\caption{\textrm{Comparison chart focusing on data splitting strategies RE}}
\footnotesize
\label{tab:7}       % Give a unique label 

\begin{tabular}{|l|l|l|l|l|l|l|}
\hline
\multicolumn{1}{|c|}{\textbf{Article}} & \multicolumn{1}{c|}{\textbf{\begin{tabular}[c]{@{}c@{}}Data splitting \\ strategy\end{tabular}}} & \multicolumn{1}{c|}{\textbf{\begin{tabular}[c]{@{}c@{}}Training set \\ proportion\end{tabular}}} & \multicolumn{1}{c|}{\textbf{\begin{tabular}[c]{@{}c@{}}Testing set \\ proportion\end{tabular}}} & \multicolumn{1}{c|}{\textbf{\begin{tabular}[c]{@{}c@{}}Validation set \\ proportion\end{tabular}}} & \multicolumn{1}{c|}{\textbf{\begin{tabular}[c]{@{}c@{}}Hyperparameter \\ tuning\end{tabular}}} & \multicolumn{1}{c|}{\textbf{Use case}}                                                  \\ \hline
{[}161{]}                              & Train-test split                                                                                 & 70-80\%                                                                                          & 10-20\%                                                                                         & 10\%                                                                                               & 20\%                                                                                           & General RS                                                                              \\ \hline
{[}162{]}                              & Time-based split                                                                                 & \begin{tabular}[c]{@{}l@{}}80\% (up to \\ a certain date)\end{tabular}                           & \begin{tabular}[c]{@{}l@{}}10\% (after \\ cutoff date)\end{tabular}                             & \begin{tabular}[c]{@{}l@{}}10\% (excluding \\ training and testing)\end{tabular}                   & \begin{tabular}[c]{@{}l@{}}20\% (training \\ and testing)\end{tabular}                         & Temporal RS                                                                             \\ \hline
{[}163{]}, {[}164{]}                   & \begin{tabular}[c]{@{}l@{}}Cross-validation \\ (k-fold)\end{tabular}                             & \begin{tabular}[c]{@{}l@{}}80\%  if 5-fold \\ (varies on each fold)\end{tabular}                 & \begin{tabular}[c]{@{}l@{}}Varies (e.g., \\ 20\% if 5-fold)\end{tabular}                        & \begin{tabular}[c]{@{}l@{}}Integrated within \\ each fold\end{tabular}                             & 10-15\%                                                                                        & \begin{tabular}[c]{@{}l@{}}Situations with \\ limited datasets\end{tabular}             \\ \hline
{[}165{]}                              & \begin{tabular}[c]{@{}l@{}}User/item \\ stratified split\end{tabular}                            & 70 to 80\%                                                                                       & 20 to 30\%                                                                                      & \begin{tabular}[c]{@{}l@{}}10 to 20\% (training \\ data)\end{tabular}                              & \begin{tabular}[c]{@{}l@{}}10 to 20\% \\ (training data)\end{tabular}                          & \begin{tabular}[c]{@{}l@{}}Ensuring no user/\\ item is over-\\ represented\end{tabular} \\ \hline
{[}166{]}                              & Validation set                                                                                   & \begin{tabular}[c]{@{}l@{}}50-70\% (excluding \\ validation set)\end{tabular}                    & 10-15\%                                                                                         & 10-15\%                                                                                            & 10-15\%                                                                                        & \begin{tabular}[c]{@{}l@{}}Complex models \\ requiring extensive \\ tuning\end{tabular} \\ \hline
{[}167{]}                              & Testing set                                                                                      & 70-85\%                                                                                          & 10-15\%                                                                                         & \begin{tabular}[c]{@{}l@{}}10-15\% (excluding \\ testing)\end{tabular}                             & \begin{tabular}[c]{@{}l@{}}10-15\% (excluding \\ testing)\end{tabular}                         & \begin{tabular}[c]{@{}l@{}}Final evaluation \\ of model's \\ performance\end{tabular}   \\ \hline
\end{tabular}
\end{table*}

\subsubsection{Training the Recommendation Model }
Train the RS model with the prepared training data. This process involves inputting the data into the model, fine-tuning the model's parameters, and enhancing the model's predictive performance. Employ techniques such as gradient descent and backpropagation to update the model's weights, as discussed previously in the model-building section.

\paragraph{Splitting the data }
The goal is to divide the dataset into training and testing sets to assess how well the model generalizes to unseen data. Choose the splitting strategy that best fits the RE and data characteristics. Additionally, it's essential to preprocess the data appropriately, handling missing values, normalizing, and encoding categorical variables, if necessary, before training the recommendation model. Here are some common techniques for splitting data for RE as shown in Table 7:\\[0.1cm]
\textit{Train-test split} [161] - Randomly divide the dataset into a training set and a testing set. For example, 80\% of the data for training and 20\% for testing may be used. Make sure that each user and item has representation in both the training and testing sets to ensure a diverse evaluation.\\[0.1cm]
\textit{Time-based split} [162] - For temporal RS, where recommendations are influenced by time, the data might be split chronologically. The training set contains data up to a certain date, and the testing set contains data after that date.\\[0.1cm]
\textit{Cross-validation} [163][164] - In situations where the dataset is limited, techniques like k-fold cross-validation may be applied. The data is divided into k subsets (folds), and the model is trained and tested k times, with each fold used as the test set once.\\[0.1cm]
\textit{User/item stratified split} [165] - Ensure that both the training and testing sets have a balanced representation of users and items. This can help prevent situations where certain users or items are not adequately represented in training or testing set.\\[0.1cm]
\textit{Validation set} [166]  - A smaller portion (e.g., 10-15\%) is used for hyperparameter tuning and model validation during training. Helps prevent overfitting by providing an independent dataset for fine-tuning.\\[0.1cm]
\textit{Testing set} [167] - Another portion (e.g., 10-15\%) is reserved for the final evaluation of the model's performance. It should not be used during training or hyperparameter tuning to ensure an unbiased evaluation.

\subsubsection{Hyperparameter tuning}
Hyperparameter tuning in an RE [168] involves optimizing the parameters that are not learned during the training process but are set prior to training. These parameters can significantly impact the performance of the recommendation algorithm. Experiment with different hyperparameters  [169] like learning rate, regularization strength, and batch size to find the optimal settings for the model. Hyperparameter tuning helps improve the model's performance and accuracy. \\[0.1cm]
\textit{Learning Rate} - while using learning-based recommendation algorithm, such as CF with matrix factorization, adjusting the learning rate can impact the convergence of the model.\\[0.1cm]
\textit{Regularization Parameters} - Regularization is crucial to prevent overfitting. Tuning parameters like L1 or L2 regularization strength can help find the right balance between fitting the training data well and generalizing to new data.\\[0.1cm]
\textit{Number of Latent Factors} - In matrix factorization-based methods, the number of latent factors determines the dimensionality of the feature space. Too few factors may lead to underfitting, while too many may result in overfitting. Experiment with different values to find the optimal number.\\[0.1cm]
\textit{Number of iterations (epochs)} -  The number of iterations during training affects how well the model converges. Too few iterations may result in an underfit model, while too many may lead to overfitting. Experiment with different numbers of epochs to find the right balance.\\[0.1cm]
\textit{Batch Size} - For models trained using stochastic gradient descent, the batch size can affect the stability of the training process. Smaller batch sizes may provide more noise in the parameter updates, while larger batch sizes may result in slower convergence. Experiment with different batch sizes to find the optimal one.\\[0.1cm]
\textit{Bias Terms} - RE often includes bias terms to account for user and item biases. Tuning the strength of these bias terms can impact the model's ability to capture user and item preferences. \par
When tuning hyperparameters, it is essential to use a validation set or cross-validation to assess the model's performance [170] under different parameter configurations. Using grid search or random search techniques enable to efficiently explore the hyperparameter space. The optimal hyperparameter values can depend on the specific characteristics of the data, so it is often necessary to experiment and iterate to find the best configuration for the RE.

\subsubsection{Validation of the recommendation model}
The trained model is validated using the testing set based on four evaluation metrics [171]-[173]: accuracy, precision, recall, and F1-score. Accuracy is used to assess the general validity of the recommendations, whereas precision and recall offer feedback about their relevance and completeness. The F1-score, which is the harmonic mean of precision and recall, provides an extensive overview of the model's performance, especially when a balance of precision and recall is required. These criteria were chosen because they are consistent with the study's research objectives and the accuracy of top-k recommendations in a RS. These metrics are particularly useful when dealing with scenarios where the algorithms recommend a fixed number (k) of items to users and to evaluate the efficiency and reliability of recommendation algorithms across a range of dataset sizes and categories.

\paragraph{Precision (P)}
It is a metric that focuses on the accuracy of the positive predictions made by a model. In the context of an RE, precision is concerned with the relevance of the recommended items. It answers the question: Out of all the items that the model recommended, how many were truly relevant to the user?
\begin{equation}
\label{eq:16} 
P = \frac{Number \ of \ Correctly \ Recommended \ Items}{Total \ number \ of \ Recommended \ Items},
\end{equation}
\begin{equation}
\label{eq:17} 
P= \frac{True \ Positives}{True \ Positives + False \ Positives}.
\end{equation}
A high precision indicates that when the model suggests items, it is often correct. However, precision alone might not give the full picture, especially if the model is missing a lot of relevant items (which recall would capture). Precision is often used in conjunction with recall, and the trade-off between these two metrics depends on the goals of the RS.

\paragraph{Recall (R)}
Recall is a metric that assesses the ability of an RE to capture all the relevant items for a user. It focuses on the fraction of relevant items that were successfully recommended. In the context of an RS, recall answers the question: Out of all the items that were truly relevant to the user, how many did the model manage to recommend?

\begin{equation}
\label{eq:18} 
R = \frac{Number \ of \ Correctly \ Recommended \ Items}{Total \ number \ of \ Relevant \ Items},
\end{equation}
\begin{equation}
\label{eq:19} 
R = \frac{True \ Positives}{True \ Positives + False \ Negatives}.
\end{equation}

A high recall indicates that the RE is effective at capturing a large proportion of the relevant items. However, a high recall might come at the cost of precision, as the system may recommend more items, including some that are not relevant. The balance between precision and recall depends on the goals of the RS. In some cases, achieving a higher precision is crucial to ensure that recommended items are highly relevant. In other cases, a higher recall might be more important, especially if it is crucial not to miss relevant items even at the expense of recommending a few irrelevant ones.

\paragraph{F1 score}
The F1 score [173] is a metric that combines both precision and recall into a single value. It provides a balance between the two metrics, offering a comprehensive evaluation of RE performance. The F1 score is particularly useful when there is an imbalance between positive and negative instances.
\begin{equation}
\label{eq:20} 
F1 = 2 \times \frac{P \cdot R}{P + R}.
\end{equation}
For tasks like recommendation, where the focus is on suggesting relevant items, metrics like [174] precision, recall, and F1 score are typically more informative, which allow us to evaluate the model's performance in capturing relevant items. While accuracy is a general metric for binary classification tasks, it might not be the most appropriate choice for evaluating RE, where the emphasis is often on the quality of recommended items rather than overall correctness.

\paragraph{Mean average precision (MAP)}
MAP is a metric commonly used to evaluate the performance of an RE [175], particularly in the context of information retrieval and ranked recommendations. MAP considers not only whether a recommendation is relevant but also the order/ranking of the recommendations. It focuses on the precision of the top-ranked items in the recommendation list, rather than assessing the ability of the system to recall all relevant items. \\[0.1 cm]
\textit{Precision at K (P\@K)} -  Measures the accuracy of the top K recommendations by calculating the ratio of relevant items to the total number of items in the top K recommendations.
\begin{equation}
\label{eq:21} 
P@K=\frac{Number \ of \ Relevant \ Items \ in \ Top \ K}{K}.
\end{equation}

\paragraph{Average precision (AP)} 
For a specific user, calculate the average precision by taking the mean of precision values at each position where a relevant item is found in the recommendation list.

\begin{equation}
\label{eq:22} 
AP=\frac{1}{No. \ of \ rel. \ item}\sum_{k=1}^{No. \ of \ rel. \ item}P@K\times Rel@K
\end{equation}

\paragraph{MAP} 
It calculates the mean of the average precision values across all users.
\begin{equation}
\label{eq:23} 
MAP=\frac{1}{Number \ of \ Relevant \ Users} \sum_{Users}AP.
\end{equation}
MAP doesn't explicitly incorporate recall into its calculation. It is more focused on assessing the precision at different positions in the recommendation list. For metrics that consider both precision and recall, it may be beneficial to consider alternatives such as the F1 score or precision-recall curves. MAP is a valuable metric for evaluating RS, particularly when the order of recommendations is crucial, but it primarily focuses on precision at different positions in the list.

\paragraph{Area under the receiver operating characteristic curve (AUC-ROC)}
AUC-ROC is commonly used in binary classification problems to evaluate the performance of a model in distinguishing between positive and negative instances. The ROC curve is a graphical representation of the trade-off between true positive rate (sensitivity or Recall) and false positive rate (1 - specificity or 1 - Precision) at various thresholds. The area under the ROC curve (AUC) [172] summarizes the overall performance across all possible thresholds. A higher AUC indicates better discrimination ability, implying that the model is better at ranking positive instances higher than negative instances.

When using AUC-ROC for RE, it's crucial to understand its limitations in the context of ranking. Other metrics like MAP or NDCG (normalized discounted cumulative gain) are often more directly applicable to ranking scenarios and provide a better understanding of the quality of recommendations. If RE involves predicting binary outcomes (e.g., whether a user will interact with a recommended item or not), AUC-ROC may be more directly applicable. While AUC-ROC can be adapted for certain aspects of RS evaluation, it's essential to recognize its limitations in the context of ranking and consider using metrics specifically designed for ranking scenarios for a more accurate assessment of the RE performance. 

A wide range of ML models, such as kNN, RF, SVM, deep matrix factorization (DeepMF), and Baseline methods such as popularity-based baseline (PBB), god-view pbb (GVPBB), feature analysis protocol (FAP), principal component analysis (PCA) were chosen for this study to combine with the various recommendation methods. Baseline methods are labeled with ($*$). These models were chosen based on their proven track record of managing high-dimensional data and their capacity to generalize across various datasets. These models were methodically paired with algorithms such that each model's advantages complimented the algorithms' features. This methodical approach was adopted to offer a balanced and comprehensive assessment of the algorithms' performance across multiple recommendations.  \par

A wide range of ML models, such as kNN, RF, SVM, deep matrix factorization (DeepMF), and Baseline methods such as popularity-based baseline (PBB), god-view pbb (GVPBB), feature analysis protocol (FAP), principal component analysis (PCA) were chosen for this study to combine with the various recommendation methods. Baseline methods are labeled with (*). These models were chosen based on their proven track record of managing high-dimensional data and their capacity to generalize across various datasets. These models were methodically paired with algorithms such that each model's advantages complimented the algorithms' features. This methodical approach was adopted to offer a balanced and comprehensive assessment of the algorithms' performance across multiple recommendations. 

\subsubsection{Results and discussion}
This section presents the outcomes from an extensive comparative analysis of various ML models and few Baseline methods employing diverse recommendation algorithms applied to MovieLens datasets. Each algorithm's performance was evaluated using four key metrics: accuracy, precision, recall, and the F1-score. These measures were chosen to provide a full assessment of the algorithms' performance and reliability. Table 8 highlights the performance comparison of various ML models that use different recommendation algorithms on the MovieLens datasets (MovieLens-100K and MovieLens-1M). This examination provides details about performance across various metrics, including accuracy, precision, recall, and F1-score. The findings offer insight into the relative strengths and limits of each ML model when combined with specific algorithms, allowing more precise evaluation of their applicability for various recommendation conditions. In the analysis using the MovieLens-100K dataset, the implementation of Matrix Factorization with the kNN model achieves balanced performance, with an accuracy of 0.760 and a recall of 0.690. In comparison, the Random Forest model outperforms kNN in every aspect, achieving a uniform score of 0.920 in accuracy, precision, recall, and F1-score, demonstrating its robust ability to handle recommendation tasks on this dataset. On the other hand, the SVM model significantly underperforms, notably in terms of accuracy (0.540), revealing its limits in this context. The evaluation of Tensor Factorization algorithm generates varied results. The Cross-Domain Multidimensional Tensor Factorization model has more uniform metrics, with an accuracy of 0.73 and a precision of 0.71, indicating its promise for cross-domain recommendation tasks. In contrast, the ATF model solely represented by an F1-score of 0.450, which is relatively low, indicating its poor performance. When moving from the smaller MovieLens-100K dataset to the bigger MovieLens-1M dataset, the Factorization Machine combined with DeepMF exhibits improved precision and recall while keeping an F1-score of 0.720. It highlights how well DeepMF scales and generalizes across larger datasets.

\begin{table*}[t!]\centering
\caption{\textrm{Comparative analysis of various machine learning algorithms on MovieLens datasets.}}
\footnotesize
\label{tab:8}       % Give a unique label 
\resizebox{\linewidth}{!}{%
\begin{tabular}{|c|c|l|l|c|c|c|c|}
\hline
\textbf{Reference}         & \textbf{Algorithm}                                                                & \multicolumn{1}{c|}{\textbf{Model}}                                                                                                                   & \multicolumn{1}{c|}{\textbf{Dataset}} & \textbf{Accuracy} & \textbf{Precision} & \textbf{Recall} & \textbf{F1 Score} \\ \hline
\multirow{3}{*}{{[}176{]}} & \multirow{3}{*}{\begin{tabular}[c]{@{}c@{}}Matrix \\ Factorization\end{tabular}}  & K-nearest neighbors (KNN)                                                                                                                             & \multirow{5}{*}{MovieLens-100K}       & 0.690             & 0.760              & 0.690           & 0.650             \\ \cline{3-3} \cline{5-8} 
                           &                                                                                   & Support vector machine (SVM)                                                                                                                          &                                       & 0.540             & 0.550              & 0.540           & 0.510             \\ \cline{3-3} \cline{5-8} 
                           &                                                                                   & Random forest (RF)                                                                                                                                    &                                       & 0.920             & 0.920              & 0.920           & 0.920             \\ \cline{1-3} \cline{5-8} 
{[}177{]}                  & \multirow{2}{*}{\begin{tabular}[c]{@{}c@{}}Tensor \\ Factorization\end{tabular}}  & \begin{tabular}[c]{@{}l@{}}Adversarial tensor factoriz-\\ ation (ATF)\end{tabular}                                                                    &                                       & -                 & -                  & -               & 0.450             \\ \cline{1-1} \cline{3-3} \cline{5-8} 
{[}178{]}                  &                                                                                   & \begin{tabular}[c]{@{}l@{}}Cross domain multidimensional \\ tensor factorization\end{tabular}                                                         &                                       & 0.730             & 0.710              & 0.680           & 0.690             \\ \hline
\multirow{2}{*}{{[}179{]}} & \multirow{2}{*}{\begin{tabular}[c]{@{}c@{}}Factorization \\ Machine\end{tabular}} & \multirow{2}{*}{\begin{tabular}[c]{@{}l@{}}Deep matrix factorization \\ (DeepMF)\end{tabular}}                                                        & MovieLens-100K                        & -                 & 0.720              & 0.690           & 0.700             \\ \cline{4-8} 
                           &                                                                                   &                                                                                                                                                       & MovieLens-1M                          & -                 & 0.740              & 0.710           & 0.720             \\ \hline
{[}180{]}                  & Random Walk                                                                       & Artificial bee colony Optimization                                                                                                                    & MovieLens-1M                          & -                 & 0.961              & 0.994           & -                 \\ \hline
\multirow{4}{*}{{[}181{]}} & \multirow{4}{*}{Linear Model}                                                     & \begin{tabular}[c]{@{}l@{}}Hybrid collaborative   \\ filtering recommend-\\ ation (HCFR)\end{tabular}                                                 & \multirow{4}{*}{MovieLens-1M}         & 0.826             & 0.834              & 0.773           & 0.802             \\ \cline{3-3} \cline{5-8} 
                           &                                                                                   & \begin{tabular}[c]{@{}l@{}}Support vector machines\\ improved ant colony \\ optimization (SVM- IACO )\end{tabular}                                    &                                       & 0.39              & 0.39               & 0.38            & 0.32              \\ \cline{3-3} \cline{5-8} 
                           &                                                                                   & \begin{tabular}[c]{@{}l@{}}Content-based collabora-\\ tive filter (CbCF)\end{tabular}                                                                 &                                       & 0.93              & 0.81               & 0.83            & 0.82              \\ \cline{3-3} \cline{5-8} 
                           &                                                                                   & \begin{tabular}[c]{@{}l@{}}Eagle deep neural based \\ movie recommender system\end{tabular}                                                           &                                       & 0.96              & 0.84               & 0.86            & 0.851             \\ \hline
\multirow{3}{*}{{[}183{]}} & \multirow{3}{*}{\begin{tabular}[c]{@{}c@{}}Ensemble \\ Method\end{tabular}}       & \begin{tabular}[c]{@{}l@{}}Voting classifier-based\\ classifier (RDAM)\end{tabular}                                                                   & \multirow{3}{*}{MovieLens-100K}       & -                 & 1.000              & 0.980           & 0.985             \\ \cline{3-3} \cline{5-8} 
                           &                                                                                   & \begin{tabular}[c]{@{}l@{}}Voting classifier-based\\ classifier (TF-IDF)\end{tabular}                                                                 &                                       & -                 & 0.750              & 1.000           & 0.856             \\ \cline{3-3} \cline{5-8} 
                           &                                                                                   & \begin{tabular}[c]{@{}l@{}}Voting classifier-based\\ classifier (RDMA-Len Var)\end{tabular}                                                           &                                       & -                 & 0.978              & 0.971           & 0.973             \\ \hline
\multirow{3}{*}{{[}184{]}} & \multirow{3}{*}{Stacking}                                                         & \begin{tabular}[c]{@{}l@{}}CF with \\ k-NN and the normalized \\ discounted cumulative gain \\ method (CF-kNN+NDCG)\end{tabular} & \multirow{2}{*}{MovieLens-100K}       & 0.912             & 0.928              & 0.909           & -                 \\ \cline{3-3} \cline{5-8} 
                           &                                                                                   & \begin{tabular}[c]{@{}l@{}}Fuzzy c-means with bat \\ algorithm (FCM-BAT)\end{tabular}                                                                 &                                       & 0.899             & 0.901              & 0.896           & -                 \\ \cline{3-8} 
                           &                                                                                   & \begin{tabular}[c]{@{}l@{}}Monarch butterfly optimiza-\\ tion (MBO) with deep belief \\ network (DBN)\end{tabular}                                    & \multirow{5}{*}{MovieLens-100K}       & 0.979             & 0.974              & 0.966           & -                 \\ \cline{1-3} \cline{5-8} 
\multirow{4}{*}{{[}185{]}} & \multirow{4}{*}{Bagging}                                                          & \begin{tabular}[c]{@{}l@{}}Random forest classifier \\ (RFC)\end{tabular}                                                                             &                                       & -                 & 0.930              & 0.780           & 0.850             \\ \cline{3-3} \cline{5-8} 
                           &                                                                                   & XGBoost                                                                                                                                               &                                       & -                 & 0.770              & 0.830           & 0.800             \\ \cline{3-3} \cline{5-8} 
                           &                                                                                   & \begin{tabular}[c]{@{}l@{}}Popularity-based baseline \\ (PBB) *\end{tabular}                                                                          &                                       & -                 & 0.910              & 0.120           & 0.210             \\ \cline{3-3} \cline{5-8} 
                           &                                                                                   & God-view PBB (GVPBB) *                                                                                                                                &                                       & -                 & 0.830              & 0.340           & 0.480             \\ \hline
\multirow{6}{*}{{[}182{]}} & \multicolumn{1}{l|}{\multirow{6}{*}{Boosting}}                                    & \begin{tabular}[c]{@{}l@{}}Degree-shilling-assault  \\ detection (SAD)\end{tabular}                                                                   & \multirow{6}{*}{MovieLens-100K}       & 0.747             & 0.786              & 0.816           & 0.780             \\ \cline{3-3} \cline{5-8} 
                           & \multicolumn{1}{l|}{}                                                             & Co-detector                                                                                                                                           &                                       & 0.747             & 0.887              & 0.816           & 0.850             \\ \cline{3-3} \cline{5-8} 
                           & \multicolumn{1}{l|}{}                                                             & Bayes-detector                                                                                                                                        &                                       & 0.932             & 0.947              & 0.941           & 0.944             \\ \cline{3-3} \cline{5-8} 
                           & \multicolumn{1}{l|}{}                                                             & \begin{tabular}[c]{@{}l@{}}Feature analysis protocol \\ (FAP) *\end{tabular}                                                                          &                                       & 0.529             & 0.949              & 0.530           & 0.680             \\ \cline{3-3} \cline{5-8} 
                           & \multicolumn{1}{l|}{}                                                             & \begin{tabular}[c]{@{}l@{}}Semi-shilling-assault \\ detection (SAD)\end{tabular}                                                                      &                                       & 0.930             & 0.900              & 0.941           & 0.920             \\ \cline{3-3} \cline{5-8} 
                           & \multicolumn{1}{l|}{}                                                             & \begin{tabular}[c]{@{}l@{}}Principal component analysis  \\ (PCA)*-select- user\end{tabular}                                                          &                                       & 0.931             & 0.921              & 0.941           & 0.931             \\ \hline
\end{tabular}
}
\end{table*}

On the MovieLens-1M dataset, the Artificial Bee Colony Optimization, a variant of the Random Walk approach, performed exceptionally well. This method is very good in capturing users' sequential and exploratory actions, which are important in applications like web navigation and sequential RS. The Eagle Deep Neural-based Movie RS and CbCF performed well on the MovieLens-1M dataset, demonstrating their ability to handle larger, more complicated datasets. This demonstrates the value of linear models in providing a solid baseline and in scenarios where complex relationships do not predominate. The Ensemble methods, exemplified by the Voting Classifier-based ML approaches such as RDAM and RDMA-Len Var, exhibit outstanding performance, particularly in terms of precision and recall. The RDMA-Len Var variant achieves near-perfect scores in these metrics, highlighting its effectiveness in delivering precise and reliable recommendations. The high F1-scores across these variants underscore the methods' reliability and their capacity to provide accurate recommendations across different scenarios. Boosting algorithms also exhibited strong performance especially when applied on the model Bayes-Detector. The Bayes-Detector, notably, achieves an F1-score of 0.944, effectively balancing precision and recall, and illustrating its proficiency in detecting relevant recommendations. Whereas the Baseline methods such as PBB and God-View PBB shows significantly lower results, with extremely low precision and recall scores, highlighting their inefficiency when compared to more advanced machine learning models.

This comprehensive analysis demonstrates how different algorithms perform across different models and datasets, with advanced ML models, particularly those utilizing ensemble and boosting techniques, consistently outperformed simpler baseline methods. Through more precise and personalized recommendations, techniques like RF with ensemble and boosting algorithms not only improve decision-making processes but also improve consumer experiences and support business growth. Through more precise and personalized recommendations, techniques like Random Forest and complex ensemble and boosting algorithms not only improve decision-making processes but also improve consumer experiences and support business growth. In addition to demonstrating the way ML models could transform RS; this analysis emphasizes how crucial it is to choose the right ML model-algorithm combination based on the requirements and features of the dataset. The evaluation results show how strategic model selections may significantly enhance the results, confirming the significant impact of ML in driving technological advancement and business growth. \\[0.1 cm]
\textit{\textbf{RQ 11}}: How do different algorithms like matrix factorization, tensor factorization, and factorization machines perform in terms of accuracy and scalability in RE with varying data sparsity levels? \\[0.1 cm]
Although all three algorithms are adept at handling sparse data, their suitability varies based on specific use cases. Matrix Factorization is highly efficient for sparse datasets and personalized recommendations. Tensor Factorization excels in multi-dimensional, context-aware scenarios but may vary in accuracy. Factorization Machines are highly accurate and efficient in large-scale, high-dimensional sparse data environments, making them suitable for diverse recommendation scenarios.\\[0.1 cm]
\textit{\textbf{RQ 12}}: What are the trade-offs in choosing between linear models and advanced algorithms like random walk or ensembling for RE with complex user-item interactions?\\[0.1 cm]
Choosing between linear models and advanced algorithms for RE, such as random walk or ensembling, requires balancing simplicity versus complexity and interpretability against accuracy. Linear models, noted for their computational speed and clarity of interpretation, are excellent for scenarios in which the relationships between user and item attributes are simple and linear. However, its simplicity may limit the ability to capture the complex and nonlinear patterns that frequently occur in user-item interactions, potentially resulting in underperformance in more complicated recommendation scenarios. On the other hand, advanced algorithms such as random walk and ensembling provide more robust methods for dealing with complicated interactions. Random Walk excels in discovering network-based similarities in graph-structured data and capturing nuanced relationships between users and items, although it may be more difficult to interpret than linear models. Ensembling approaches combine different models to increase prediction performance, provide higher accuracy and robustness in a variety of settings. However, this comes at the expense of greater computational complexity and reduced interpretability. In essence, the choice is based on the RE specific requirements: linear models for simpler, more interpretable results, or complicated algorithms for dealing with complex relationships at the expense of increasing computational demands and decreased transparency.\\[0.1 cm]
\textit{\textbf{RQ 13}}: How does the choice of data splitting strategy (train-test split, time-based split, cross-validation, etc.) impact the model's ability to generalize to unseen data in RE?\\[0.1 cm]
Each strategy affects the model's capacity to generalize. A train-test split provides a fundamental division but does not consider time-based trends or user/item balance. Time-based splits are required to capture temporal dynamics, although they may not be appropriate for all types of RE. While computationally complex, cross-validation provides a full evaluation across diverse subsets of data, which may lead to greater generalization. User/Item stratified splits are critical for ensuring the suggestions are fair and representative. Validation sets make it possible to fine-tune models without overfitting, which is critical for complicated models that require considerable tuning and validation. To achieve effective model training and optimal generalization to new, unseen data, the data splitting approach should be adapted to the RE specific characteristics and requirements.\\[0.1 cm]
\textit{\textbf{RQ 14}}: What are the effects of using User/Item splits compared to traditional random splits on the diversity and representation of users and items in the training and testing datasets?\\[0.1 cm]
User/Item splits compared to traditional random splits, improve the diversity and representation of users and items in training and testing datasets. Traditional random splits carry the risk that specific user profiles or item categories will be underrepresented or overrepresented in either the training or testing group. This can result in a model that is biased toward more often represented groups, thus compromising its effectiveness and fairness. User/Item splits solve this problem by guaranteeing that each user and item category represents proportionally in both databases. This method is especially useful when the dataset includes many occasionally occurring users or items. This method matches the dataset more closely with real-world distributions, resulting in better model performance and suggestions.\\[0.1 cm]
\textit{\textbf{RQ 15}}: How does the adjustment of hyperparameters like learning rate, regularization strength, and number of latent factors influence the performance of matrix factorization-based recommendation models?\\[0.1 cm]
The adjustment of hyperparameters such as learning rate, regularization strength, and the number of latent components is critical to the effectiveness of matrix factorization-based recommendation models. The learning rate determines how quickly the model learns; too high a rate may cause the model to overshoot ideal solutions, whereas too low a rate may result in slower convergence. Regularization strength helps in preventing overfitting by penalizing higher parameter values, hence preserving a balance between model complexity and generalization ability. The number of latent components determines the dimensionality of the feature space; an appropriate number captures the complexity of user-item interactions without complicating the model. Adjusting these hyperparameters can have a considerable impact on the model's accuracy, convergence time, and capacity to generalize to new data. Fine-tuning these parameters is therefore critical for optimizing the model's performance, achieving a balance between accuracy, overfitting, and computing efficiency.\\[0.1 cm]
\textit{\textbf{RQ 16}}: In what scenarios is AUC-ROC a suitable metric for RE, and what are its limitations compared to metrics like MAP or NDCG?\\[0.1 cm]
AUC-ROC is ideal for binary classification problems in RE such as predicting whether a user would like or dislike an item. It performs well in situations where distinguishing between positive and negative instances is critical, offering an indicator of a model's capacity to rank positive occurrences higher than negative ones. However, its shortcomings become obvious in ranking or relevance scenarios, which are characteristic of RE. Unlike MAP or NDCG, AUC-ROC does not take into consideration the sequence of recommendations or the various degrees of relevance between items. MAP and NDCG give better evaluations in these circumstances, focusing on the precision of top-ranked items and considering the decline in relevance.  AUC-ROC is useful for interpreting binary results, MAP and NDCG provide a more complete assessment of an RE ability to provide recommendations.

\subsection{Recommendation model deployment}
Once the model is trained and evaluated, it is deployed to the production environment where it can generate real-time recommendations for users. The true value of an RE is realized when it's deployed into a production environment and starts influencing user behavior and business metrics. Transitioning from a development environment to a production environment  [186] is a pivotal step in realizing the practical impact of RE. It allows us to gather meaningful insights, optimize performance, and make informed decisions that contribute to the achievement of business goals. Choose a technology stack for deployment (e.g., AWS, Azure, Google Cloud, or on-premises servers).  Selecting a technology stack for deploying RE [187]  involves considering factors such as scalability, flexibility, ease of deployment, and the specific requirements of the application.

\subsubsection{Containerization and cloud deployment orchestration}
To make the RE portable and consistent across environments, it can be packaged along with its dependencies into Docker containers [188]. These containers ensure that the application works the same way in any environment. Kubernetes can be used to deploy and manage these containers, automating tasks such as deployment, scaling, and lifecycle management. By using Kubernetes, the RE can dynamically scale to handle varying workloads and ensure availability and failover. Cloud service providers like AWS, Azure, or GCP can host Kubernetes-managed containers. Kubernetes helps by monitoring container health and automatically replacing unhealthy ones. It also uses load balancers to distribute incoming requests evenly across instances, preventing overload and maintaining performance. The web application frontend can be developed using frameworks such as React, Angular, or Vue.js, and integrated with a backend framework. A database should be connected to the application to store and retrieve user and recommendation data. The trained recommendation model is integrated into the application for real-time or near-real-time predictions [189]. Kubernetes also supports rolling updates for the model or inference code, ensuring that updates are applied gradually without downtime. This keeps the service continuously available, even during updates, while maintaining consistent performance. Figure 14 displays the random container orchestration methods.    

\begin{figure}[b!] 
	\centering 
	\includegraphics[width=0.53\textwidth]{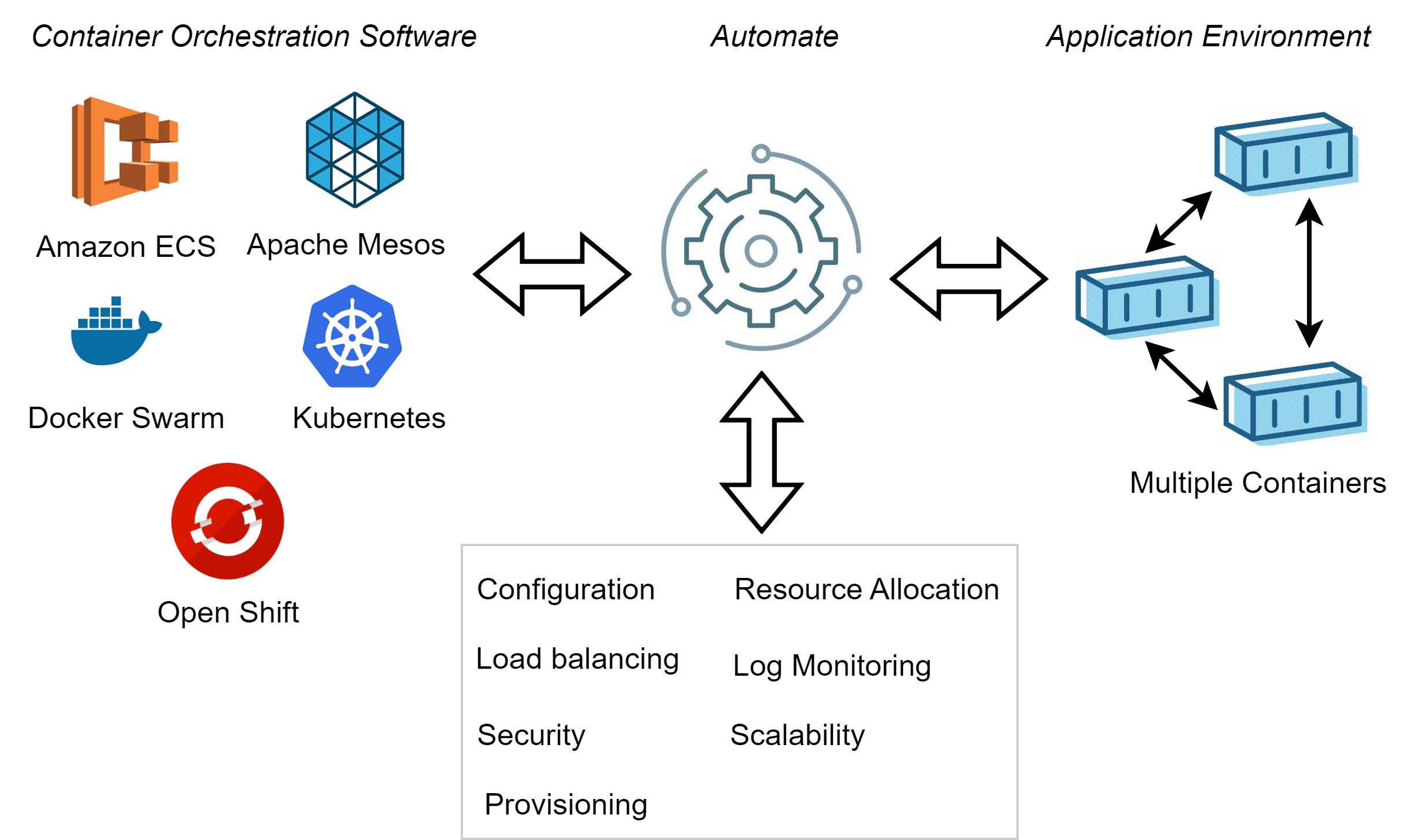}
	\\
	\caption{\textrm{Container orchestration}}
	\label{fig:14}       
\end{figure}

\subsubsection{API Integration}
In the deployment of RE, API integration plays an essential role for several reasons, significantly enhancing their effectiveness, scalability, and versatility. RE can access real-time user data from a variety of sources to make sure suggestions are accurate and relevant by enabling seamless data exchange. APIs facilitate interoperability between a wide range of services and platforms, from payment gateways to CRM tools, allowing for unified user experience. Moreover, APIs enable more personalized recommendations by integrating detailed user data into their applications [190], resulting in deeper personalization. They also enhance scalability by managing increased data loads and enabling dynamic adjustments to user demand. Moreover, APIs facilitate the development and deployment processes, allowing for the incorporation of new features and data sources with ease. By eliminating the need for custom solutions, this not only fosters innovation and flexibility but also ensures cost efficiency. Lastly, API integration allows RS to be embedded within a broader ecosystem, allowing them to create value and collaborate with others. Define API endpoints to handle user requests, process recommendations, and update user interactions. Creating an API to expose the RE model's functionality is a crucial step in integrating the model with the website or application. Select a framework that aligns with the technology stack and the language in which RE is implemented. Common choices include Flask (Python), Express (Node.js), Django (Python), or FastAPI (Python). Define the endpoints that the API will expose. For a RE, typical endpoints may include:
\begin{itemize}
    \item /recommend: Get personalized recommendations for a user.
    \item /feedback: Collect user feedback on recommendations.
    \item /update: Update user preferences or interactions.
\end{itemize} 
Implement authentication and authorization mechanisms [191] to secure the API. AWS Cognito, JWT (JSON Web Tokens), or OAuth are commonly used for this purpose. Define the formats for API requests and responses. Typically, JSON is used for data interchange. Clearly document the expected structure of requests and the format of the returned recommendations.

\subsubsection{Model loading and inference}
Implement code to load the trained recommendation model into memory. Ensure the model is ready to make predictions based on incoming requests. The implementation details can vary based on the specific recommendation model, the ML library used, and the requirements of the application. Ensure that the model loading process is efficient and consider optimizing for deployment scenarios where models need to be loaded quickly and handle multiple concurrent requests. User interface (UI) integration with a recommendation model is a critical aspect that dictates how users interact with and experience the RE. An effective recommendation model captures user inputs, such as preferences, ratings, or browsing behavior, and passes that information on to the RE. The user interface serves as a bridge between the user and the complex algorithms running behind the scenes. These recommendations are then sent back to the UI, where they are displayed in a meaningful format that is accessible to the user. Based on the model, it processes this data and applies algorithms. UI integration [192] involves designing and implementing the visual elements and interactions that allow users to interact with the recommendations generated by the system. Interactive Elements allow users to click on recommendations for more details. Implement a way for users to provide feedback on recommendations. It is common for users to experience a brief delay when interacting with the recommendation engine, especially in web or mobile applications. This occurs while the recommendation engine processes the user's data, preferences, and other relevant information to provide personalized recommendations. The loading spinner is often displayed during this time to indicate that the recommendations are being retrieved. This visual feedback is essential since it lets the user know that their request is being handled and helps them control how quickly they expect a response. By preserving engagement and lowering perceived latency, integrating a loading spinner efficiently reduces user irritation related to waiting and improves the overall user experience and shows user-friendly messages if there are issues fetching recommendations.

\subsubsection{Monitoring and logging}
Regularly review and analyze monitoring and logging data to identify trends, patterns, and potential optimizations. Conduct post-incident reviews to learn from any outages or critical issues and update monitoring and logging strategies accordingly. Periodically reassess the relevance of metrics and logs [193] to ensure they align with changing business requirements and system architecture. By implementing a robust monitoring and logging strategy, gain visibility into the health and performance of the RE enabling proactive identification of issues and continuous improvement.\\[0.1 cm]
\textit{\textbf{RQ 17}}: How does the implementation of containerization and cloud deployment orchestration technologies, like Docker and Kubernetes, impact the scalability, performance, and manageability of RE in a production environment?\\[0.1 cm]
The implementation of containerization and cloud deployment orchestration technologies like Docker and Kubernetes dramatically improves the scalability, performance, and manageability of RE in production. Containerization separates the system and its dependencies into isolated contexts (Docker containers), assuring consistency across many deployment platforms and reducing conflicts between system components. This increases the system's portability and simplifies deployment procedures. Kubernetes  [194], as an orchestration platform, successfully manages these containers by automating scaling, load balancing, and self-healing. It ensures that the RS can react dynamically to changing loads while preserving performance and availability without user intervention. Furthermore, Kubernetes automates the updating process with rolling updates, ensuring minimal downtime.\\[0.1 cm]
\textit{\textbf{RQ 18}}: What role do API integration, model loading, inference, monitoring, and logging play in enhancing the efficiency, security, and user experience of these systems?\\[0.1 cm]
API integration facilitates seamless communication between the RE and client applications, allowing for real-time personalized recommendations and efficient handling of user requests and feedback. This ensures a responsive and interactive user experience. Model loading and inference are key to the system's performance; efficient loading and optimized inference processes ensure that the system can handle multiple concurrent requests swiftly, minimizing latency. Additionally, robust authentication and authorization mechanisms within the API safeguard against unauthorized access [195], thereby enhancing security. Monitoring and logging, on the other hand, are vital for maintaining system health and performance. Regular analysis of logs and monitoring data helps in identifying trends, diagnosing issues, and optimizing the system. This proactive approach to identifying and resolving issues before they impact users significantly boosts the overall reliability and effectiveness of the RS [193] thereby enhancing user satisfaction.

\subsection{Best practices for optimizing the RE}
Best practices for optimizing a RE [196] involve maintaining data quality, personalization, scalability, continuous improvement, transparency, and seamless integration. By prioritizing these elements, RE can consistently deliver high-quality suggestions, enhancing performance, accuracy, and user satisfaction.

\subsubsection{Update user information }
Regularly update the user data [197] to capture changes in user preferences and behavior. This helps ensure that the recommendations remain relevant and up to date. User preferences and behavior can change over time and incorporating these changes into the model ensures that recommendations stay accurate and reflect the evolving interests of users. By incorporating user lifecycle changes, data versioning, and granular user segmentation into the RE data update strategy, a system can be maintained that adapts to changes in user behavior, ensuring relevant and timely recommendations over the long term.

\subsubsection{Consistently retrain the model}
As user preferences evolve and new data becomes available  [198], retraining ensures that the model stays up-to-date and continues to provide relevant recommendations. This will help improve the accuracy and effectiveness of the recommendations. Implement incremental training to update the model with new data efficiently. Train the model on the most recent data while retaining knowledge from previous training sessions to avoid unnecessary computational overhead. Determine an appropriate retraining schedule based on the rate of change in the data and the nature of the application. Consider factors such as seasonality, promotional events, or changes in user behavior that might impact the effectiveness of the model. Establish an automated pipeline for model retraining to streamline the process. Set up triggers for retraining based on specific events or performance metrics. Retrain the model when there is a significant shift in user behavior or when performance metrics indicate a decline in recommendation quality. Automation ensures that the model  [199] is consistently updated without manual intervention, reducing the risk of overlooking necessary updates. A dynamic and adaptive system can be maintained by following the retaining strategies that continue to deliver accurate and relevant recommendations to users as their preferences evolve.

\subsubsection{User feedback loop implementation}
Feedback provides valuable insights into user preferences, helping to refine the model and improve the overall user experience. This feedback can provide insights into user preferences [200], identify any issues or biases in the recommendations, and make necessary improvements. There are two types of feedback: explicit and implicit feedback. Explicit Feedback allows users to provide direct feedback on recommendations, such as ratings, likes, dislikes, or comments. Implicit Feedback Captures user behavior indirectly, such as clicks, purchases, or dwell time, to infer preferences. Design an intuitive and user-friendly interface for collecting feedback. Prompt users to provide feedback at relevant points in the user journey, such as after viewing a recommendation or completing a purchase. Offer multiple channels for feedback, including in-app surveys, email, or dedicated feedback forms. Consider leveraging social media platforms or other communication channels where users are active. Provide incentives  [201], such as discounts, loyalty points, or exclusive content, to encourage users to provide feedback. Incentives can increase the quantity and quality of feedback received. Develop a robust system for analyzing and interpreting feedback data. Use NLP and sentiment analysis to understand the sentiment behind textual feedback. Incorporate feedback data into the recommendation model training process. Use feedback  [202] to update user preferences, improve item rankings, and adjust model parameters accordingly. Regularly analyze feedback data for biases that may affect the recommendation model. Implement strategies to mitigate bias, such as reweighting feedback from underrepresented user groups.
\subsubsection{Regular observation of performance metrics}
Monitoring performance metrics is an important aspect of maintaining and optimizing RE. Regularly assessing key metrics [203] helps ensure that the system continues to provide high-quality recommendations and allows for timely identification and resolution of any issues. Clearly define the key performance metrics relevant to the RS. Common metrics include precision, recall, F1-score, mean squared error, CTRs, conversion rate, and user engagement. Establish baseline values for each metric [204] [205] based on historical performance or industry benchmarks. Baselines provide a reference point for evaluating the impact of changes and improvements. Implement real-time monitoring  [206] to quickly detect anomalies or performance degradation. Utilize monitoring tools and alerts to receive notifications when key metrics deviate from expected values. Conduct error analysis to understand the types of mistakes the RS makes. Identify patterns and commonalities in mispredictions to guide improvements.

\paragraph{A/B testing}
A/B testing is a powerful method for comparing the performance of different versions of RE, algorithms, or user interfaces. By conducting A/B tests, Data-driven decisions can be made, optimizing the user experience, and identify which changes lead to improved outcomes. Clearly define the objective of the A/B test. For RE, objectives could include improving click-through rates [207], increasing user engagement, or enhancing conversion rates. Develop hypotheses based on the changes to be tested. For example, it could be hypothesized that a new recommendation algorithm will lead to higher user engagement. Implement the changes to be tested in the test group while keeping the control group unchanged. This could involve deploying a new algorithm, tweaking parameters, or modifying the user interface. Run the A/B test for a predetermined duration to collect sufficient data. The duration depends on factors such as the volume of user interactions and the time required to observe meaningful trends. Evaluate the results of the A/B test, considering both statistical significance and practical significance. Determine whether the changes have a meaningful impact on the chosen metrics and align with the initial hypotheses. Ensuring that A/B testing [208] is conducted ethically, respecting user privacy, and obtaining informed consent where necessary. Communicate transparently with users about the purpose of the test and the changes being tested. Embrace a culture of continuous improvement by regularly conducting A/B tests to refine and optimize the RE based on evolving user needs and preferences.

\paragraph{Dashboard reporting}
Creating a dashboard for reporting is an effective way to monitor the performance of the RE and present key metrics and insights in a clear and accessible manner. A well-designed dashboard [209] can provide stakeholders with real-time information and help them make informed decisions. Clearly define the objectives of the dashboard. Understand the key questions stakeholders want to answer and the metrics that are most relevant to them. Determine the KPIs [210] that align with the goals of the RE. These may include click-through rates, conversion rates, user engagement, and algorithm performance metrics. Choose appropriate visualization tools based on the nature of the data and the preferences of the audience. Common tools include data visualization libraries (e.g., D3.js, Matplotlib, or Plotly) or dashboard platforms (e.g., Tableau, Power BI, or Google Data Studio). Implement user access control [211] to restrict access to sensitive data and ensure that each stakeholder has access only to the relevant information. A well-defined dashboard that serves as a valuable tool for monitoring and reporting on the performance of the RE. A well-designed dashboard can facilitate data-driven decision-making and contribute to the continuous improvement of the RE.

\paragraph{Periodic audits}
Conduct periodic audits of RE to assess their overall performance and identify areas for improvement. Use audit findings to guide ongoing optimization efforts. Periodic audits of the RE are essential to ensure their ongoing effectiveness, reliability, and alignment with the business goals. Audits  [212] help identify potential issues, assess the impact of changes, and verify that the system operates as intended. It provides valuable insights that can guide continuous improvement efforts and help maintain a high-quality RE.

\subsubsection{Incorporating diversity into RE}
Consider incorporating diversity in the recommendations to tailor a wider range of user preferences. This helps prevent the RE from becoming too focused on a specific set of items and ensures a more balanced user experience. Ensure that the item catalog represents a broad range of genres, categories, or attributes. Incorporate various types of items [213] to accommodate different user tastes and interests. Consider user diversity when generating recommendations. Ensure that the system caters to a variety of user-profiles and preferences. Avoid over-recommending popular items to create a more inclusive experience. Find the right balance between exploring new items and exploiting the knowledge gained from user interactions. Implement algorithms that dynamically adjust the exploration-exploitation tradeoff based on user behavior. Allow users to set preferences for the level of diversity they prefer in their recommendation and provide user-controlled diversity settings to tailor the experience to individual preferences. Diversity  [214] ensures that users are exposed to a variety of items, reducing the risk of creating filter bubbles and enhancing the potential for serendipitous discoveries.

\subsubsection{Scaling strategies in recommendations}
Scaling Strategies in RE increase user loads, grow item catalogs, and evolve user behavior. The scalability of an RE ensures that it can deliver timely and accurate suggestions even as the user base expands. 
\begin{itemize}
    \item Vertical scaling (scaling up) [215] - Increase the capacity of a single server by upgrading its resources, such as CPU, RAM, or storage. It is suitable for small to medium-sized RE with modest traffic.
    \item Horizontal scaling (scaling out) [216] - Add more machines or nodes to RS to distribute the workload. It is effective for large RE with high traffic and diverse user interactions.
    \item Load balancing [217] - Distribute incoming recommendation requests across multiple servers to balance the load. Ensures even distribution of requests, preventing overload on specific servers.
    \item Caching [218] - Cache frequently accessed recommendation results to reduce the need for recompilation. It improves response times, particularly for popular and stable recommendations.
    \item Auto-scaling [219] - Automatically adjust the number of recommendation service instances based on demand or predefined metrics. It ensures that the system dynamically adapts to fluctuating user loads.
    \item Efficient data pipelines [220] - Optimize data processing pipelines for RE to handle large volumes of user interactions and updates. Ensures timely and accurate data processing, supporting real-time recommendations. 
\end{itemize} 
Implementing a combination of these scaling strategies can help ensure that an RS remains responsive, reliable, and efficient as it scales to meet the demands of a growing user base and expanding content catalog. It is essential to choose strategies that align with the specific requirements and characteristics of the recommendation application. \\[0.1 cm]
\textit{\textbf{RQ 19}}: How do various scaling strategies like vertical scaling, horizontal scaling, load balancing, caching, auto-scaling, and efficient data pipelines individually and collectively influence the performance, scalability, and user response time of large-scale RE?\\[0.1 cm]
Each of these strategies contributes significantly to the performance, scalability, and responsiveness of RE. The impact of each technique varies depending on the system architecture, the type of data, and the application's specific requirements. Implementing a combination of these strategies can assist in ensuring that RE remains robust and efficient as it scales to meet increased demand. They ensure high availability, scalability for high query volumes [221], and improved latency, addressing the main concerns of performance, reliability, and responsiveness in large-scale RS.

\section{tools for businesses}
RE has become essential tools for businesses, driving numerous benefits such as improved customer experience, increased sales and conversion rates, and enhanced revenue. They facilitate efficient marketing spending, enable cross-selling and up-selling, and help reduce abandoned carts. These engines also streamline the checkout process, optimize multi-channel sales, implement dynamic pricing strategies, and encourage subscription and repeat business. The symbiotic relationship between businesses and customers is strengthened through personalized shopping experiences for customers and data-driven, actionable insights for businesses. With ongoing technological advancements, the functionalities and effectiveness of RE are expected to further advance, providing increasingly effective tools and solutions across a variety of industries.

\subsection{Improved customer experience}
RE can significantly contribute to improving customer experience across various touchpoints in a business. Personalization in RE analyzes customer behavior and preferences to provide personalized product suggestions. This level of personalization enhances the overall customer experience by showing relevant items that align with individual interests. Enhanced recommendations [222] keeps customers engaged with the platform, leading to longer sessions, increased page views, and a higher likelihood of conversions. Improved Customer Retention develops enhanced user loyalty and encourages repeat business by suggesting products that align with the customer's interests and needs. Customers are more likely to find value, satisfaction, and convenience, contributing to increased loyalty and positive word-of-mouth referrals. 

\subsection{Increase in sales and conversions}
RE give users extremely relevant and customized suggestions, which promptly boost sales and conversions. Businesses can improve their RS to better understand and anticipate client preferences and behavior by utilizing ML algorithms and data analytics. By presenting customers with products or content that they are more likely to find interesting, this specific approach improves user experience and improves their chances that they will engage or make a purchase. Higher sales right away are the outcome, but over time, customer loyalty and retention are also enhanced. By maintaining the recommendations' relevance and efficacy, ongoing optimization based on user feedback and interaction data enables dynamic adaptation to shifting consumer trends, guaranteeing a steady increase in sales and conversions.

\subsubsection{Cross-selling and up-selling}
Cross-selling in RE analyzes user behavior and purchase history to suggest complementary products that align with the customer's current selection. For example, if a customer is buying a camera, the RE might suggest related accessories like a camera bag, lenses, or a tripod. Up-selling works by recommending higher-priced or premium versions of products, RE [223] aim to increase the average order value. For instance, if a customer is looking at a particular smartphone, the RE suggests an upgraded model with additional features.

\subsubsection{Reduced abandoned carts }
When users abandon their shopping carts  [224], it often indicates hesitations or uncertainties. RE can intervene by displaying relevant products, offering discounts, or providing additional information that addresses the customer's concerns. To encourage users to complete their purchases, RE can provide incentives such as limited-time discounts, free shipping, or exclusive offers on the items in their cart.

\subsubsection{Optimized checkout process}
RE works in tandem with efforts to streamline and simplify the checkout process. Implementation of one-click actions reduces friction [225], enhancing the overall user experience and reducing the likelihood of abandonment.

\subsection{Efficient marketing spend}
Product recommendations enable businesses to target their marketing efforts more effectively. By understanding customer preferences [110], companies can allocate resources to the most relevant audience segments. By delivering personalized content and promotions, businesses can achieve a higher return on investment (ROI) for their marketing campaigns. Ad spending  [226] is directed toward audiences more likely to convert. Achieving efficiency in marketing spending involves strategic planning, data-driven decision-making, and continuous optimization.  

\subsection{Enhanced revenue}
RE plays a pivotal role in enhancing revenue by providing personalized, relevant, and timely suggestions to users. The ability to guide customers through their journey, address hesitations, and promote strategic sales opportunities contributes to increased sales, customer loyalty, and overall revenue growth. Continuous optimization and adaptation [227] based on data-driven insights are essential for sustained success in leveraging RE for revenue enhancement.

\subsubsection{Multi-channel sales optimization}
Consistent Cross-Channel Experience [228] ensures that recommendations provide a consistent experience across online and offline channels. In-Store Recommendations Extend RE to in-store displays or personalized shopping experiences to enhance the overall customer journey.

\subsubsection{Dynamic pricing strategies}
Personalized pricing, also known as dynamic or individualized pricing [229], is a strategy that adjusts the cost of a product or service to an individual customer based on various factors such as their purchasing behavior, demographics, location, and other relevant data. This strategy aims to optimize revenue by setting prices that reflect the perceived value to each customer. Businesses should approach personalized pricing ethically, ensuring compliance with legal standards and building trust with customers through clear communication and value-driven strategies. Discount optimization is the strategic process of determining and applying discounts in a way that maximizes the impact on sales, profitability, and customer satisfaction. Effective discount optimization involves careful consideration of various factors, including customer behavior, pricing strategies, and business goals. Offer personalized discounts on recommended products to incentivize purchases without compromising profitability. The key is to continuously analyze data  [230], stay attuned to customer preferences, and iteratively refine discount strategies for long-term success.

\subsubsection{Subscription and repeat business}
Subscription and repeat business strategies focus on cultivating long-term relationships with customers by encouraging them to subscribe to services or products and making repeat purchases. These strategies are vital for sustaining revenue, building customer loyalty, and maximizing customer lifetime value. RE [231] can promote subscription-based services or products, leading to recurring revenue. By recommending products based on past purchases, businesses can encourage repeat buying behavior.

\begin{table*}[t!]\centering
\caption{\textrm{Summary of case study}}
\footnotesize
\label{tab:9}       % Give a unique label 
\resizebox{\linewidth}{!}{%
 \begin{tabular}{|l|l|l|l|l|l|}
\hline
\multicolumn{1}{|c|}{\textbf{Article}} & \multicolumn{1}{c|}{\textbf{Company}} & \multicolumn{1}{c|}{\textbf{Implementation}}                                                                                                    & \multicolumn{1}{c|}{\textbf{Impact}}                                                                                                          & \multicolumn{1}{c|}{\textbf{Challenges}}                                                                                     & \multicolumn{1}{c|}{\textbf{Solutions}}                                                                                     \\ \hline
{[}232{]}                              & Amazon                                & \begin{tabular}[c]{@{}l@{}}CF \\ and deep learning for \\ personalized product \\ recommendations.\end{tabular}            & \begin{tabular}[c]{@{}l@{}}Drives up to 35\% of \\ total sales, enhancing \\ loyalty and satisfaction.\end{tabular}                           & \begin{tabular}[c]{@{}l@{}}Managing vast inventory \\ and engaging new users.\end{tabular}                                   & \begin{tabular}[c]{@{}l@{}}Hybrid models combining \\ content-based filtering \\ with CF.\end{tabular} \\ \hline
{[}233{]}                              & Netflix                               & \begin{tabular}[c]{@{}l@{}}Viewing history, search \\ patterns, and temporal \\ data for content \\ suggestions.\end{tabular}                   & \begin{tabular}[c]{@{}l@{}}Recommendations lead \\ to over 80\% of watched \\ content, boosting user \\ retention.\end{tabular}               & \begin{tabular}[c]{@{}l@{}}Serving a global audience \\ with diverse preferences.\end{tabular}                               & \begin{tabular}[c]{@{}l@{}}R\&D investment, like the \\ dynamic optimizer, for \\ personalized streaming.\end{tabular}      \\ \hline
{[}234{]}                              & Spotify                               & \begin{tabular}[c]{@{}l@{}}CF, \\ NLP, and audio analysis \\ for music and playlist \\ recommendations.\end{tabular}       & \begin{tabular}[c]{@{}l@{}}\enquote{Discover Weekly} and \\ similar features significantly \\ boost engagement and \\ subscriber growth.\end{tabular} & \begin{tabular}[c]{@{}l@{}}Blending new music \\ discovery with user \\ preferences.\end{tabular}                            & \begin{tabular}[c]{@{}l@{}}Introducing discovery \\ features and adjusting the \\ recommendation engine.\end{tabular}       \\ \hline
{[}235{]}                              & YouTube                               & \begin{tabular}[c]{@{}l@{}}User interaction data \\ leveraged through machine \\ learning for tailored \\ content suggestions.\end{tabular}     & \begin{tabular}[c]{@{}l@{}}Over 70\% of viewing \\ time is driven by the \\ recommendation engine,\\ enhancing user engagement.\end{tabular}  & \begin{tabular}[c]{@{}l@{}}Addressing echo \\ chambers and ensuring \\ content diversity.\end{tabular}                       & \begin{tabular}[c]{@{}l@{}}Algorithm adjustments \\ and user feedback \\ incorporation.\end{tabular}                        \\ \hline
{[}236{]}                              & LinkedIn                              & \begin{tabular}[c]{@{}l@{}}Data on user activities \\ and profile information \\ for job, connections, and \\ content suggestions.\end{tabular} & \begin{tabular}[c]{@{}l@{}}Enhances professional \\ networking and content \\ discovery, contributing to \\ platform engagement.\end{tabular} & \begin{tabular}[c]{@{}l@{}}Accurately matching job \\ opportunities and content \\ with professional interests.\end{tabular} & \begin{tabular}[c]{@{}l@{}}Advanced matching \\ algorithms and continuous \\ learning from user interaction.\end{tabular}   \\ \hline
\end{tabular}
}
\end{table*}

\subsection{Case Study: Implementation of RE}
Recommendation Engines have completely transformed user experiences across various industries by increasing consumer engagement and enhancing sales. Many studies have been conducted on the deployment and effects of recommendation algorithms on well-known services including YouTube, LinkedIn, Netflix, Amazon, and Spotify. These platforms have greatly improved user engagement, consumer satisfaction, and their capacity to forecast user preferences over time by utilizing advanced algorithms to produce personalized user experiences. Table 9 illustrates the case study of RE on digital platforms. The recommendation algorithms' ongoing development and improvement to handle scalability, diversity, and the personalization-versus-privacy conundrum is what unites these case studies. These businesses have successfully enhanced the promise of RE to drive growth, improve consumer experience, and maintain a competitive edge in their respective industries by using creative approaches to data analysis and ML.

\textit{\textbf{RQ 20}}: What are the key strategies and ethical considerations associated with the use of RE in various industries, and how can businesses influence ongoing technological advancements to further enhance their effectiveness? \\[0.1 cm]
RE employs several key methods, such as dynamic pricing, cross-selling, personalization, checkout process optimization, decreasing cart abandonment, and subscription schemes. Ethical considerations encompass transparency, data privacy, fairness and bias mitigation, user control, avoiding manipulation, avoiding over-personalization, and maintaining a feedback loop with users. Businesses can influence technological innovations to increase efficacy using sophisticated algorithms, big data, artificial intelligence, real-time and contextual recommendations, integrating RE across channels, investigation of voice and conversational interfaces, adherence to AI ethics, experimentation, and A/B testing to continuously refine recommendations. These strategies and considerations apply across various industries to improve customer experience and drive revenue growth.

\section{How RE propels business expansion?}
Building a business model [237] with data-driven insights from the core recommendation involves advancing data to make informed decisions, optimizing processes, and drive strategic initiatives. Regularly reassess and iterate on the model to adapt to changing business environments and evolving data landscapes. These can be approached through three key strategies, as given below.

\subsection{Personalization and targeted marketing}
Personalization and targeted marketing [238] are crucial components of a data-driven strategy, allowing businesses to improve their interactions with customers based on individual preferences, behaviors, and characteristics. Enhance user engagement and boost conversion rates by delivering highly personalized experiences. Personalization involves customizing experiences, products, or services to meet the specific needs and preferences of individual users or customers. Targeted Marketing involves directing marketing efforts and messages to specific segments of the audience that are more likely to be interested in a product or service. The implementation strategies [239] followed are leveraging RE insights to create personalized user profiles, marketing communications, emails, and promotions based on user preferences and implementing targeted advertising campaigns that align with individual user behaviors. By combining personalization and targeted marketing strategies [240], businesses can create more meaningful and effective interactions with their audience, leading to user engagement, Increased customer satisfaction and loyalty, Higher conversion rates due to relevant and timely offers, Improved customer retention and repeat business, and ultimately, improved business outcomes. 

\subsection{Optimizing product recommendations}
Optimizing product and content recommendations is crucial for maximizing sales and enhancing user satisfaction. This process involves ensuring that the data driving recommendations is accurate, up-to-date, and of high quality. Regular analysis of user interactions with recommended products or content allows for continuous refinement of recommendation strategies, incorporating user feedback and adapting to evolving market dynamics. Testing and optimization of algorithms play a key role in improving relevance, while A/B testing is used to evaluate the effectiveness of various strategies. The benefits include increased average transaction values through cross-selling and up-selling, reduced cart abandonment rates, and enhanced user engagement with the platform.

\subsection{Data monetization and partnerships}
Data monetization involves leveraging the data that a business collects to generate revenue or create new business opportunities. Partnerships enable businesses to collaborate with other organizations to extract value from their data. Data Monetization [241] recognize the types of data that the business collects have potential value. Assess customer data, transaction data, behavioral data, or any unique datasets the business possesses. Then conduct market research to understand industries or sectors that could benefit from the data. Explore opportunities to monetize RE data through partnerships with third-party advertisers or complementary businesses. Offer premium subscription models that provide enhanced personalization or exclusive content. Collaborate with other platforms or businesses [242] to share recommendation insights for mutual benefit. Invest in secure and scalable data storage and processing systems and explore cloud-based solutions for flexibility and scalability. \par 
Robust security measures must be followed while sharing data with partners. Establish secure data transfer protocols and Implement access controls to protect sensitive information. Regularly assess the effectiveness of partnerships by establishing KPIs for partners. Conduct regular reviews [243] to ensure alignment with business goals. Adhere to data protection regulations by clearly communicating the ethical principles governing data sharing. By effectively implementing data monetization strategies and building partnerships, businesses can unlock new revenue streams, gain valuable insights, and foster collaborative innovation in the evolving landscape of data-driven opportunities. Each of these strategies contributes to the overall goal of building a business model that maximizes the value derived from data-driven insights. By focusing on personalization, optimizing recommendations, and exploring opportunities for data monetization and partnerships, businesses can create a holistic approach that enhances user experience, drives revenue growth, and establishes a sustainable advantage in the market.\\[0.1 cm]
\textit{\textbf{RQ 21}}: What are the measurable impacts on customer engagement, revenue growth, and market positioning?\\[0.1 cm]
The measurable impacts of RE on customer engagement include increased session duration, higher page views, improved click-through rates (CTRs), and enhanced user retention. On revenue growth, RE led to an increase in conversion rates, raise average order values (AOV), reduce cart abandonment, and promote subscription and repeat business. In terms of market positioning, businesses get a competitive edge by cultivating brand loyalty and utilizing data-driven insights to make well-informed strategic decisions, enhancing their market position. These effects are quantifiable through KPIs like as session duration, page visits, CTRs, conversion rates, AOV, cart abandonment rates, user retention rates, and revenue growth, which can be continuously analyzed and optimized.

\section{Future of RS}
There is plenty of scope for RS in various fields due to the advancements in AI. Some of them are discussed below.
\begin{itemize}
    \item Continued advancements in AI and ML will enhance the sophistication of recommendation algorithms. Deep learning techniques and neural networks will contribute to more accurate predictions.
    \item RS will become more context-aware, considering factors such as user location, time of day, and current activities to provide even more relevant suggestions.
    \item Future RS may extend beyond traditional content recommendations to include multi-modal suggestions, incorporating images, videos, and other interactive elements.
    \item There will be a growing emphasis on making recommendation algorithms more explainable and transparent, helping users understand why specific recommendations are made and building trust.
    \item Future RS will likely incorporate privacy-preserving techniques to balance personalization with data protection.
    \item Integrating RS with voice-activated and conversational interfaces will become more prevalent, allowing users to receive recommendations through natural language interactions.
    \item Real-time recommendations will be optimized for immediacy, providing users with up-to-the-moment suggestions based on their current context and preferences.
    \item CF, a key technique in RS, will see advancements with the integration of more sophisticated algorithms and hybrid models that combine various recommendation approaches.
    \item The application of RS will expand across diverse industries, including healthcare, education, finance, and more, tailoring personalized experiences in different domains.
    \item RS will likely integrate with emerging technologies such as augmented reality (AR) and virtual reality (VR) to provide immersive and personalized experiences.
    \item RS works on strategies to handle the cold start problem, where there is limited data for new users or items. Implement strategies, such as content-based recommendations, CF, or hybrid RE based on user demographics, to provide meaningful suggestions for new users. 
\end{itemize} 

\section{Conclusion}
This research underscores the pivotal role of ML in shaping the evolution and effectiveness of RS, particularly in commercial business environments. It illuminates the multifaceted contributions of ML in crafting and refining these systems, from data sourcing to feature engineering and crucial evaluation metrics. Furthermore, this research highlights the profound impact of advanced RE across diverse domains, enhancing user experiences, streamlining information discovery, and driving business success through increased sales and revenue. As users increasingly expect personalized and intuitive online experiences, the study emphasizes the significance of understanding and harnessing ML in RS to unlock the potential of personalized recommendations. This study also highlighted several encouraging recommendations for future research, such as improvements in deep learning models, ethical issues with RS, and scalability challenges to ensure the continued growth and ethical deployment of RS in the digital environment. The objective of this research is to serve as a valuable resource for both researchers and practitioners seeking to navigate the dynamic world of RS and ML techniques for commercial business prospects.

\section*{Acknowledgement} 
This research was supported by the Digital Innovation Hub project supervised by  the Daegu Digital Innovation Promotion Agency(DIP) grant funded by the Korean government (MSIT and Daegu Metropolitan City) in 2024 (No. DBSD1-04, Smart management system for preventing lonely deaths of elderly people living alone based on automatic meter reading information and CCTV access information), ※ MSIT: Ministry of Science and ICT and was also supported by the BK21 FOUR project (AI-driven Convergence Software Education Research Program) funded by the Ministry of Education, School of Computer Science and Engineering, Kyungpook National University, Korea (4199990214394).

 \vskip -2\baselineskip plus -1fil 
\begin{IEEEbiography}[{\includegraphics[width=1in,height=1.25in,clip,keepaspectratio]{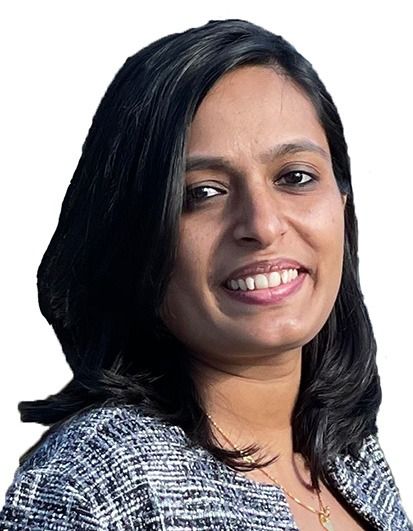}}]{Kapilya Gangadharan} received her Ph.D. in Computer Science and Engineering from Saveetha Institute of Medical and Technical Sciences, India. She is currently an adjunct faculty in the department of research and development at the same institution.  Additionally, she is an independent researcher engaged in various research and development with Talhance research laboratory. Previously, she was working with Fidelity National Information Services, Durham, USA. Her current research includes object detection and classification, image segmentation, gesture recognition, interactive recommender systems, and deep learning.
\end{IEEEbiography} 

\vskip -2\baselineskip plus -1fil
\begin{IEEEbiography}[{\includegraphics[width=1in,height=1.25in,clip,keepaspectratio]{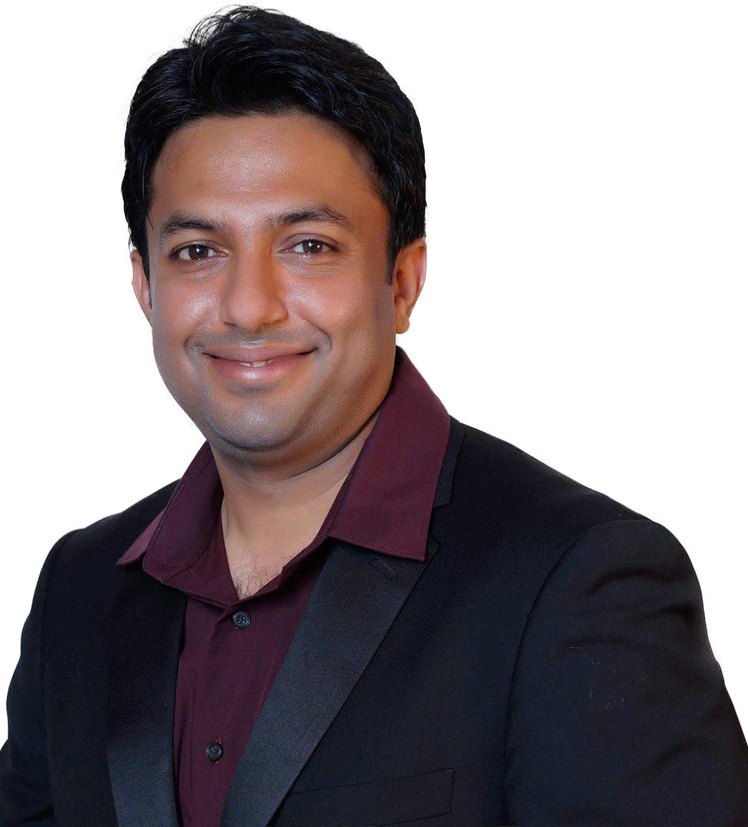}}]{Anoop Purandaran} is a Lead Software Engineer at Lowe's, NC, USA with an extensive career spanning 16 years in the Information Technology. He holds a Bachelor's degree in Computer Science and Technology from the Cochin University of Science and Technology, Kerala, India. His research interests have been centered around search and recommendation systems, particularly in understanding how well search results provided by engines align with user queries and intent. His work aims at improving search relevance and performance through linguistic analysis, ranking algorithms, and the consideration of contextual factors such as user behaviour analytics, location data, popularity metrics, and search history. Currently, he is dedicated to advancing Large Language Models, exploring their potential to transform natural language understanding and generation in complex, real-world applications.
\end{IEEEbiography} 

 \vskip -2\baselineskip plus -1fil
\begin{IEEEbiography}[{\includegraphics[width=1in,height=1.25in,clip,keepaspectratio]{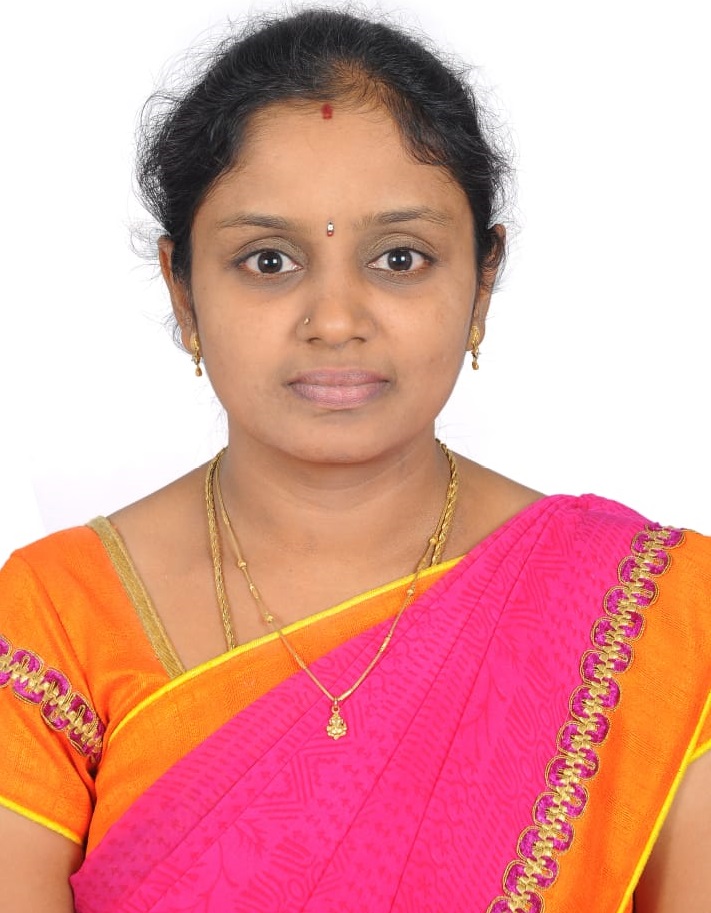}}]{K.Malathi} is associated with Saveetha Institute of Medical and Technical Science as Associate Professor in the Department of Computer Science and Engineering. She received her doctorate in Saveetha Institute of Medical and Technical Sciences, Chennai. She has 14 years of teaching and 1.5 years of industry experience as a Database Administrator. She has published more than 60 research journals.
\end{IEEEbiography}

\vfill

\vskip -2\baselineskip plus -1fil
\begin{IEEEbiography}[{\includegraphics[width=1in,height=1.25in,clip,keepaspectratio]{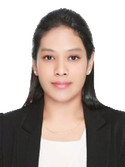}}]{Barathi Subramanian}  received her Ph.D. in Computer Science and Engineering from Kyungpook National University, Daegu, South Korea. Her research focuses on image segmentation, object classification, anomaly detection, and gesture recognition using advanced machine learning techniques. Barathi has significantly contributed to these fields by developing lightweight models for efficient computing and applying deep learning to various practical applications, from medical image analysis to real-time gesture-based systems.
\end{IEEEbiography}

\vskip -2\baselineskip plus -1fil
\begin{IEEEbiography}[{\includegraphics[width=1in,height=1.25in,clip,keepaspectratio]{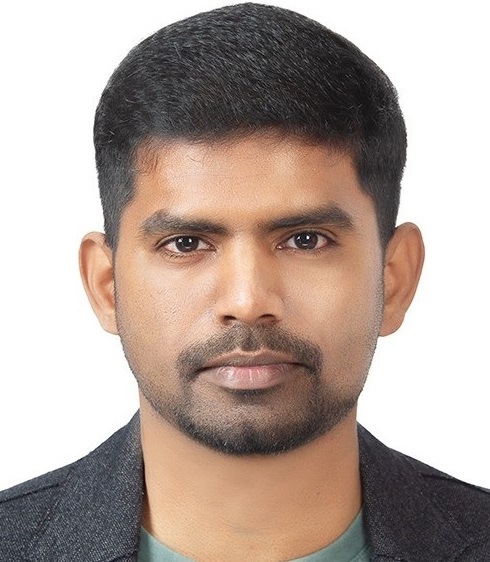}}]
{Rathinaraja Jeyaraj} (Senior Member, IEEE) earned a Ph.D. in Information Technology from the National Institute of Technology Karnataka in India. His current research interests include developing machine and deep learning models for computer vision, natural language processing, and cyber-physical systems applications. He has contributed significant journals and books to these fields. Home page: \url{https://jrathinaraja.co.in/}
\end{IEEEbiography} 

\vskip -2\baselineskip plus -1fil
\begin{IEEEbiography}[{\includegraphics[width=1in,height=1.25in,clip,keepaspectratio]{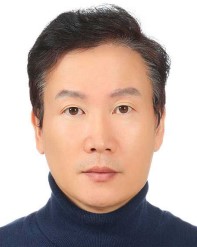}}]
{Soon Ki Jung} (Senior Member, IEEE) received the Ph.D. degree in computer science from KAIST, in 1997. From 1997 to 1998, he was a Research Associate with the University of Maryland Institute for Advanced Computer Studies (UMIACS). Since 1998, he has been with the School of Computer Science and Engineering, Kyungpook National University (KNU), Daegu, South Korea, where he is currently a professor. From 2001 to 2002, he was a Research Associate, and from 2008 to 2009, he was a Visiting Faculty with the IRIS Computer Vision Laboratory, University of Southern California. He is the author of over 200 articles on computer vision and graphics. He holds more than 20 patents deriving from his research. His research interests include improving the understanding and performance of intelligent vision systems and VR/AR systems, mainly through the application of 3D computer vision, computer graphics, visualization, and HCI. He serves as the Vice President for the Korean Computer Graphics Society, the Korean HCI Society, and the Korean Multimedia Society.
\end{IEEEbiography}  

\vfill
\EOD
\end{document}